\documentclass[preprint2]{aastex}
\usepackage{amsmath}

\newcommand{\hi}{H\,{\sc i}}

\newcommand{\co}{$^{12}$CO\,$J$=3-2}
\newcommand{\cooz}{$^{12}$CO\,$J$=1-0}
\newcommand{\corat}{$^{12}$CO\,$J$=3-2/$J$=1-0}

\newcommand{\moltotsd}{${\Sigma}_{\rm H_2 (CO\,J=3-2)}/{\Sigma}_{\rm HI + H_2 (CO\,J=1-0)}$}
\newcommand{\molsd}{${\Sigma}_{\rm H_2 (CO\,J=3-2)}$}

\newcommand{\himolsd}{${\Sigma}_{\rm HI + H_2 (CO\,J=1-0)}$}

\newcommand{\MB}{$M_{\rm B}$}

\newcommand{\Msun}{~${\cal M}_{\sun}$}

\newcommand{\MHI}{${\cal M}_{\rm HI}$}

\newcommand{\kms}{~km\,s$^{-1}$}
\newcommand{\kkms}{km\,s$^{-1}$}
\newcommand{\kelkms}{~K\,km\,s$^{-1}$}

\newcommand{\vsys}{$v_{\rm sys}$}
\newcommand{\pI}{Paper~{\rm I}}

\shorttitle{\co{} in Nearby Spiral Galaxies}
\shortauthors{Warren, et al.}

\slugcomment{{\sc Accepted to ApJ:} March 2, 2010}

\begin{document}
\title{The JCMT Nearby Galaxies Legacy Survey {\sc II}: Warm Molecular Gas and Star Formation in Three Field Spiral Galaxies}
\author{B. E. Warren\altaffilmark{1}\footnotemark{}\footnotetext{BEW is currently at the International Centre for Radio Astronomy Research, M468, University of Western Australia, 35 Stirling Highway, Crawley WA 6009, Australia} , C. D. Wilson\altaffilmark{1}, 
F. P. Israel\altaffilmark{2}, 
S. Serjeant\altaffilmark{3},
G. J. Bendo\altaffilmark{4},
E. Brinks\altaffilmark{5}, 
D. L. Clements\altaffilmark{4}, 
J. A. Irwin\altaffilmark{6}, 
J. H. Knapen\altaffilmark{7},
J. Leech\altaffilmark{8}, 
H. E. Matthews\altaffilmark{9}, 
S. M\"uhle\altaffilmark{10}, 
A. M. J. Mortimer\altaffilmark{11},
G. Petitpas\altaffilmark{12}, 
E. Sinukoff\altaffilmark{1}, 
K. Spekkens\altaffilmark{13},
B. K. Tan\altaffilmark{8}, 
R. P. J. Tilanus\altaffilmark{14,15},
A. Usero\altaffilmark{16}, 
P. P. van der Werf\altaffilmark{2}, 
C. Vlahakis\altaffilmark{2}, 
T. Wiegert\altaffilmark{17}, 
and M. Zhu\altaffilmark{18}
}

\altaffiltext{1}{Department of Physics and Astronomy, McMaster University, Hamilton, ON, Canada, L8S 4M1; bradley.warren@icrar.org, wilson@physics.mcmaster.ca, sinukoej@muss.cis.mcmaster.ca}
\altaffiltext{2}{Sterrewacht Leiden, Leiden University, PO Box 9513, 2300 RA Leiden, The Netherlands; israel@strw.leidenuniv.nl, pvdwerf@strw.leidenuniv.nl, vlahakis@strw.leidenuniv.nl}
\altaffiltext{3}{Department of Physics \& Astronomy, Open University, Milton Keynes, Mk7 6AA England}
\altaffiltext{4}{Astrophysics Group, Imperial College London, Blackett Laboratory, Prince Consort Road, London SW7 2AZ United Kingdom; g.bendo@imperial.ac.uk, d.clements@imperial.ac.uk}
\altaffiltext{5}{Centre for Astrophysics Research, University of Hertfordshire, College Lane, Hatfield  AL10 9AB, United Kingdom; E.Brinks@herts.ac.uk}
\altaffiltext{6}{Department of Physics, Engineering Physics and Astronomy, Queen's University, Kingston, ON, Canada; irwin@astro.queensu.ca}
\altaffiltext{7}{Instituto de Astrof\'isica de Canarias, E-38200 La Laguna, Spain; jhk@iac.es}
\altaffiltext{8}{Department of Physics, University of Oxford, Keble Road, Oxford OX1 3RH, UK; jxl@astro.ox.ac.uk, tanbk@astro.ox.ac.uk}
\altaffiltext{9}{National Research Council Canada, Herzberg Institute of Astrophysics, DRAO, P.O. Box 248, White Lake Road, Penticton, BC V2A 6J9, Canada; henry.matthews@nrc-cnrc.gc.ca}
\altaffiltext{10}{Joint Institute for VLBI in Europe, Postbus 2, 7990 AA Dwingeloo, The Netherlands; muehle@jive.nl}
\altaffiltext{11}{Scottish Universities Physics Alliance, Institute for Astronomy, Royal Observatory Edinburgh, Royal Observatory, Blackford Hill, Edinburgh, EH9 3HJ, UK; ajm@roe.ac.uk}
\altaffiltext{12}{Harvard-Smithsonian Center for Astrophysics, Cambridge, MA 02138; gpetitpa@cfa.harvard.edu}
\altaffiltext{13}{Department of Physics, Royal Military College of Canada, PO Box 17000, Station Forces, Kingston, K7K 4B4 Ontario, Canada; Kristine.Spekkens@rmc.ca}
\altaffiltext{14}{Joint Astronomy Centre, 660 N. A'ohoku Pl., Hilo, Hawaii, 96720, USA; r.tilanus@jach.hawaii.edu, m.zhu@jach.hawaii.edu}
\altaffiltext{15}{Netherlands Organisation for Scientific Research, The Hague}
\altaffiltext{16}{Observatorio Astron\'omico Nacional, C/ Alfonso XII 3, Madrid 28014, Spain; a.usero@oan.es}
\altaffiltext{17}{Department of Physics and Astronomy, University of Manitoba, Winnipeg, Manitoba R3T 2N2, Canada; wiegert@physics.umanitoba.ca}
\altaffiltext{18}{National Astronomical Observatories, Chinese Academy of Sciences, A20 Datun Road, Chaoyang District, Beijing, China; mz@bao.ac.cn}

\begin{abstract}
We present  the results of large-area \co{} emission mapping of three nearby field galaxies, NGC\,628, NGC\,3521, and NGC\,3627, completed at the James Clerk Maxwell Telescope as part of the Nearby Galaxies Legacy Survey.  These galaxies all have moderate to strong \co{} detections over large areas of the fields observed by the survey, showing resolved structure and dynamics in their warm/dense molecular gas disks.  All three galaxies were part of the Spitzer Infrared Nearby Galaxies Survey sample, and as such have excellent published multi-wavelength ancillary data.  These data sets allow us to examine the star formation properties, gas content, and dynamics of these galaxies on sub-kiloparsec scales.  We find that the global gas depletion times for dense/warm molecular gas in these galaxies is consistent with other results for nearby spiral galaxies, indicating this may be independent of galaxy properties such as structures, gas compositions, and environments.  Similar to the results from the THINGS \hi{} survey, we do not see a correlation of the star formation efficiency with the gas surface density consistent with the Schmidt-Kennicutt law.  Finally, we find that the star formation efficiency of the dense molecular gas traced by \co{} is potentially flat or slightly declining as a function of molecular gas density, the \corat{} ratio (in contrast to the correlation found in a previous study into the starburst galaxy M83), and the fraction of total gas in molecular form.

\end{abstract}


\section{Introduction}
\label{sec:intro}

Gas and dust in disk galaxies are intimately linked to almost all aspects of their evolution, dynamics, and appearance.  For example, the  interstellar medium (ISM) influences star formation and evolution, while these processes in turn have an effect on the distribution and turbulence of the ISM.  One of the most important links is that between the molecular clouds, which provide the fuel for star formation, and the location and rate of star formation.  However the underlying physical process that drives this relation is still poorly understood \citep{sch08}.  A recent detailed study by \citet{ler08}, using the high resolution \hi{} maps from THINGS \citep{wal08}, found that many of the commonly used prescriptions do not predict the star formation well.  Tracing {\em all} components of the ISM across a wide variety of galaxies can lead to a greater understanding of the physics of the star formation process, its dependence on the properties of the host galaxy's ISM, the evolution of galaxies, and the distribution of dark matter.

One of the limiting factors in understanding the physics of star formation is our ability to trace the molecular gas clouds within galaxies.  Much work has been done recently to examine the molecular gas within individual and small samples of nearby galaxies using various molecular transition lines that trace different gas densities and temperatures.  The most extensive observations thus far have used the millimeter \cooz{} transition to map molecular gas, including large surveys of nearby galaxies with single-dish telescopes and interferometers \citep[e.g.][]{you95,hel03,kun07}.  This transition is most sensitive to cold, lower density molecular gas, which does not appear to trace star formation on a one-to-one basis \citep{gre05}. Recent studies such as \citet{ion09} have found that higher order transitions such as the \co{} line more closely follow global star formation, nearly linearly over five orders of magnitude.

The James Clerk Maxwell Telescope (JCMT) is a 15~meter single dish submillimeter telescope operated by the Joint Astronomy Centre (JAC) on Mauna Kea, Hawai'i, which after a recent major upgrade is undertaking the JCMT Legacy Survey\footnote{http://www.jach.hawaii.edu/JCMT/surveys/} (JLS) a series of seven surveys from planetary to cosmological scales.  One of these, the JCMT Nearby Galaxies Legacy Survey (NGLS), is the first large submillimeter survey of nearby galaxies at $\sim$15\arcsec{} spatial resolution.  Taking advantage of new instrumentation on the JCMT, the survey will observe a well-selected sample of 155 nearby galaxies (within 25~Mpc) at 850~$\mu$m and 450~$\mu$m and in the \co{} line.  The NGLS data set will be a powerful tool for studying the physics of the dusty ISM in galaxies as well as the interplay between star formation and the ISM.  With a large, well-selected sample, we will be able to search for variations in the physical properties of the ISM as a function of galaxy type, metallicity, star formation rate, mass, and environment.  By having limited our sample galaxies to distances closer than 25~Mpc, we will be able to study spatial scales of 0.2-2~kpc and to search for variations inside a single galaxy as well as between galaxies.

\begin{deluxetable}{llccccccccc}
\tabletypesize{\scriptsize}
\tablecaption{Summary of Previously Measured Galaxy Properties.
  \label{tab:prop}}
\tablewidth{0pt}
\tablehead{\colhead{Name} & \colhead{$\alpha$(J2000.0)} & \colhead{$\delta$(J2000.0)} & \colhead{Type} & \colhead{\vsys{}} & \colhead{$d$} & \colhead{\MB{}} & \colhead{\MHI{}} & \colhead{${\cal M}_{\rm H_{2}}$} & \colhead{$\log L_{{\rm H}\alpha}$} & \colhead{$i$} \\

    &  &  &  & \colhead{(\kkms{})} & \colhead{(Mpc)} & \colhead{(mag)} & \colhead{($10^{8}$\Msun{})} & \colhead{($10^{8}$\Msun{})} & \colhead{($\log$\,erg\,s$^{-1}$)} & \colhead{(\degr{})} \\
 
 \colhead{(1)} & \colhead{(2)} & \colhead{(3)} & \colhead{(4)} & \colhead{(5)} & \colhead{(6)} & \colhead{(7)} & \colhead{(8)} & \colhead{(9)} & \colhead{(10)} & \colhead{(11)} }
\startdata
 NGC\,628  & $01\,36\,41.77$ & $+15\,47\,00.5$ & SA(s)c    & 648 &  7.3 & -19.58 & 38.0 & 6.3 & 40.87 & 7.0 \\

 NGC\,3521 & $11\,05\,48.88$ & $-00\,02\,05.7$ & SAB(rs)bc & 778 & 10.7 & -19.85 & 80.2 & 19.5 & 41.00 & 72.7 \\

 NGC\,3627 & $11\,20\,15.03$ & $+12\,59\,29.6$ & SAB(s)b   & 697 &  9.4 & -20.44 & 8.18 & 41.2 & 41.11 & 61.8 \\

\enddata
\tablecomments{Col. (1): Galaxy name. 
Col. (2) and (3): J2000.0 right ascension and declination as given in RC3 \citep{dev91}. Units of right ascension are hours, minutes, and seconds, and units of declination are degrees, arcminutes, and arcseconds.
Col. (4): Galaxy morphology type from NED.
Col. (5): Systemic velocity from the \hi{} line (Heliocentric).
Col. (6): Adopted distance \citep[][respectively]{kar04,wal08,fre01}.
Col. (7): Total absolute {\em B}-band magnitude \citep{ken08}.
Col. (8): The \hi{} mass from THINGS \citep{wal08}.
Col. (9): The H$_{2}$ mass from the NRAO 12~m \cooz{} on-the-fly observations for BIMA SONG \citep{hel03}.
Col. (10): The H$\alpha$ luminosity from \citet{ken08}.
Col. (11): The adopted inclination \citep{wal08,deb08}.
}
\end{deluxetable}

Since November 2007 the first phase of the survey has been complete, observing 57 galaxies of the sample in the \co{} line;  21 overlap with the Spitzer Infrared Nearby Galaxies Survey \citep[SINGS,][]{ken03} sample, and 36 are members of the Virgo Cluster.  \co{} observations for the remaining SINGS and field sample galaxies are currently underway.  Following on from \citet[][ hereafter \pI{}]{wil09}, which investigated four SINGS subsample galaxies that are members of the Virgo Cluster, we present the \co{} results for three SINGS galaxies outside the cluster environment.

For our initial examination of the \co{} data in non-cluster NGLS galaxies, we have chosen to look at three well known local spiral galaxies that were completed in the early stages of the survey: NGC\,628, NGC\,3521, and NGC\,3627.  All three of these galaxies have extensive multi-wavelength ancillary data, most recently from SINGS and the follow-up surveys associated with it.  Having been included in SINGS, all three have publicly available infrared observations (seven bands from 3.6 to 160~$\mu$m, plus some spectroscopy).  All three were also included in The \hi{} Nearby Galaxy Survey \citep[THINGS,][]{wal08}, the VLA \hi{} follow up survey to SINGS. Additionally, they were all observed in the \cooz{} line on the Berkeley-Illinois-Maryland Association (BIMA) millimeter interferometer as part of the BIMA Survey of Nearby Galaxies \citep[BIMA SONG,][]{hel03}, and two (NGC\,3521 and NGC\,3627) have single dish \cooz{} observations taken with the Nobeyama 45-m telescope by \citet{kun07}.  Comparing some of the abundant ancillary data available with our JCMT \co{} observations provide us with a powerful tool for exploring star formation within these galaxies.  Table~\ref{tab:prop} outlines some of the previously published properties of these three galaxies.

NGC\,628 (M\,74) is an isolated, near face-on spiral of type SA(s)c.  It is known to have an extended \hi{} disk, extending to approximately three times the optical Holmberg radius \citep{kam92}.  \cooz{} observations of NGC\,628 for BIMA SONG \citep{hel03} only cover the inner region of the galaxy, but show that the molecular gas is clumpy and distributed mostly along the spiral arms.  It is slightly less luminous than the other two galaxies we are examining here, and also has the lowest H$\alpha$ and \cooz{} luminosity of the three.  We adopt the same distance estimates in this paper as in \citet{wal08}.  For NGC\,628 the distance we use is 7.3~Mpc, derived by \citet{kar04} using the luminosity of the brightest stars.

NGC\,3521 is a highly inclined flocculent spiral galaxy (type SAB(rs)bc) that shows a ring-like \cooz{} structure \citep{hel03}.  It is associated with the nearby Leo Triplet (of which another of our galaxies, NGC\,3627, is a member).  It has a moderately high \hi{} mass \citep[$8 \times 10^9$\Msun{},][]{wal08}, the highest of these three galaxies, and the \hi{} disk extends well beyond the stellar disk.  The adopted distance to NGC\,3521 is 10.7~Mpc \citep[][ Hubble flow distance using the velocity listed on NED]{wal08}.

NGC\,3627 (M\,66) is an active, asymmetric barred spiral galaxy (type SAB(s)b), a member of the Leo Triplet well known for its unusual kinematics that are influenced by both its bar and external interactions \citep{zha93,reu96,reg02,che03}.  It is the most luminous and most active of these three galaxies, and has the highest molecular gas content of the three as deduced from \cooz{} observations \citep[${\cal M}_{\rm H_{2}} = (4.1\pm0.4) \times 10^{9}$\Msun{},][]{hel03}.  In contrast, the \hi{} mass is the lowest of the three galaxies, such that the molecular gas dominates the gas content of the galaxy.  The adopted distance to NGC\,3627 is 9.3~Mpc, derived by \citet{fre01} using Cepheid variables.  

In this paper we focus on the molecular gas kinematics and detailed star formation properties of these three galaxies. \S~\ref{sec:data} describes our observations at the JCMT and data reduction process, and the processing of ancillary data.  \S~\ref{sec:resmaps} presents the resulting moment maps from our \co{} observations.  \S~\ref{sec:resratio} compares our new \co{} observations to existing \hi{} and \cooz{} data.  \S~\ref{sec:resdyne} deals with the dynamics of the molecular gas, looking at differences in the velocity fields from molecular and atomic gas, and comparing the rotation curves we derive from our \co{} observations with those from the \hi{} data of THINGS \citep{wal08}.  In \S~\ref{sec:ressf}, we discuss several aspects of star formation within these galaxies.  Finally we present our conclusions in \S~\ref{sec:conclusions}.

\section{Observations and Data Reduction \\ Overview}
\label{sec:data}

\subsection{JCMT \co{} Data}

\begin{deluxetable}{llcccccccc} 
\tabletypesize{\scriptsize}
\tablecaption{Summary of Observing Parameters.
  \label{tab:obs}}
\tablewidth{0pt}
\tablehead{\colhead{Name} & \colhead{Obs. Dates} & \colhead{PA} & \colhead{$D_{25,maj}/2$} & \colhead{$D_{25,min}/2$} & \colhead{$\overline{T_{sys}}$} & \colhead{$\overline{\tau}$} & \colhead{$t_{int}$} & \colhead{$\Delta T$} & \colhead{Ref. Pos.}  \\

    &  & \colhead{(deg)} & \colhead{(arcsec)} & \colhead{(arcsec)} & \colhead{(K)} & \colhead{(225 GHz)} & \colhead{(s)} & \colhead{(mK)} & \colhead{(arcsec)} \\
 
 \colhead{(1)} & \colhead{(2)} & \colhead{(3)} & \colhead{(4)} & \colhead{(5)} & \colhead{(6)} & \colhead{(7)} & \colhead{(8)} & \colhead{(9)} & \colhead{(10)} }
\startdata
 NGC\,628  & 071123, 071124, 071125 &  25 & 312 & 288 & 388 & 0.097 & 80 & 15.8 & 432 \\
           & 080107, 080113 \\
 \\
 NGC\,3521 & 080310, 080311         & 163 & 324 & 162 & 402 & 0.109 & 65 & 16.8 & 444 \\
 \\
 NGC\,3627 & 080112, 080113         & 173 & 264 & 120 & 328 & 0.084 & 70 & 13.2 & 384 \\
\enddata
\tablecomments{Col. (1): Galaxy name. 
Col. (2): Date of observations in 2007 and 2008, YYMMDD.
Col. (3): Position angle of the major axis from the optical component, and orientation of the observed field.
Col. (4): Major axis length of the target scan region, $D_{25,maj}/2$.
Col. (5): Minor axis length of the target scan region, $D_{25,min}/2$.
Col. (6): Median system temperature over all observations.
Col. (7): Median atmospheric opacity at 225~GHz as measured with the JCMT water vapor radiometer.
Col. (8): Typical integration time per point in the map.
Col. (9): Rms noise ($T^{*}_{A}$) in line-free regions of the spectra at 20\kms{} resolution.
Col. (10): Location of the shared off position, in arcseconds east of the center.
}
\end{deluxetable}

The HARP-B \co{} line (rest frequency 345.79~GHz) observations for the three galaxies we are examining here, NGC\,628, NGC\,3521, and NGC\,3627, were taken over multiple runs between November 2007 and March 2008 as part of the JLS program.  For all three galaxies we mapped a rectangular area corresponding to $D_{25}/2$ using a basket weave raster scanning method (see Appendix~\ref{sec:dred} for details).  We set up the backend spectrometer, the Auto-Correlation Spectrometer Imaging System (ACSIS), with a bandwidth of $\sim$1~GHz and a resolution of 488.28 kHz (0.423\kms{} at the frequency of the observations).  In each scan, we integrated for 10~seconds per pointing within the target field (if all 16 receptors were in working order, see Appendix~\ref{sec:dred}), and scans were repeated until we reached the target RMS for the survey, 19~mK ($\rm T_A^*$) after binning to 20\kms{} resolution.  Individual setup and observing details for the three galaxies are listed in Table~\ref{tab:obs}.

Data reduction and most of the analysis were carried out with recent versions (Humu and Lehuakona) of the Starlink\footnote{available for download from http://starlink.jach.hawaii.edu/} software package that is maintained by the JAC.  Additionally, analysis tasks in MIRIAD were used after the creation of cubes and moment maps.  The general reduction procedure was to inspect the raw data files, flag poor data where necessary, combine the raw data into a cube, subtract a baseline from the cube, and collapse it into moment maps.  We converted the \co{} data from $\rm T_A^*$ to the main beam temperature scale by dividing by $\eta_{MB}=0.6$.  Full details of the observations and data reduction methods are described in Appendix~\ref{sec:dred}.

\subsection{Ancillary Data}
\label{sec:anc}

All three galaxies in this paper are well studied members of the SINGS sample, and as a result have a large quantity of published ancillary data, from the far ultraviolet to centimeter radio observations.  We obtained 24~$\mu$m and H$\alpha$ maps from the SINGS survey itself \citep{ken03} at the SINGS website\footnote{http://sings.stsci.edu}.  \hi{} data cubes and 0th moment maps came from the THINGS survey \citep{wal08} website\footnote{http://www.mpia-hd.mpg.de/THINGS/Overview.html}.  
There were two sources for \cooz{} maps, single dish Nobeyama maps from \citet{kun07}\footnote{http://www.nro.nao.ac.jp/~nro45mrt/COatlas/} for NGC\,3521 and NGC\,3627, and the BIMA SONG\footnote{Obtained from NED, http://nedwww.ipac.caltech.edu/ level5/March02/SONG/SONG.html} \cooz{} maps for NGC\,628 as there were no Nobeyama data available.  Although BIMA SONG data was available for all three, they only have good sensitivity in the central part of the field covered by the NGLS, so we preferred the single dish Nobeyama data where available as they cover a similar field size, and the Nobeyama beam size for \cooz{} data is well matched to the JCMT at \co{} (so no convolution is required).  
In all cases, the available ancillary data maps were of similar or higher resolution than the 14\farcs5 beam of the NGLS \co{} observations, and they cover a field-of-view of similar size or larger.

The procedures for processing the SINGS 24~$\mu$m and  H$\alpha$ data are outlined in \pI{}, and follows the notes and conversions supplied the SINGS User's Guide \citet{sin07}.  To compare these data to the NGLS \co{} maps, we first had to convolve them to give a smoothed image with a 14\farcs5 circular beam to match the JCMT resolution at 345~GHz.  The H$\alpha$ images contained a significant (and uneven) background sky level that needed to be subtracted by fitting a surface to the image.  The fit was limited to a first order polynomial as any higher orders would begin to trace low level H$\alpha$ emission from the galaxy.  Several artifacts caused by over-exposed stars were also removed from these images using upper and lower thresholds prior to smoothing.  The 24~$\mu$m data did not require any special processing prior to convolving the map to the JCMT resolution.
The same smoothing procedure was performed on the BIMA SONG \cooz{} map for NGC\,628; however, the Nobeyama \cooz{} data for the other two galaxies were not convolved to this beam as the angular resolution already closely matches the JCMT. 

For the \hi{} data, we use the naturally weighted zeroth moment maps that are available on the THINGS Public Data Repository web site \linebreak (http://www.mpia-hd.mpg.de/THINGS/Data.html) for comparison to our \co{} data.  These \hi{} data are well matched in resolution to the NGLS observations, with our single dish \co{} observations being at only slightly poorer resolution than the naturally weighted interferometric \hi{} data (14\farcs5 for the JCMT compared to 9\arcsec{} to 14\arcsec{} for the VLA data).  As with the other ancillary data, we convolved the \hi{} intensity (0th moment) maps to match the JCMT resolution.

Unlike the \hi{} intensity maps, we cannot directly convolve the beams of the velocity fields (1st moment) and velocity dispersion (2nd moment) maps from the THINGS survey for comparison to our \co{} moment maps.  So instead, we obtained the naturally weighted THINGS \hi{} data cubes for the three galaxies, convolved the cubes to a 14\farcs5 circular beam, and then created moment maps from the resulting cubes (using a similar procedure to the creation of the \co{} maps).  While this also produced 0th moment maps, we prefer to use the convolved version of the THINGS maps in this case as their moment map production procedure has been refined for their data set.  After convolving to the appropriate resolution, all ancillary data were aligned to the JCMT \co{} map and resampled to the same pixel scale as the \co{} maps ($\sim$7\farcs3 pixels).

\section{Results: \co{} Images}
\label{sec:resmaps}

\begin{figure*} 
  \includegraphics[width=85mm]{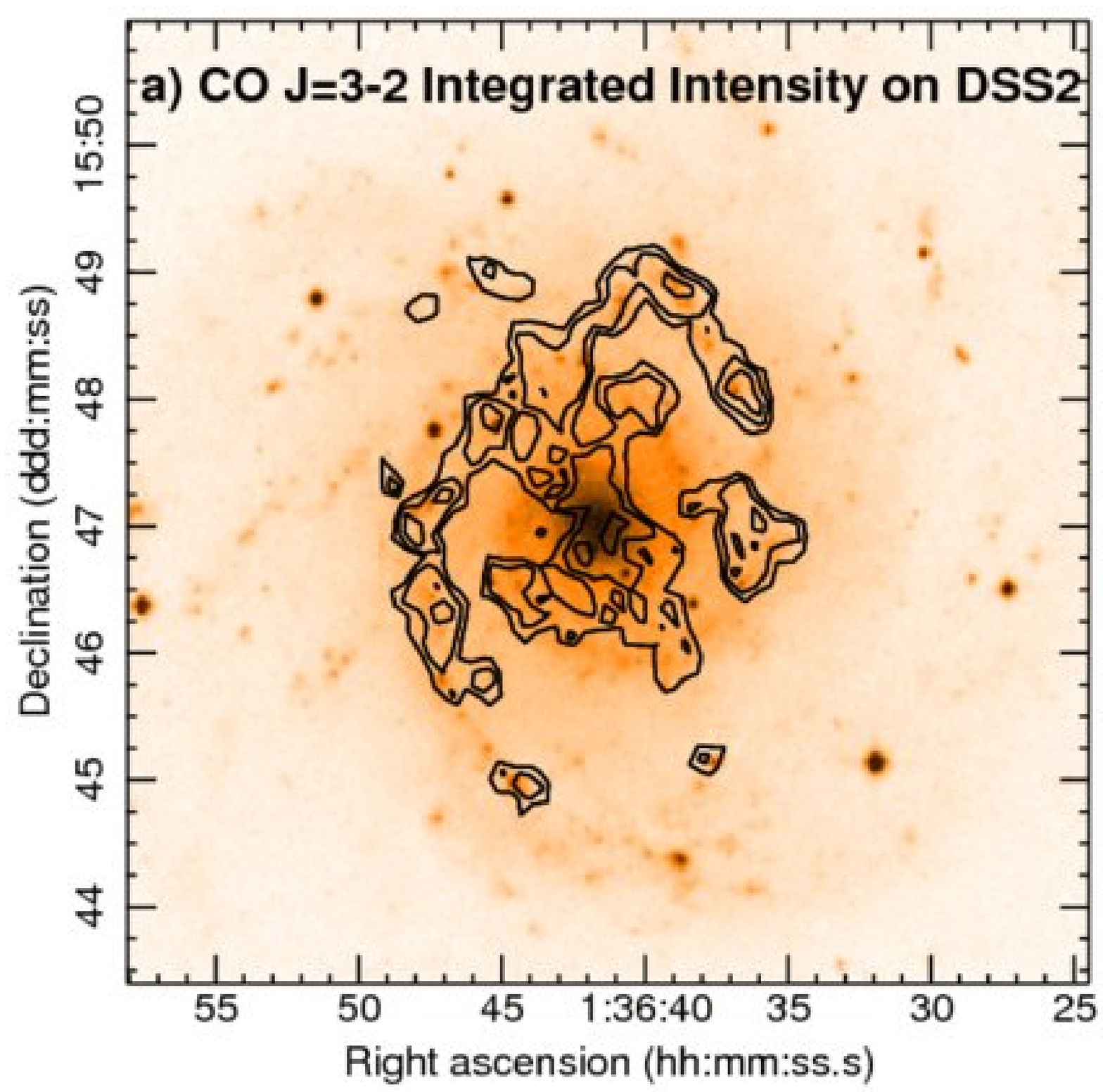}
  \includegraphics[width=85mm]{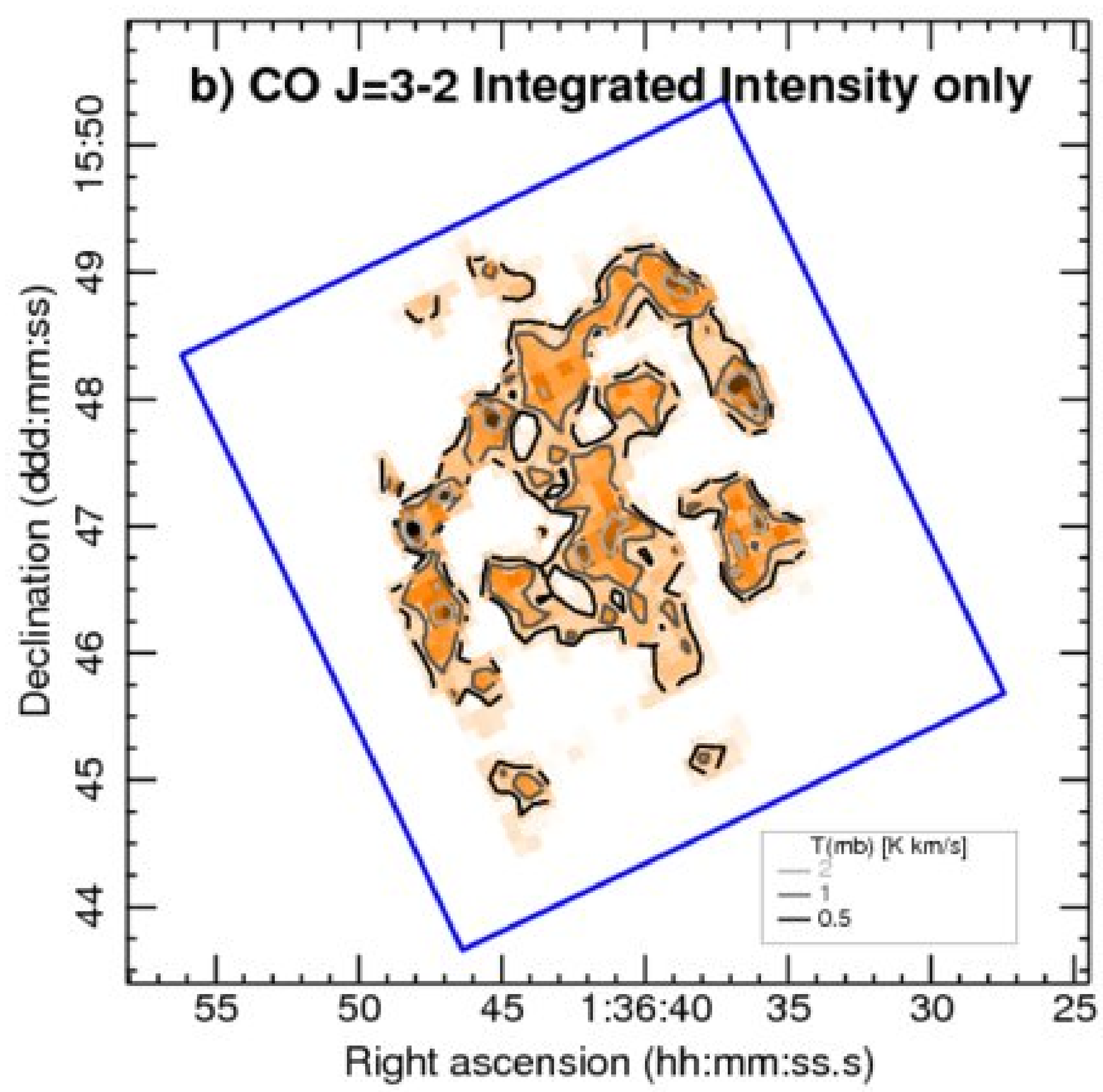}
  \includegraphics[width=85mm]{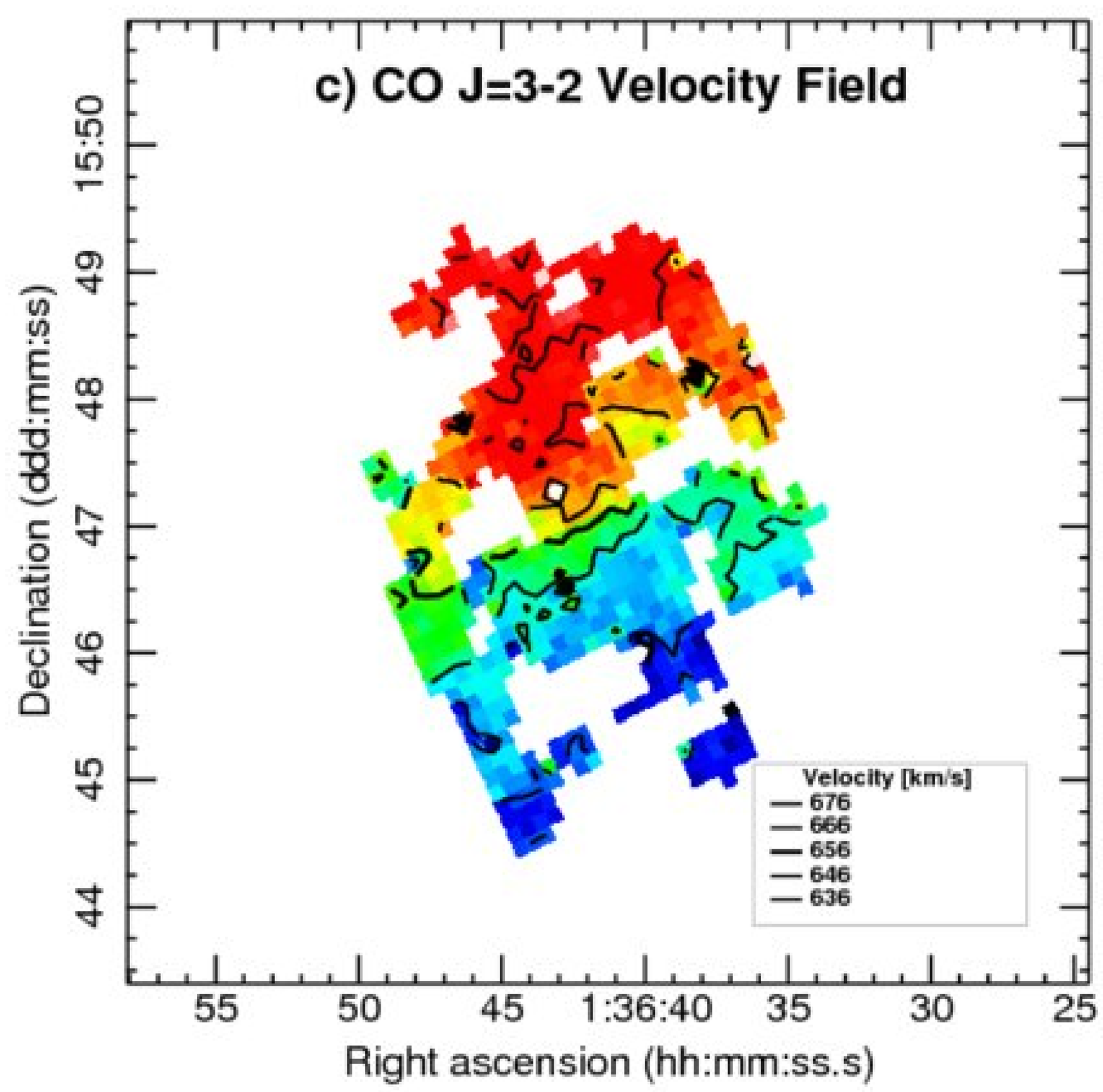}
  \includegraphics[width=85mm]{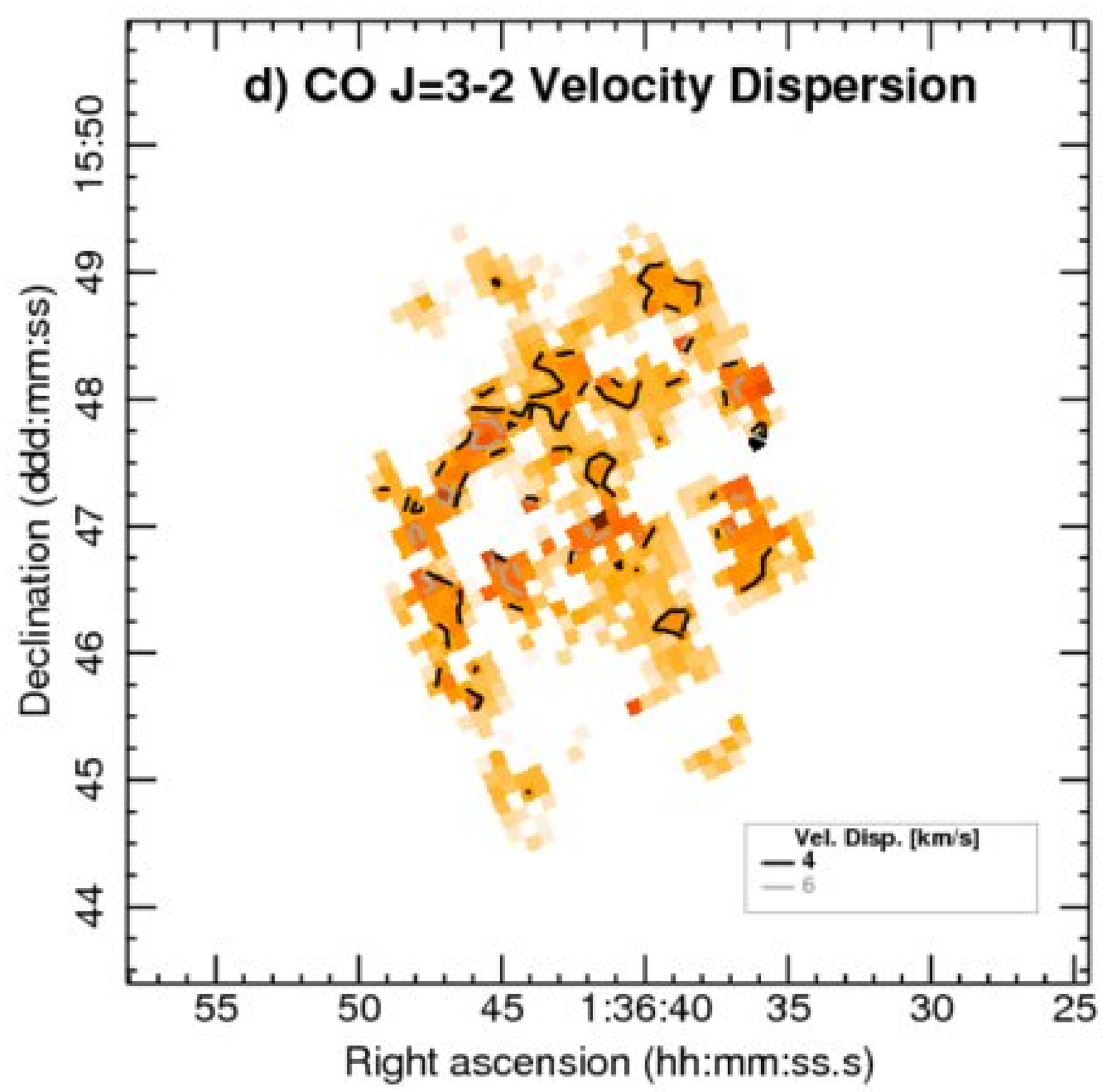}
\caption{\co{} moment maps of NGC\,628. a) Integrated \co{} intensity distribution (moment 0) contours for NGC\,628 overlaid onto an optical DSS\,{\sc ii} {\em R} band image.  Contour levels are 0.5, 1, and 2\kelkms{} (temperatures in all moment 0 maps are $T_{mb}$).  b) Integrated \co{} intensity contours as in first panel but with our \co{} map in the background instead of the DSS\,{\sc ii} image.  White is low (or blank), and black is high.  The blue box in this panel shows our target mapping region (note that the useful region of the final image is slightly larger than these boxes, see Appendix~\ref{sec:dred}).  c) \co{} velocity field (moment 1) for NGC\,628.  Contour levels (from blue to red end) are 636, 646, 656 (thick contour, systemic velocity), 666, and 676\kms{}.  d) \co{} velocity dispersion map (moment 2) for NGC\,628.  Contour levels are 4, and 6\kms{}.
\label{fig:ngc0628}}
\end{figure*}

\begin{figure*} 
  \includegraphics[width=85mm]{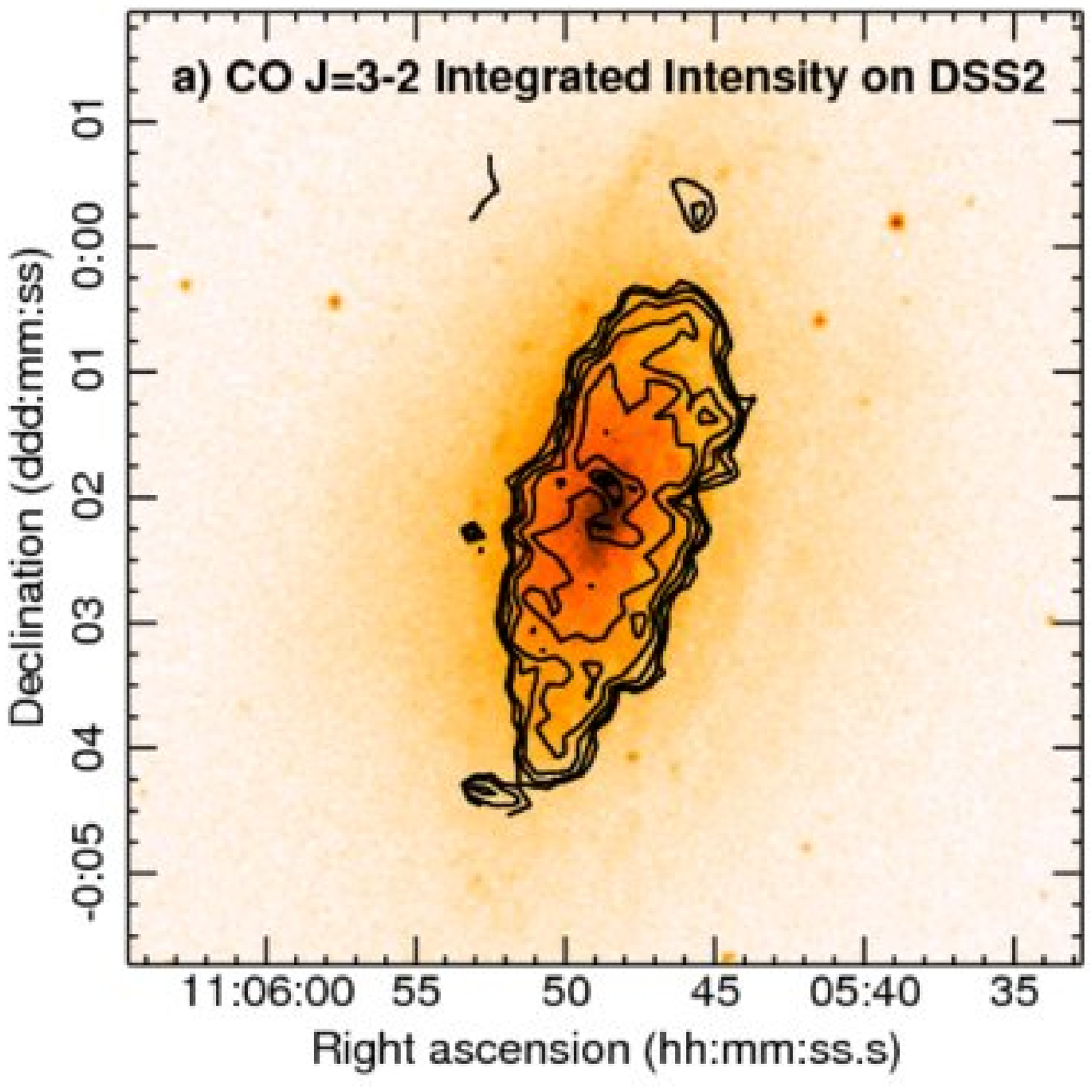}
  \includegraphics[width=85mm]{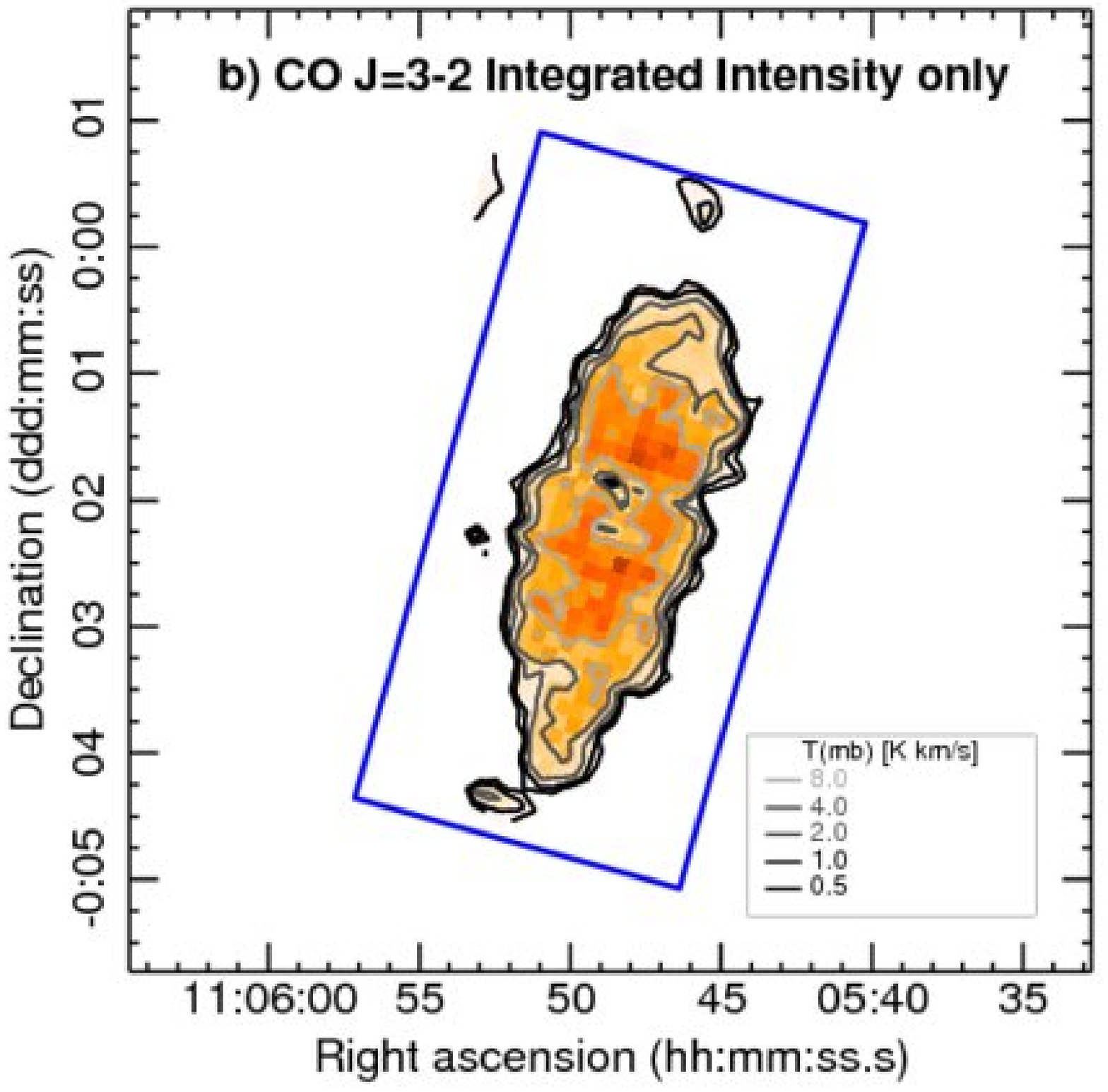}
  \includegraphics[width=85mm]{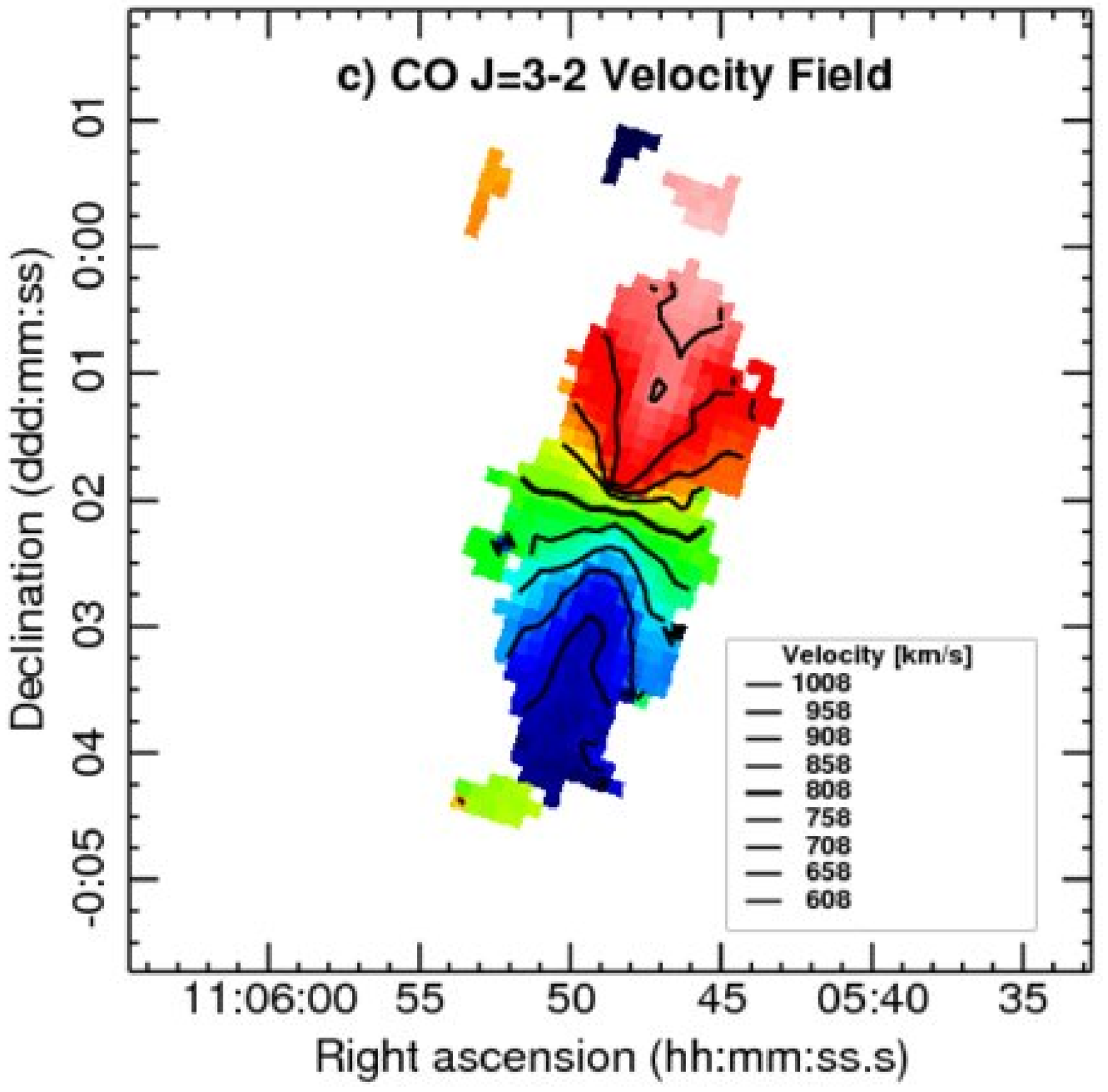}
  \includegraphics[width=85mm]{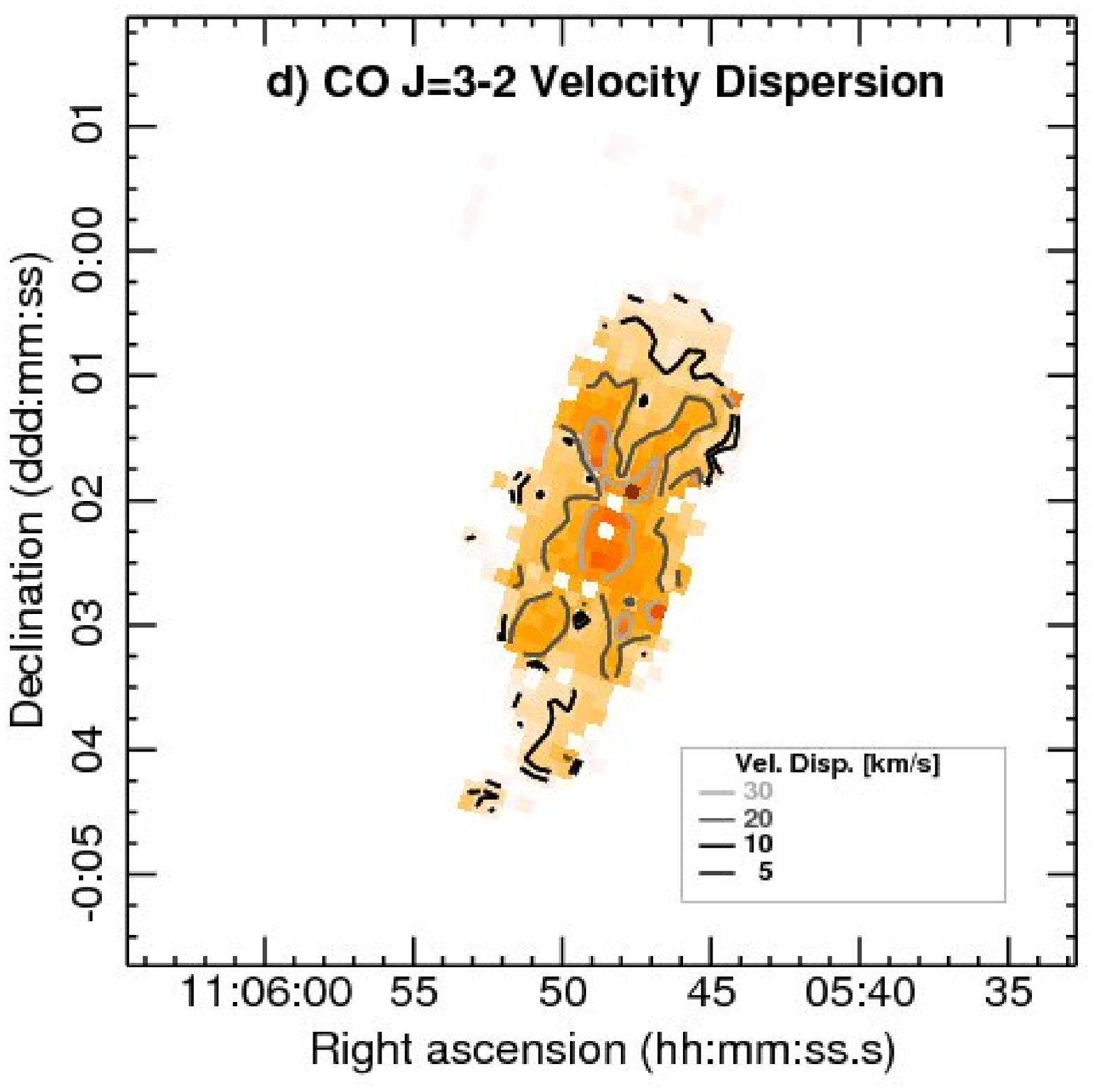}
\caption{\co{} moment maps of NGC\,3521. a) Integrated \co{} intensity distribution (moment 0) contours for NGC\,3521 overlaid onto an optical DSS\,{\sc ii} {\em R} band image.  Contour levels are 0.5, 1, 2, 4, and 8\kelkms{}.  b) Integrated \co{} intensity contours as in first panel but with our \co{} map in the background instead of the DSS\,{\sc ii} image.  White is low (or blank), and black is high.  The blue box in this panel shows our target mapping region.  c) \co{} velocity field (moment 1) for NGC\,3521.  Contour levels (from blue to red end) are 608, 658, 708, 758, 808 (thick contour, systemic velocity), 858, 908, 958, and 1008\kms{}.  d) \co{} velocity dispersion map (moment 2) for NGC\,3521.  Contour levels are 5, 10, 20, and 30\kms{}.
\label{fig:ngc3521}}
\end{figure*}

\begin{figure*} 
  \includegraphics[width=85mm]{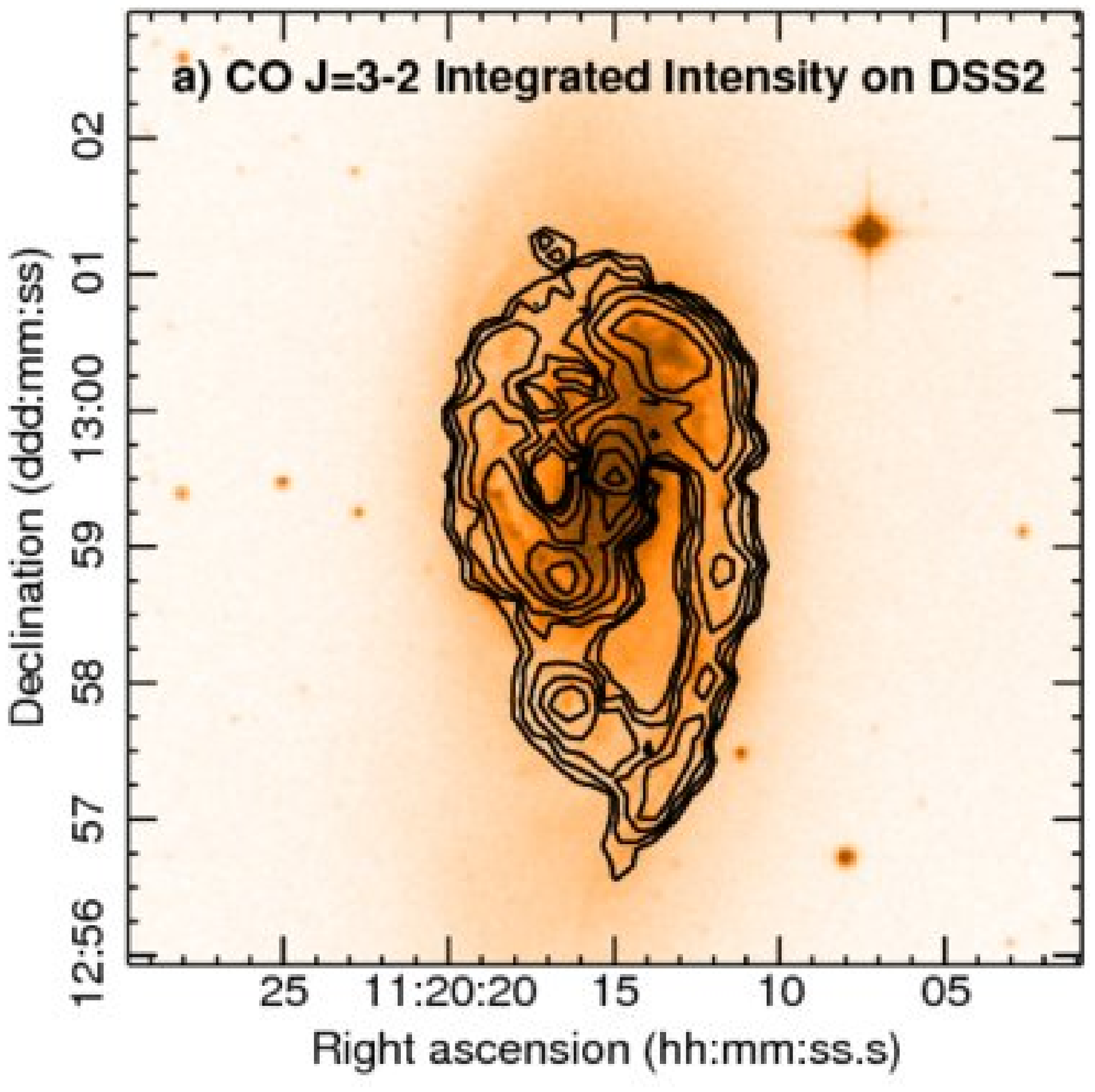}
  \includegraphics[width=85mm]{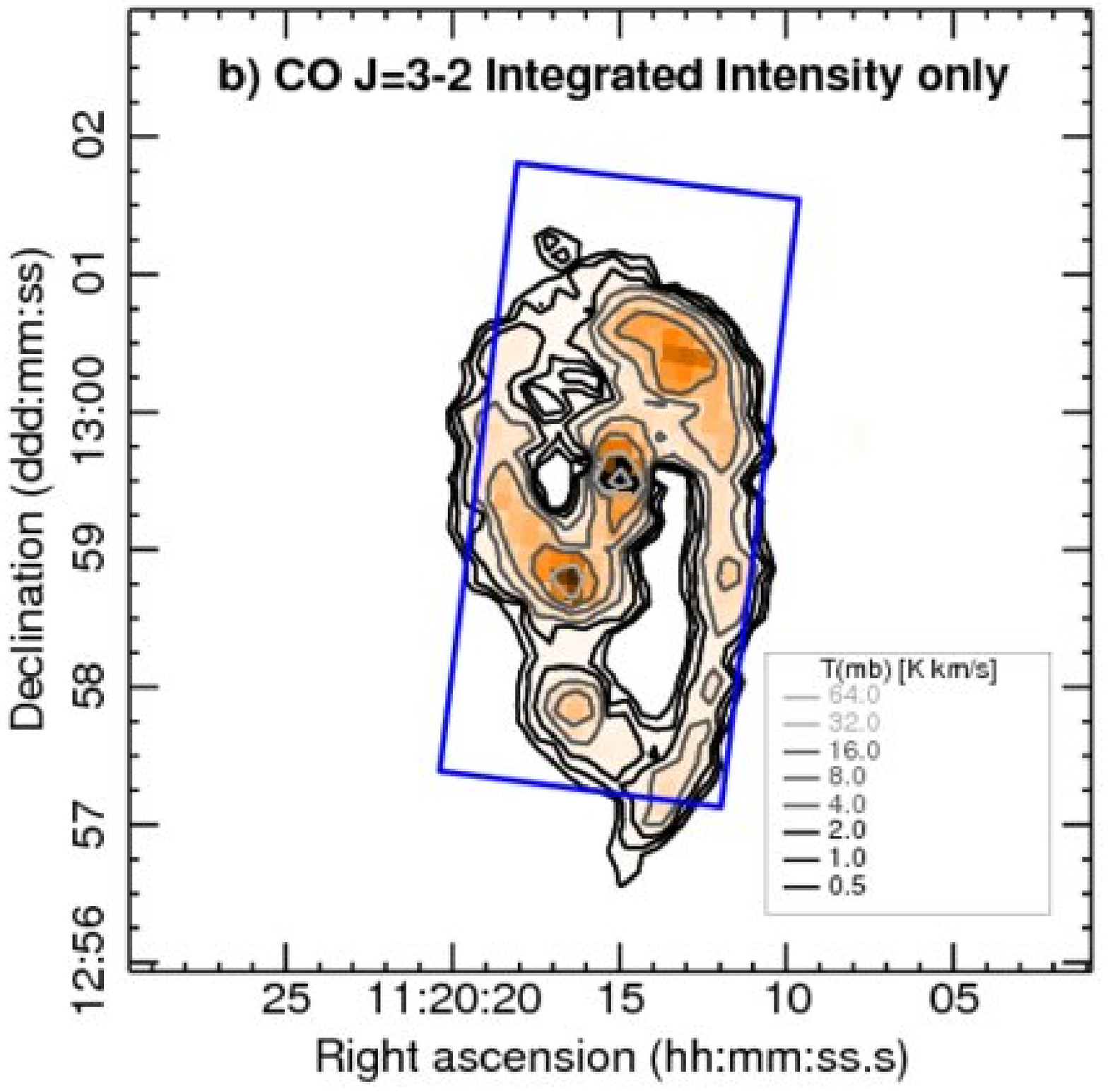}
  \includegraphics[width=85mm]{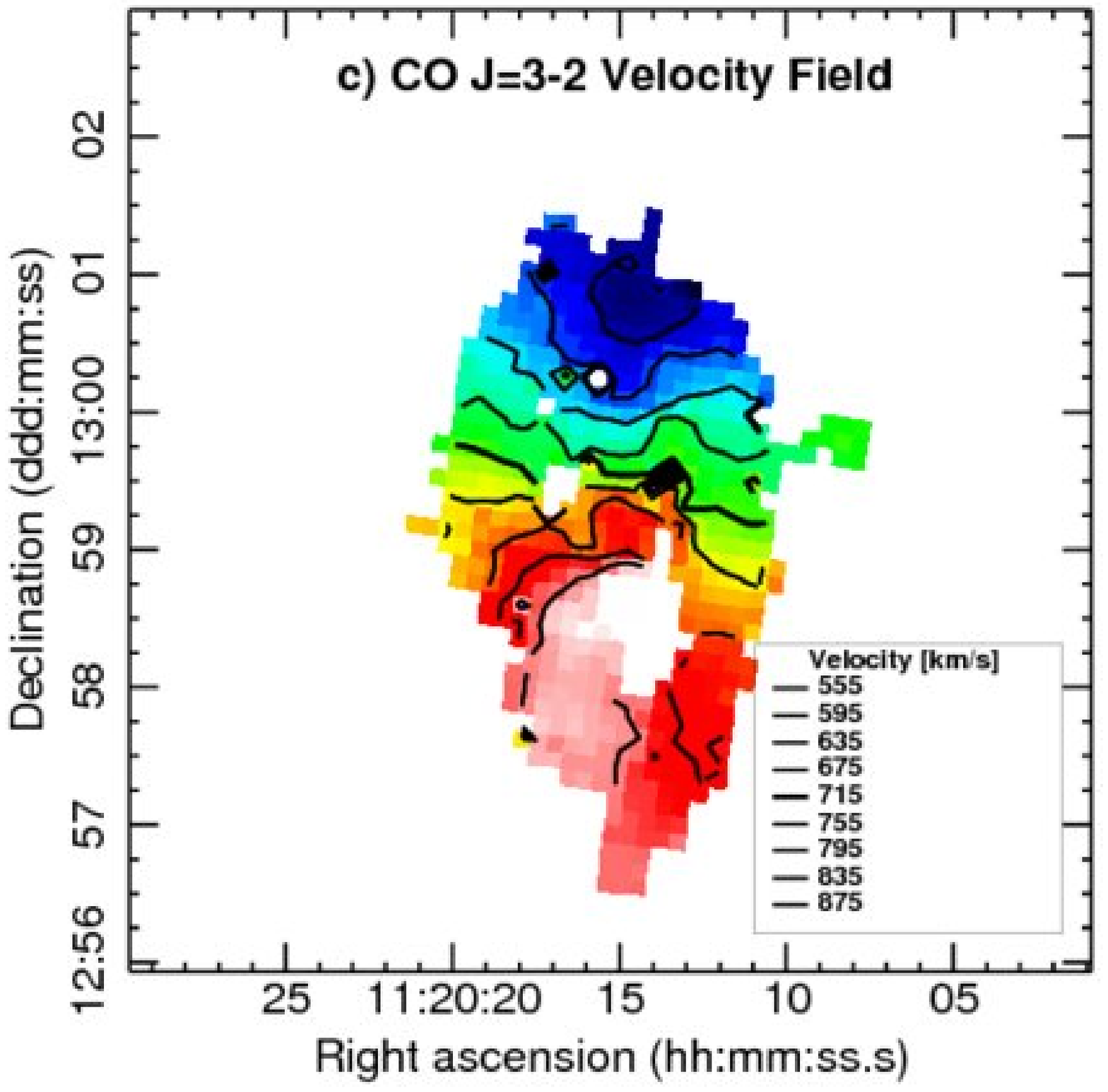}
  \includegraphics[width=85mm]{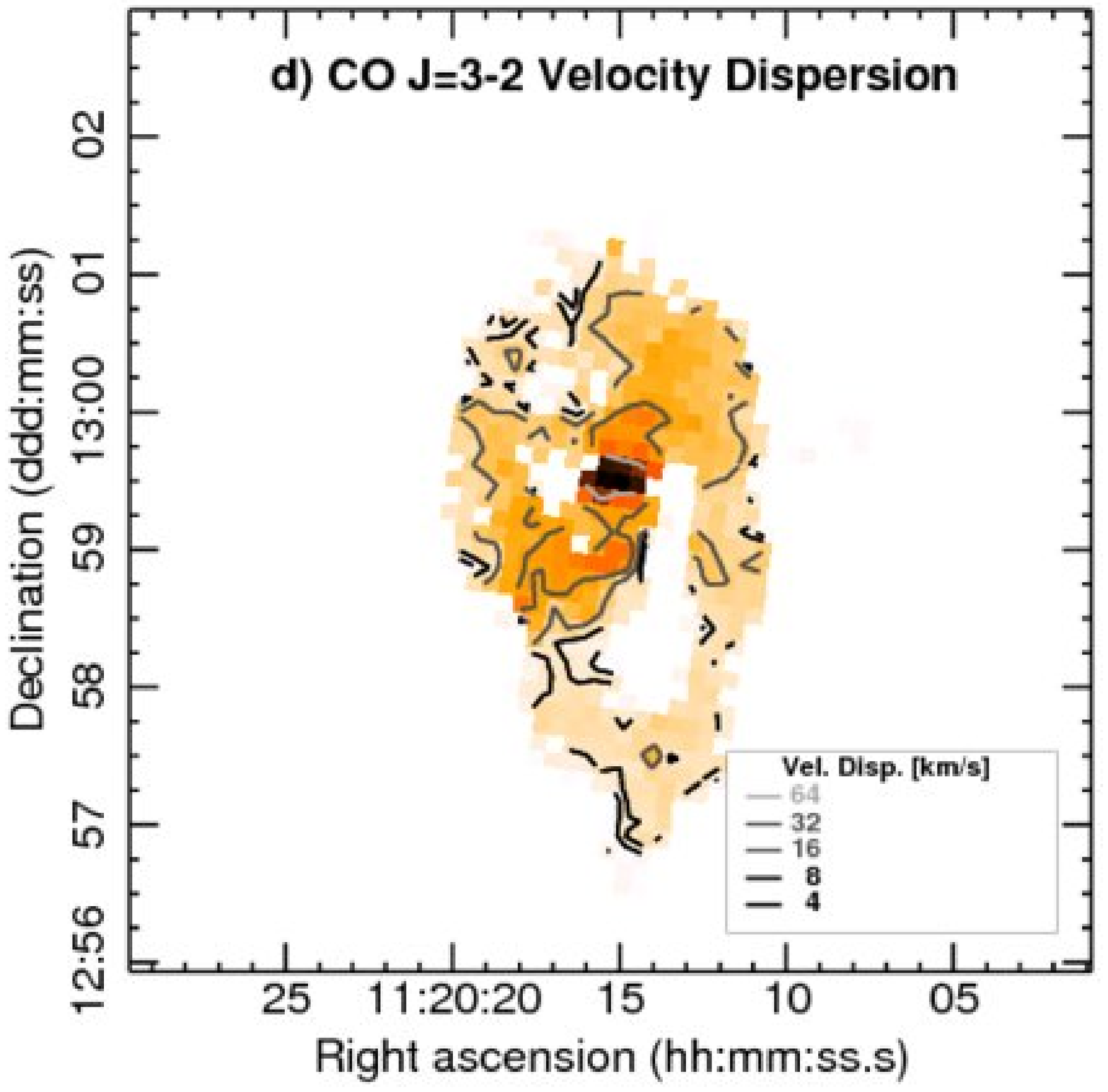}
\caption{\co{} moment maps of NGC\,3627. a) Integrated \co{} intensity distribution (moment 0) contours for NGC\,3627 overlaid onto an optical DSS\,{\sc ii} {\em R} band image.  Contour levels are 0.5, 1, 2, 4, 8, 16, 32, and 64\kelkms{}.  b) Integrated \co{} intensity contours as in first panel but with our \co{} map in the background instead of the DSS\,{\sc ii} image.  White is low (or blank), and black is high.  The blue box in this panel shows our target mapping region.  c) \co{} velocity field (moment 1) for NGC\,3627.  Contour levels (from blue to red end) are 555, 595, 635, 675, 715 (thick contour, systemic velocity), 755, 795, 835, and 875\kms{}.  d) \co{} velocity dispersion map (moment 2) for NGC\,3627.  Contour levels are 4, 8, 16, 32, and 64\kms{}.
\label{fig:ngc3627}}
\end{figure*}

The \co{} moment maps we obtained from our HARP-B raster scan mapping are shown in Figures~\ref{fig:ngc0628} (NGC\,628), \ref{fig:ngc3521} (NGC\,3521), and \ref{fig:ngc3627} (NGC\,3627).  Each figure contains four panels, showing the integrated \co{} intensity overlaid on the optical DSS\,{\sc ii} ({\em R} band) image, the integrated \co{} intensity by itself, the velocity field, and the velocity dispersion map.  The second panel of each figure shows the target field for the survey in blue, a D$_{25,maj}$/2 by D$_{25,min}$/2 box, where theoretically we have complete sensitivity coverage to our target depth.  In practice, signal-to-noise variations between receptors, and missing/flagged receptors in the array, make the sensitivity coverage of the field more complex.  Additionally, the region with good sensitivity is slightly larger than the target field due to the over-scanned area (see Appendix~\ref{sec:dred}).  At the adopted distances of the three galaxies, the 14\farcs{}5 beam of the JCMT \co{} observations gives us resolutions of 510, 750, and 660 pc for NGC\,628, NGC\,3521, and NGC\,3627, respectively.

The integrated intensity maps of NGC\,628 (Fig.~\ref{fig:ngc0628}a,b) show that the warm molecular gas detected in the NGLS \co{} line observations is patchy, of relatively low intensity, and mostly follows the spiral arms of the galaxy.  The center region shows neither a strong concentration nor a strong dip in the molecular gas.  The few higher peaks in the map are mostly distributed along the spiral arms, the strongest at $3.7 \pm 0.7$\kelkms{} being $\sim$1\farcm5 east of the galaxy center (though there are several others of similar strength).  The total \co{} luminosity for NGC\,628 is $(3.1 \pm 1.3)\times 10^{7}$\kelkms{} pc$^{2}$.

NGC\,628 is almost face on \citep[inclination of about 7\degr{},][]{wal08}, so the velocity field (Fig.~\ref{fig:ngc0628}c) shows only a narrow velocity range across the galaxy ($\sim$50\kms{}).  As our target field only covers the center of the galaxy, the coverage is patchy, and the inclination is so low, we cannot tell from the \co{} data alone if we have data far enough out to reach the flat portion of the rotation curve, though as we only map out to D$_{25}$/2 that is unlikely.  
Fig.~\ref{fig:ngc0628}d shows the \co{} velocity dispersion map (moment 2 map) for NGC\,628.  The velocity dispersion is generally low, around 2 to 6\kms{}, with the higher velocity dispersion around the integrated \co{} intensity peaks.  The median velocity dispersion is 3.2\kms{}.  These velocity dispersions are similar to the internal velocity dispersions seen in giant molecular clouds (GMC) in the Milky Way \citep{sol87}.  This suggests that the main contribution to velocity dispersion in the molecular gas for NGC\,628 is internal dispersion within GMCs, rather than inter-cloud velocity differences.  Alternatively, \citet{com97} also looked at the velocity dispersion of molecular gas in this galaxy, using IRAM 30~m telescope observations at \cooz{} and $^{12}$CO\,$J$=2-1 (23\arcsec{} and 12\arcsec{} resolution, respectively), and finding a near constant velocity dispersion as a function of radius of $\sim$6\kms{} they conclude that cloud-cloud velocity dispersions dominate over internal dispersions.  Without resolving the clouds themselves however we cannot tell for certain which components contribute most to the velocity dispersion measured within our beam.  We would expect contributions from the rotation velocity gradient across our beam to be minimal beyond the center of the galaxy given the low inclination.

The integrated \co{} intensity maps for NGC\,3521 (Fig.~\ref{fig:ngc3521}a,b) show that its molecular gas is concentrated towards the inner regions of the optical disk, with a slight central dip forming a torus-like structure in the inner $\sim$1\arcmin{} ($\sim$3~kpc along the major axis).  The high inclination angle of the galaxy makes it difficult to identify other structures within the molecular gas disk.  The detected \co{} disk extends along the major axis out to a radius of $\sim$130\arcsec, or $\sim$6.7~kpc, which is about one third of the optical radius $D_{25,min}$.
There are two peaks in the map on either side of the central dip ($<$30\arcsec{} from the center along the major axis in both directions) that are of about equal intensity in our \co{} data, with the northern peak at $14.6 \pm 0.6$\kelkms{} and the southern at $15.1 \pm 0.7$\kelkms{}.  The total \co{} luminosity for NGC\,3521 is $(2.8 \pm 0.3)\times 10^{8}$\kelkms{} pc$^{2}$.

The \co{} velocity field for NGC\,3521 (Fig.~\ref{fig:ngc3521}c) shows the general characteristics of a moderate to highly inclined rotating disk galaxy (the well known spider diagram pattern).  Over the region where we detect \co{} there do not appear to be any significant distortions in the velocity field.  A brief examination of position-velocity slices through the cube for NGC\,3521, along and parallel to the major axis, showed no significant detections of molecular gas with anomalous velocities, and that the rotation curve appears to flatten out within the radius of the disk detected in \co{}.  The former suggests that the \co{} velocity field is tracing the bulk motions of the gas disk.  Fig.~\ref{fig:ngc3521}d shows the \co{} velocity dispersion map (moment 2 map) for NGC\,3521.  It has a distinctive `X' pattern through the center of the galaxy (best seen with the 20\kms{} contour), which seems to be aligned with the velocity field contours.  The highest velocity dispersions are seen towards the center of the galaxy, where the velocity field contours are spaced closest together.  Given the high inclination of this galaxy, these may indicate that the velocity dispersion map is partially indicating regions where there is a large component of the rotational velocity gradient in the direction of the line of sight.  The median velocity dispersion is 15.0\kms{}.

The integrated intensity maps of NGC\,3627 (Fig.~\ref{fig:ngc3627}a,b) shows a distinct pattern of an inclined barred spiral.  Molecular gas is found throughout much of the optical disk, with \co{} detected right up to and sightly beyond the edges of our target field (the blue box in the upper right panel).  It is mostly constrained to the bar and along the spiral arms, with no gas detected in the intervening regions.  The spiral arms are asymmetrical in \co{}, with the arm that extends to the south from the western side of the galaxy at least 50\% longer than the arm that extends north from the eastern side.  The center and the ends of the bar have particularly strong concentrations of \co{}, with the peak in the bar/galaxy center at $81 \pm 1$\kelkms{}.  The peak intensity at the southern bar end is $43.0 \pm 0.6$\kelkms{}, while in the northern bar end the peak is at $30.2 \pm 0.6$\kelkms{}.  The total \co{} luminosity for NGC\,3627 is $(3.1 \pm 0.2)\times 10^{8}$\kelkms{} pc$^{2}$.

Fig.~\ref{fig:ngc3627}c shows the \co{} velocity field for NGC\,3627.  The field shows some distortion between the inner bar and spiral arms, with the gradient of the field in the inner bar region aligned close to north-south (the receding side is $\sim$180\degr{} East of North), while further out the alignment is closer to $\sim$170\degr{}.  This resembles streaming motions associated with spiral arms \citep{bos81,kna93}.  Additionally, the kinematical major and minor axes are not orthogonal, characteristic of oval distortions (such as bars) or warping \citep{bos81}.  Fig.~\ref{fig:ngc3521}d shows the \co{} velocity dispersion map (moment 2 map) for NGC\,3627.  The velocity dispersion is moderate to low in the spiral arms ($\sim$7 to 20\kms{}), but much higher in the bar region, particularly in the center of the galaxy where the velocity dispersion is over 80\kms{}, though there is a large velocity gradient across the 14\farcs5 JCMT beam in this region.  The median velocity dispersion is 13.0\kms{}.

\section{The \corat{} Ratio and the Distributions of Molecular and Atomic Gas}
\label{sec:resratio}

\begin{figure*} 
  \includegraphics[width=85mm]{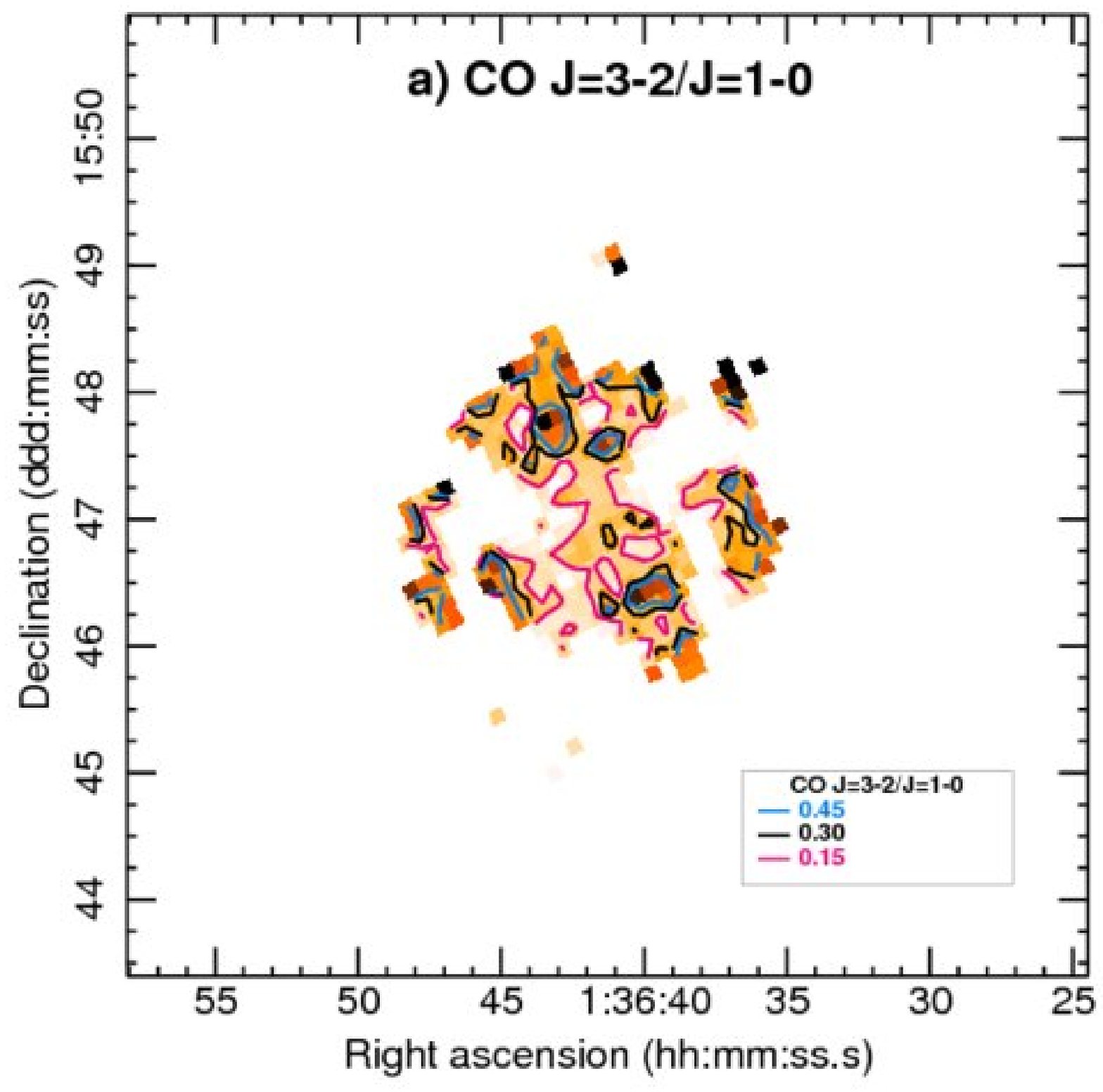}
  \includegraphics[width=85mm]{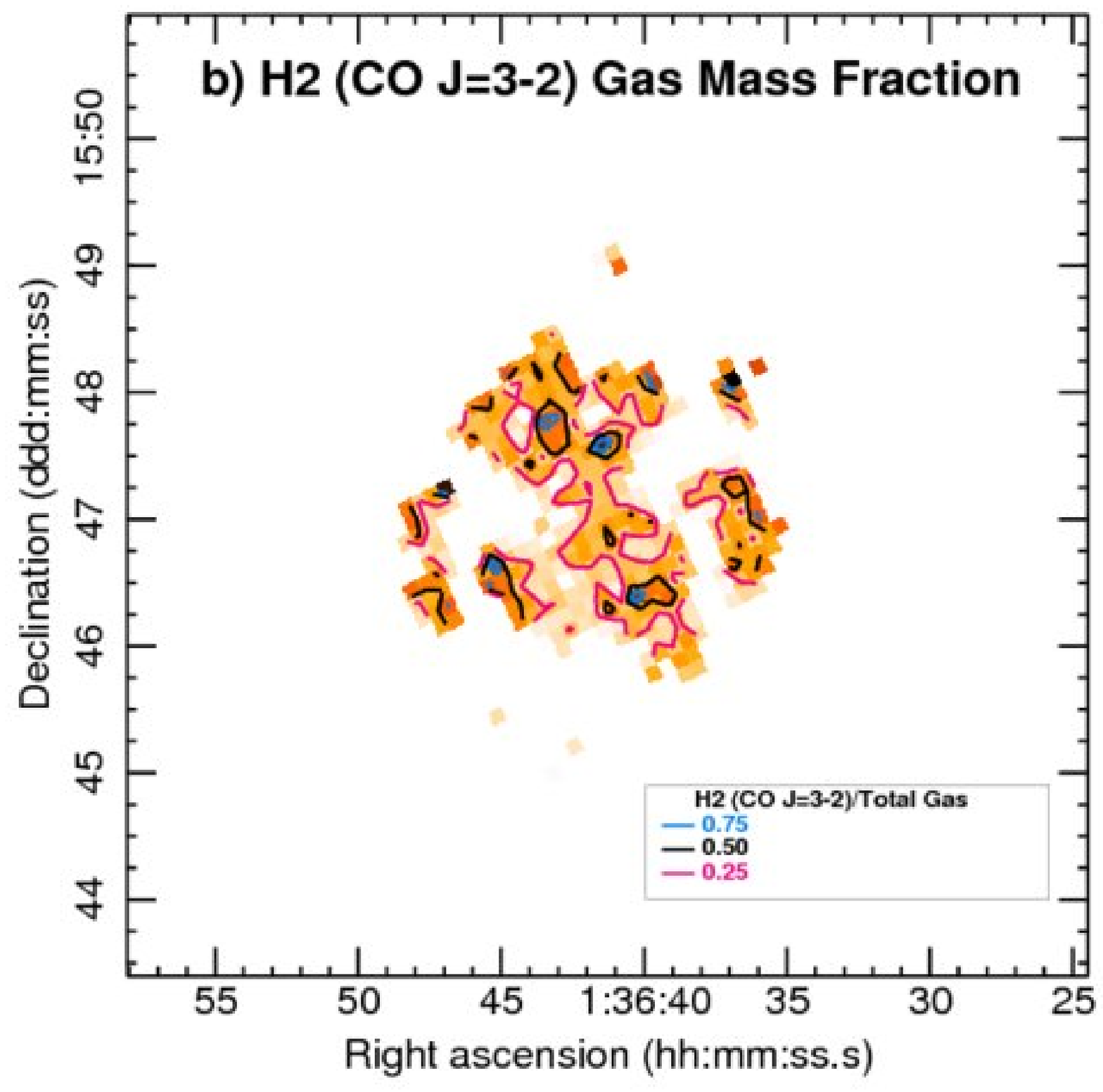}
  \includegraphics[width=85mm]{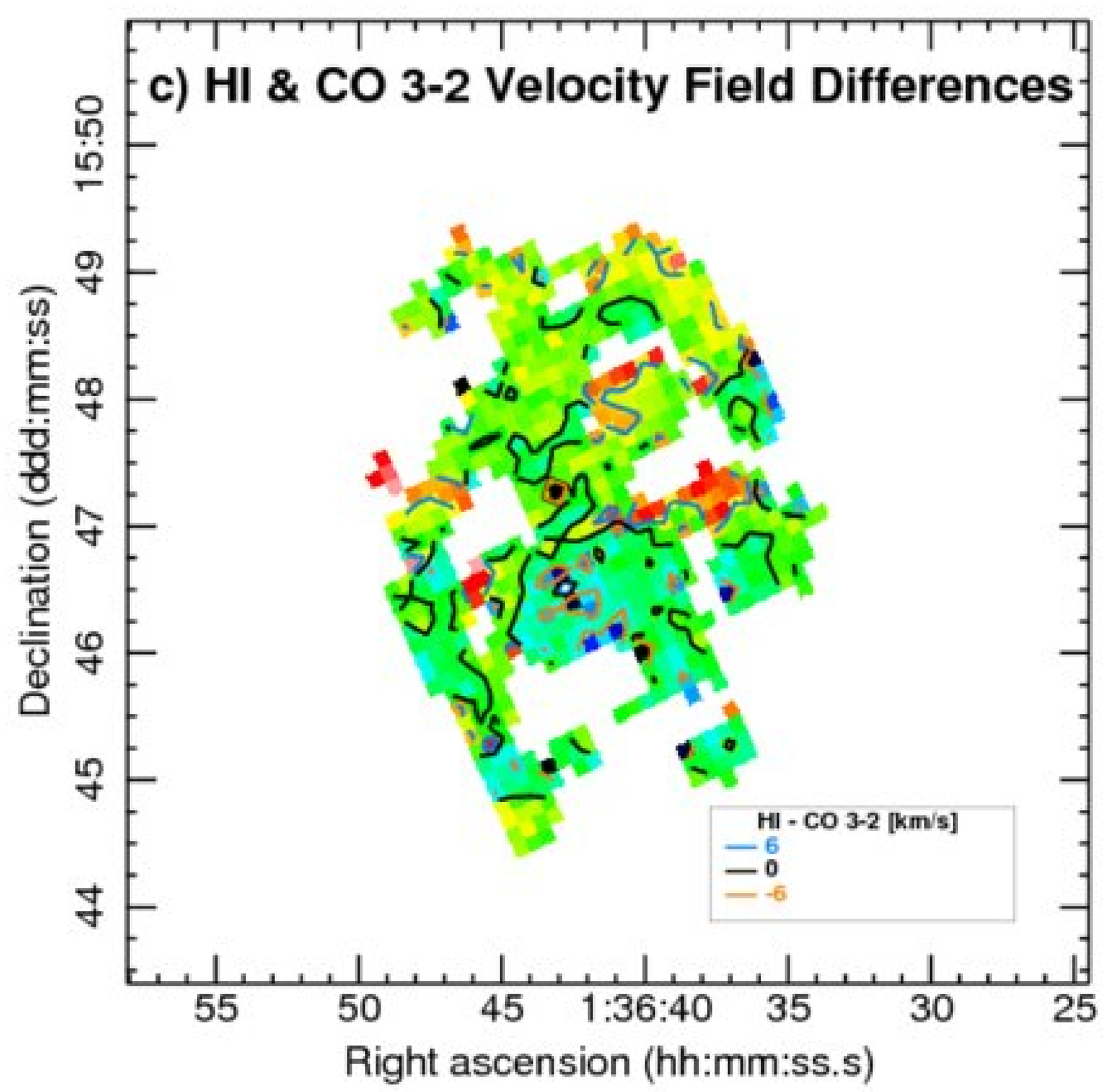}
  \includegraphics[width=85mm]{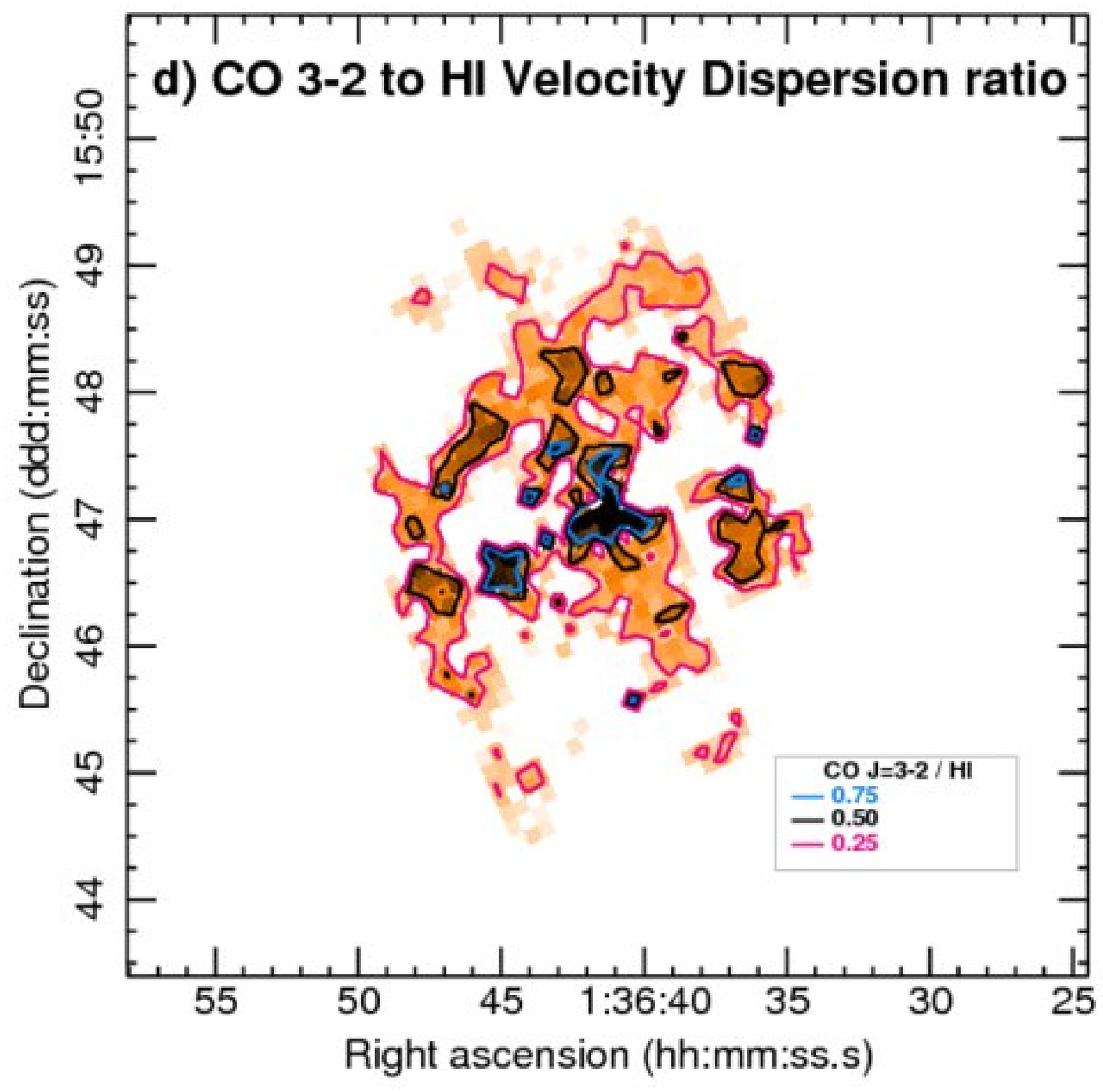}
\caption{Comparisons of the NGLS \co{} data to other atomic and molecular gas tracers for NGC\,628.  a) The ratio of the NGLS integrated \co{} intensity to the BIMA SONG integrated \cooz{} intensity.  The magenta, black, and cyan contours mark where the ratio is 0.15, 0.30, and 0.45, respectively.  b) The H$_{2}$ (CO\,$J$=3-2) molecular gas mass from the NGLS observations as a fraction of the total gas mass (H$_{2}$ [CO\,$J$=1-0] mass plus \hi{} mass from THINGS).  The magenta, black, and cyan contours mark where H$_{2}$ (CO\,$J$=3-2) makes up 25\%, 50\%, and 75\% of the total gas mass, respectively.  c) The velocity field differences between the THINGS \hi{} maps and the NGLS \co{} maps (\hi{} velocity field minus \co{} velocity field).  Redder values represent positive differences (\hi{} velocity greater than the \co{} velocity), while bluer values represent negative differences (\hi{} velocity less than the \co{} velocity).  The cyan, black, and orange contours represent differences of $+6$, 0 and $-6$\kms{}, respectively.  d) The ratio of the NGLS \co{} velocity dispersion to the THINGS \hi{} velocity dispersion.  The magenta, black, and cyan contours mark where the \co{} velocity dispersion is 25\%, 50\%, and 75\% of the \hi{} velocity dispersion, respectively.
\label{fig:n0628rat}}
\end{figure*}

\begin{figure*} 
  \includegraphics[width=85mm]{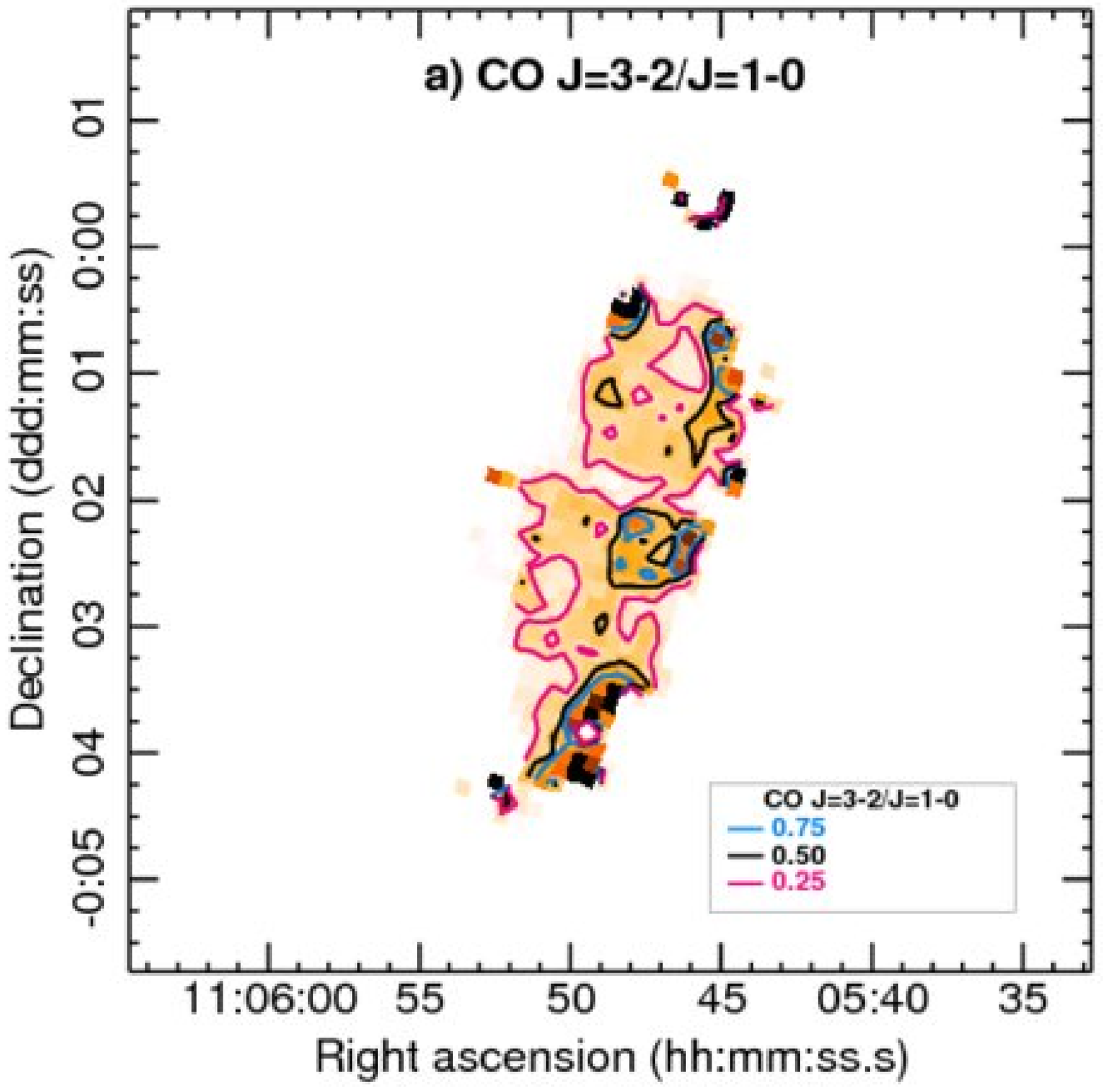}
  \includegraphics[width=85mm]{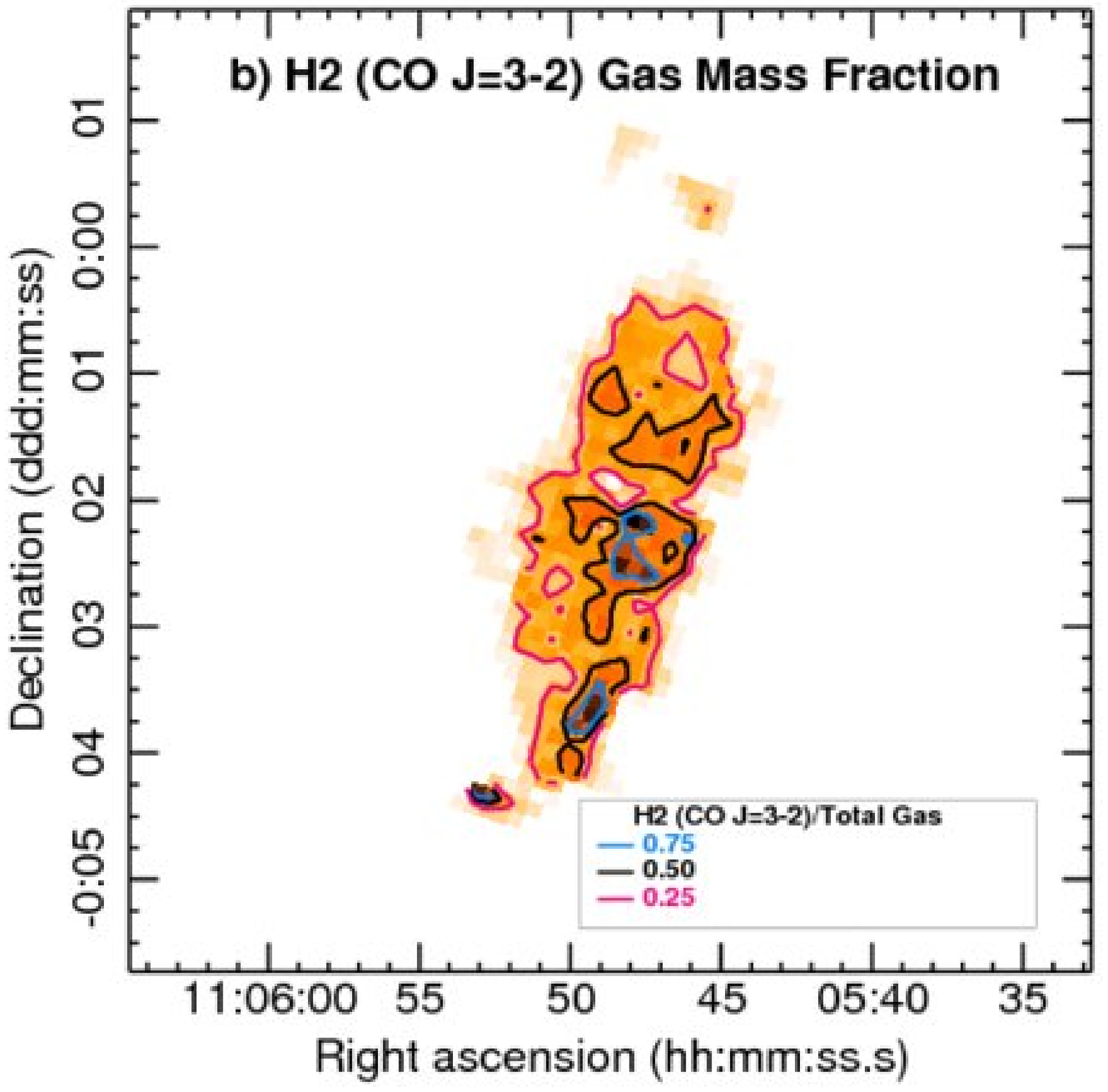}
  \includegraphics[width=85mm]{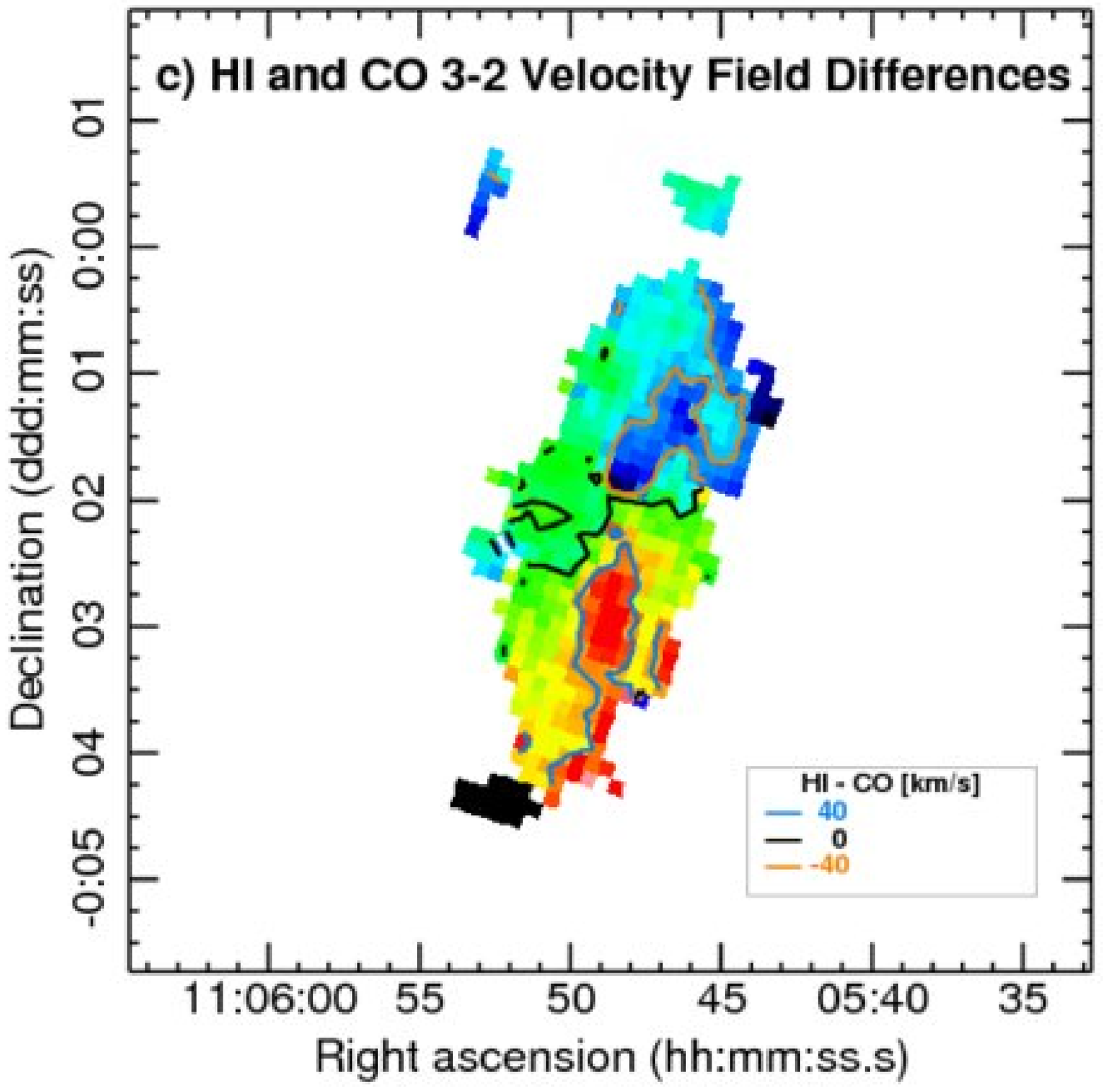}
  \includegraphics[width=85mm]{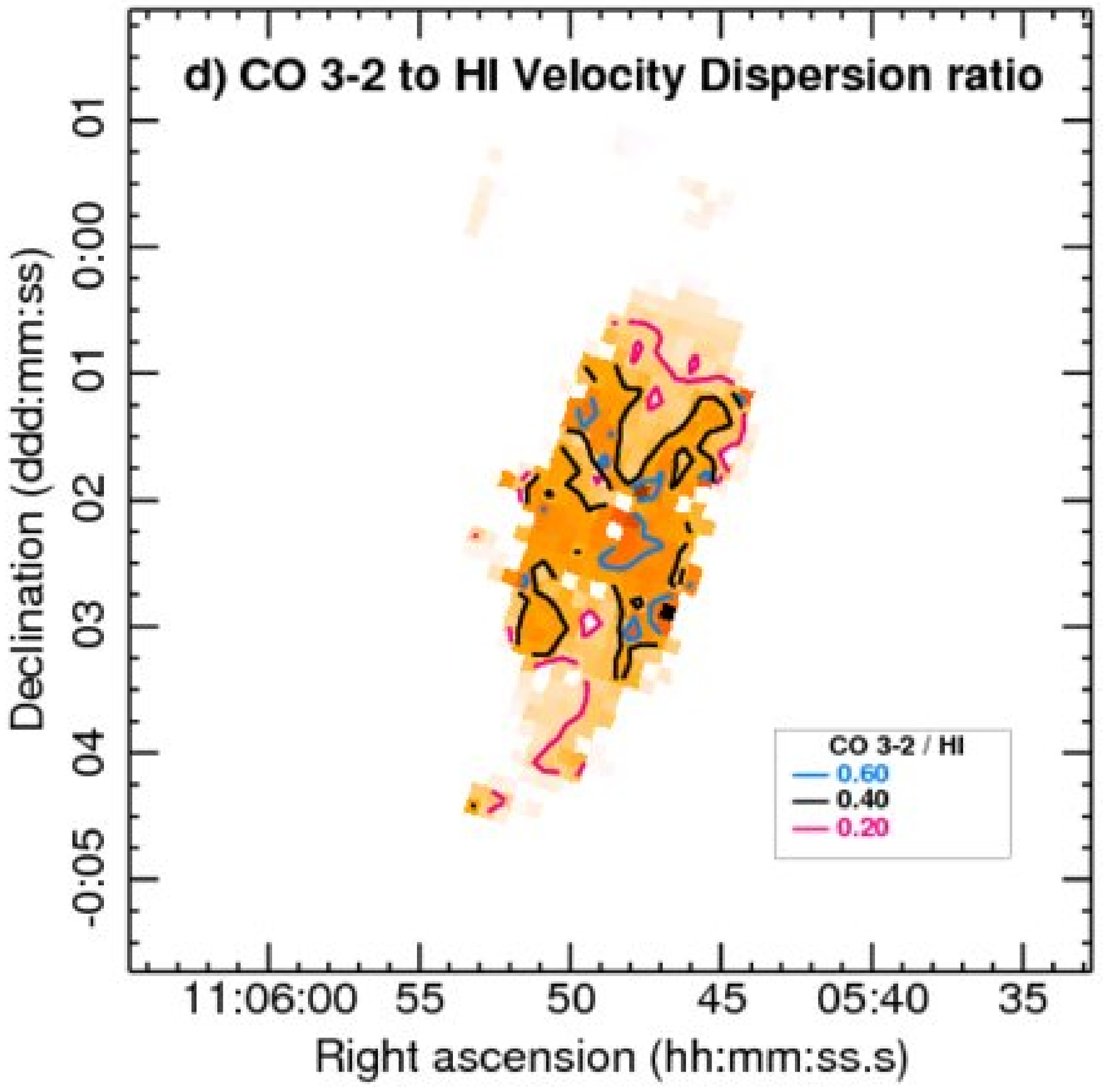}
\caption{Comparisons of the NGLS \co{} data to other atomic and molecular gas tracers for NGC\,3521.  a) The ratio of the NGLS integrated \co{} intensity to the \citet{kun07} integrated \cooz{} intensity.  The magenta, black, and cyan contours mark where the ratio is 0.25, 0.50, and 0.75, respectively.  b) The H$_{2}$ (CO\,$J$=3-2) molecular gas mass from the NGLS observations as a fraction of the total gas mass (H$_{2}$ [CO\,$J$=1-0] mass plus \hi{} mass from THINGS).  The magenta, black, and cyan contours mark where H$_{2}$ (CO\,$J$=3-2) makes up 25\%, 50\%, and 75\% of the total gas mass, respectively.  c) The velocity field differences between the THINGS \hi{} maps and the NGLS \co{} maps (\hi{} velocity field minus \co{} velocity field).  Redder values represent positive differences (\hi{} velocity greater than the \co{} velocity), while bluer values represent negative differences (\hi{} velocity less than the \co{} velocity).  The cyan, black, and orange contours represent differences of $+40$, 0 and $-40$\kms{}, respectively.  d) The ratio of the NGLS \co{} velocity dispersion to the THINGS \hi{} velocity dispersion.  The magenta, black, and cyan contours mark where the \co{} velocity dispersion is 20\%, 40\%, and 60\% of the \hi{} velocity dispersion, respectively.
\label{fig:n3521rat}}
\end{figure*}

\begin{figure*} 
  \includegraphics[width=85mm]{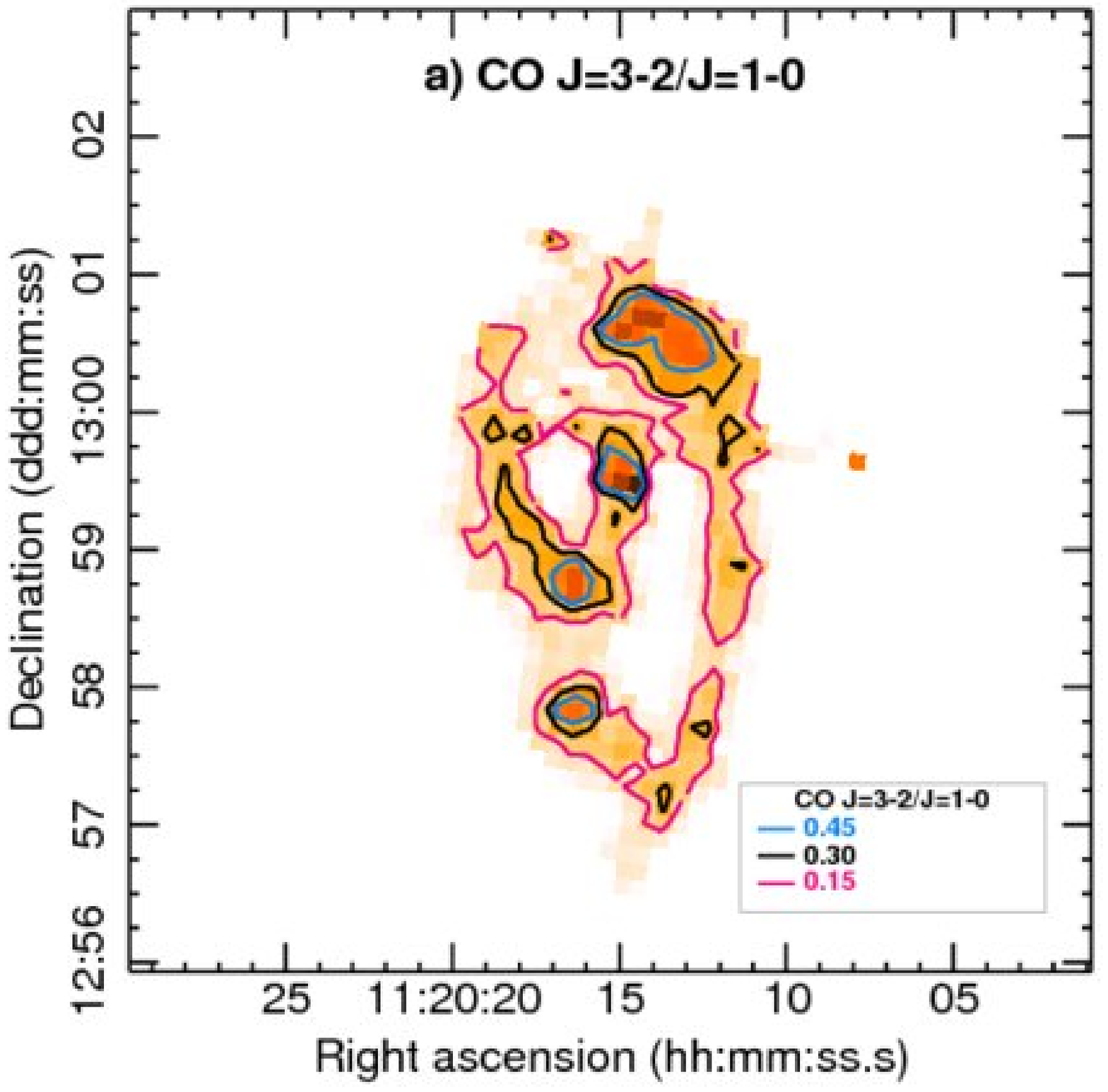}
  \includegraphics[width=85mm]{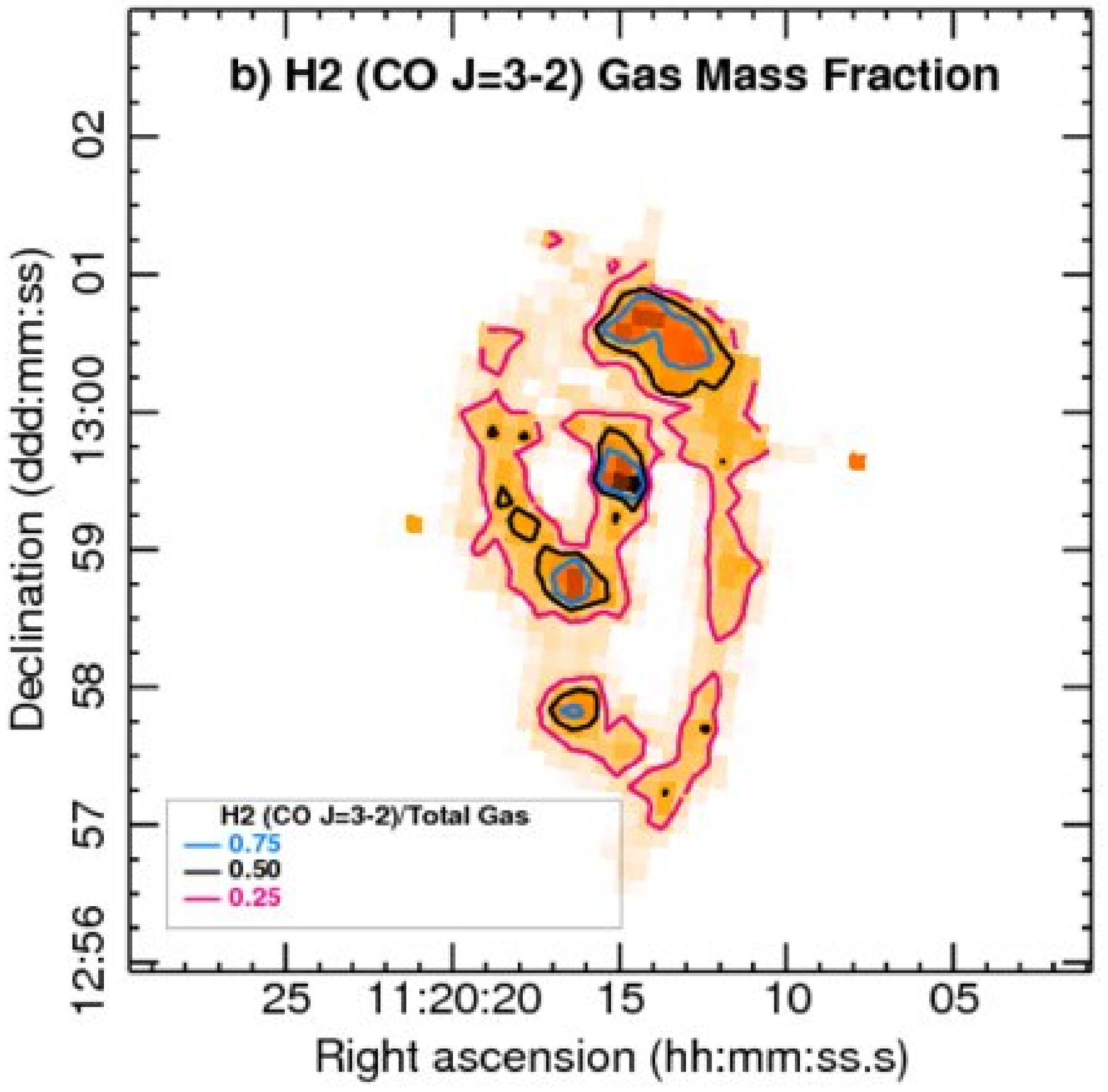}
  \includegraphics[width=85mm]{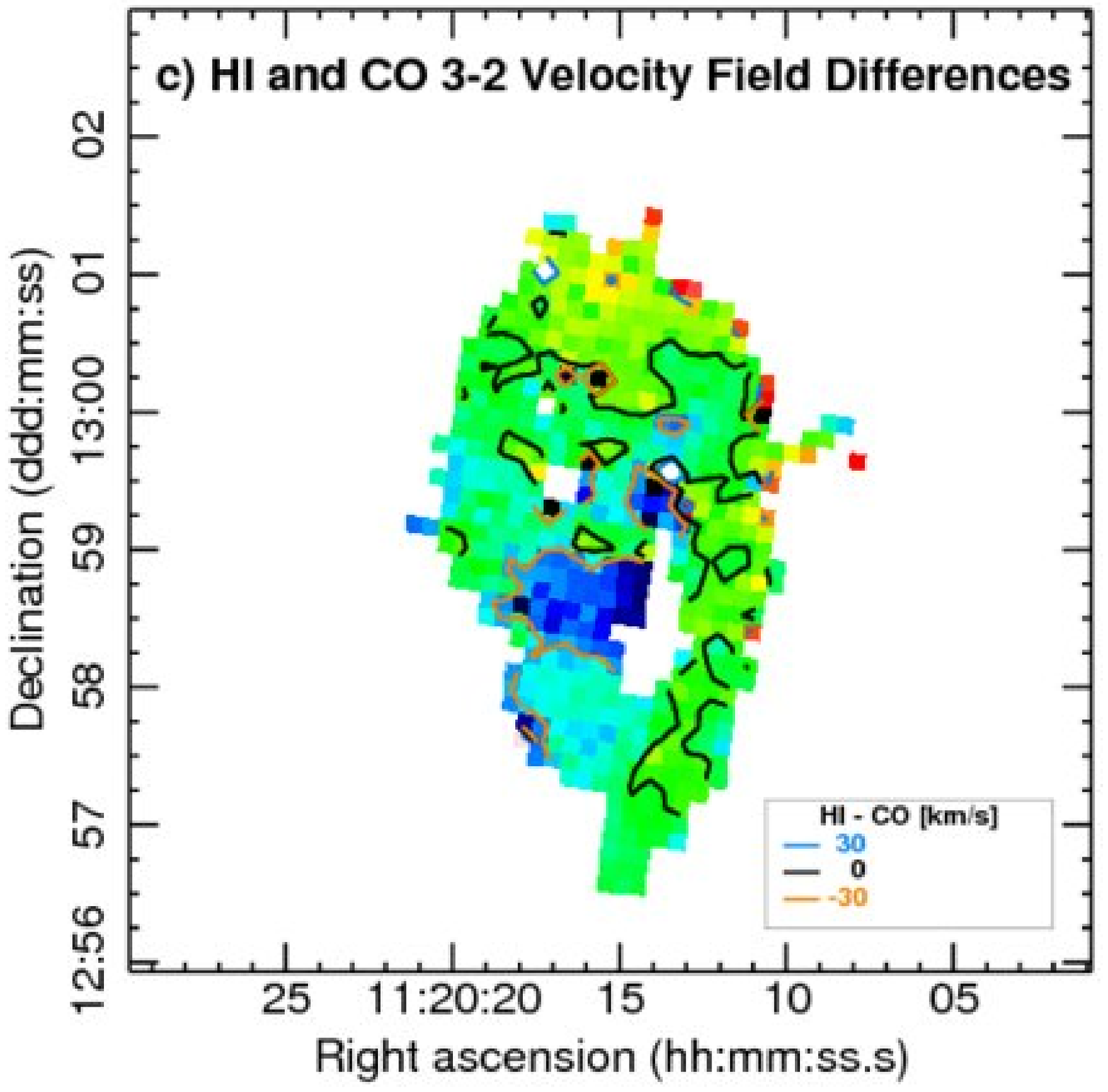}
  \includegraphics[width=85mm]{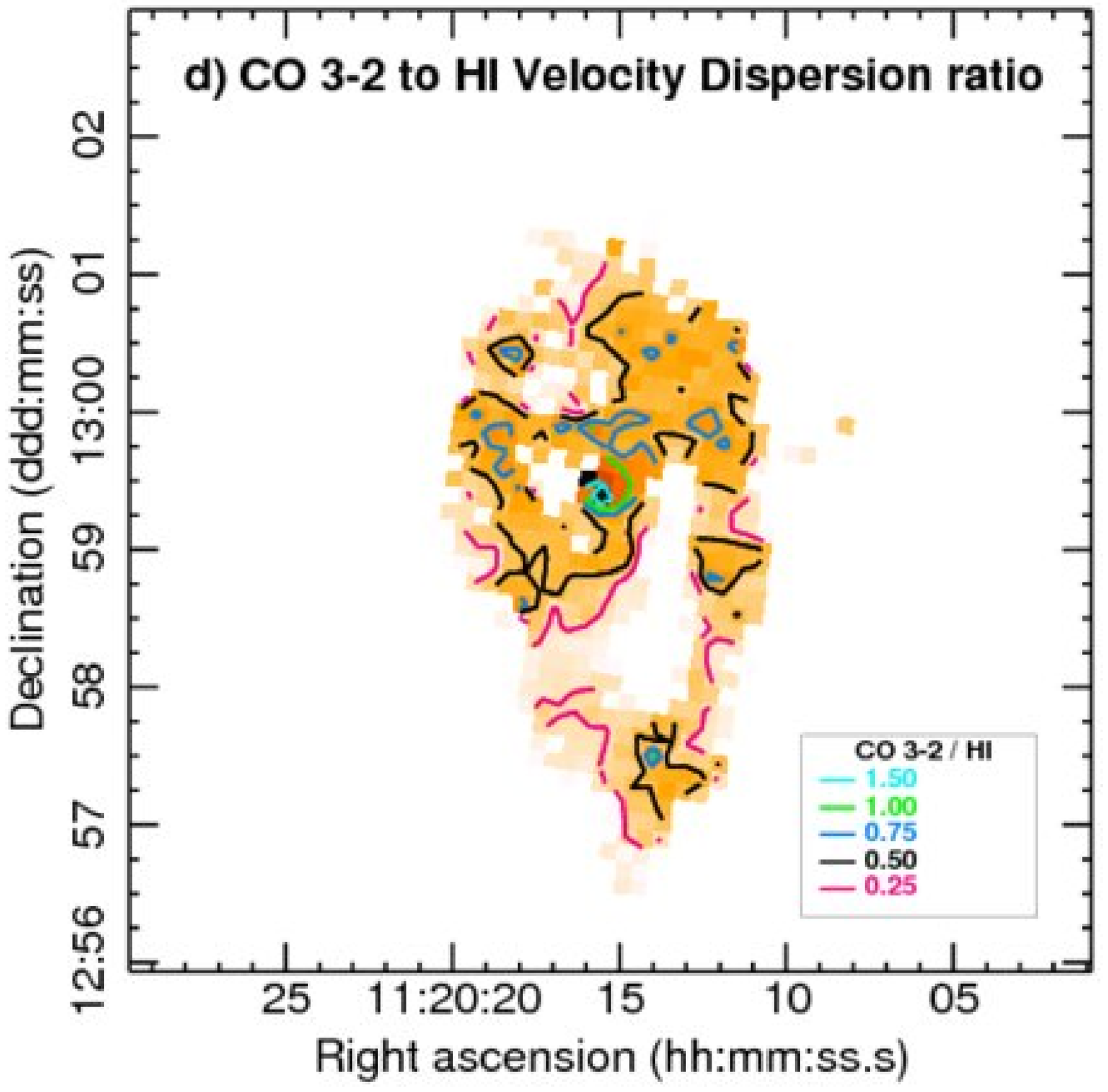}
\caption{Comparisons of the NGLS \co{} data to other atomic and molecular gas tracers for NGC\,3627.  a) The ratio of the NGLS integrated \co{} intensity to the \citet{kun07} integrated \cooz{} intensity.  The magenta, black, and cyan contours mark where the ratio is 0.15, 0.30, and 0.45, respectively.  b) The H$_{2}$ (CO\,$J$=3-2) molecular gas mass from the NGLS observations as a fraction of the total gas mass (H$_{2}$ [CO\,$J$=1-0] mass plus \hi{} mass from THINGS).  The magenta, black, and cyan contours mark where H$_{2}$ (CO\,$J$=3-2) makes up 25\%, 50\%, and 75\% of the total gas mass, respectively.  c) The velocity field differences between the THINGS \hi{} maps and the NGLS \co{} maps (\hi{} velocity field minus \co{} velocity field).  Redder values represent positive differences (\hi{} velocity greater than the \co{} velocity), while bluer values represent negative differences (\hi{} velocity less than the \co{} velocity).  The cyan, black, and orange contours represent differences of $+30$, 0 and $-30$\kms{}, respectively.  d) The ratio of the NGLS \co{} velocity dispersion to the THINGS \hi{} velocity dispersion.  The magenta, black, cyan, green, and pale blue contours mark where the H$_{2}$ velocity dispersion is 25\%, 50\%, 75\%, 100\%, and 150\% of the \hi{} velocity dispersion, respectively.
\label{fig:n3627rat}}
\end{figure*}

The \corat{} ratio is an important indicator of the portion of warm/dense molecular gas relative to the total molecular gas.  As noted in \pI{} it is also important in the calculation of the molecular gas mass, as to date conversions from CO luminosity have been done with the \cooz{} line.  We calculated \corat{} maps for all three galaxies by the same method as \pI{}, taking the NGLS \co{} integrated intensity maps (in main beam temperature) and dividing by the available \cooz{} maps of the same resolution (Nobeyama maps for NGC\,3521 and NGC\,3627, and the smoothed BIMA SONG \cooz{} maps for NGC\,628).  The resulting \corat{} ratio maps are shown in the upper left panels of Figs.~\ref{fig:n0628rat}, \ref{fig:n3521rat}, and \ref{fig:n3627rat}, for NGC\,628, NGC\,3521, and NGC\,3627, respectively.

The \co{} detection in NGC\,628 is relatively weak compared to the other two sources, and the rapid drop off in the BIMA SONG mapping sensitivity beyond $\sim$1\farcm5 ($\sim$3.2~kpc) from the galaxy center reduces the useful area of the \corat{} ratio map.  Within the useful region, the gas distribution traced by both $^{12}$CO transitions follow each other closely.  The \corat{} ratio is generally low (mostly between 0.15 and 0.45) with the exception of a few peaks (some of which may be due to poor sensitivity in one tracer or the other).  It is difficult to discern a trend with the limited coverage we have for this galaxy.  For NGC\,3521 the \corat{} ratio map shows that throughout most of the galaxy the \corat{} is in the range 0.25 to 0.5.  In the center of the galaxy, where the NGLS \co{} map shows a strong dip (see Fig.~\ref{fig:ngc3521}), and the \citet{kun07} \cooz{} also shows a central depression, the ratio drops below 0.25.  Just to the southwest of the center the ratio peaks over 0.75, with some stronger peaks around the edges of the map.  For NGC\,3627, the general \co{} structure matches well with both the BIMA SONG and \citet{kun07} \cooz{} maps of the galaxy, though there is more inter-arm gas in those maps than in the NGLS map.  The \corat{} ratio is low throughout most of the galaxy where we detect \co{}, only going above 0.45 in four locations (the center, the bar ends, and the concentration to the south of the southern bar end).  The median \corat{} ratios for NGC\,628, NGC\,3521, and NGC\,3627 are 0.35, 0.33, and 0.18, respectively (for pixels where the \co{} signal-to-noise is 3 or greater).

The standard method to calculate the column density of H$_2$ gas using \cooz{} as a tracer is to adopt a CO-to-H$_2$ conversion factor ($X_{CO}$).  As in \pI{} we are adopting a value of $2\times 10^{20}$ cm$^{-2}$ (K km s$^{-1}$)$^{-1}$ \citep{str88}.  In order to calculate the molecular gas mass from \co{} measurements we need to adopt a \corat{} value to change our \cooz{}-to-H$_2$ conversion factor in to a \co{}-to-H$_2$ conversion factor.  The choice of this conversion ratio depends strongly on where we assume the molecular gas that we are tracing with \co{} is located.  Our assumption is that \co{} traces pre-star formation molecular gas that is in general denser than \cooz{}.  In \pI{} we used a constant ratio of 0.6, representative of ratios seen in Galactic and extragalactic GMCs \citep[based on the results of ][]{ler08,wil99,isr09a,isr09b}, essentially focusing on the molecular gas closest to the star formation (unlike the above median \corat{} ratios for the three galaxies, which are more representative of the global molecular gas population).  The limitation of this assumption is that the cause of the differences between different molecular gas tracers is degenerate, thus higher \co{} values could indicate warmer rather than denser gas \citep[as found by][]{oka07}.  It has also been suggested \citep{tos07} some \co{} emission may be due to post-star formation heating of the molecular gas.  In this paper, where we calculate an H$_2$ mass using this assumption (\corat{} = 0.6) we will refer to `H$_2$ (CO\,$J$=3-2).'

If we instead wish to examine the more diffuse general molecular gas content in a galaxy then the available \cooz{} data is a more appropriate tracer.  So in cases where we require the total gas content we use this instead, using the adopted $X_{CO}$ value to obtain the H$_2$ gas column density, and refer to the `H$_2$ (CO\,$J$=1-0)' gas mass.  As the only \cooz{} observations we have for NGC\,628 come from BIMA SONG, and given the limitations of those data described above (see \S\ref{sec:anc}), we have only used parameters needing H$_2$ (CO\,$J$=1-0) for the other two galaxies in our later analysis.

In order to compare the dense molecular gas content to the total gas (\hi{} plus H$_2$ [CO\,$J$=1-0]), we produced maps of the H$_2$ (CO\,$J$=3-2) gas mass surface density fraction (\molsd{} divided by \himolsd{}), where the H$_2$ surface densities were derived as above, and the \hi{} surface density came from the THINGS integrated \hi{} intensity maps smoothed to the JCMT beam.  The resulting  H$_2$ gas mass fraction maps are shown in the upper right panel of Figs.~\ref{fig:n0628rat}, \ref{fig:n3521rat}, and \ref{fig:n3627rat}.  Since \molsd{} is not in the denominator for this calculation it should be noted here that there are some pixels in these maps where this fraction is greater than 1.

As for the \corat{} map of NGC\,628, the H$_2$ (CO\,$J$=3-2) gas mass fraction map for this galaxy is limited by the BIMA SONG \cooz{} map.  The THINGS moment maps for NGC\,628 show the \hi{} intensity dips in the central $\sim$1\arcmin{} ($\sim$2.1~kpc) and there is little correlation between the \co{} and \hi{} in this area; beyond this radius they both trace the spiral structure (though this goes beyond the useful area of the \cooz{} data).  In the limited regions where we have all three gas tracers we find the dense molecular gas is only the dominant gas phase at a few peaks in the map.  It is difficult to draw any conclusions from this map given the limited overlap between the three gas tracers.  

In NGC\,3521, the THINGS \hi{} map shows a larger central depression, that encompasses most of the region with molecular gas detections.  Molecular gas in general dominates the inner regions of the galaxy (out to $\sim$1\farcm5 from the center along the major axis, or about 4.7~kpc at the adopted distance).  Dense/warm molecular gas appears to be the dominant gas phase in several large patches of the gas disk, with a strong peak just south of the galaxy center where the dense molecular gas is more than three quarters of the total gas along the line-of-sight, and a similar region further out to the south.  The most notable location with little dense molecular gas is a small region slightly north of the center, in approximately the same location we see a dip in the \co{} distribution map (Fig.~\ref{fig:ngc3521}b).

Molecular gas is in general the dominant form of gas in NGC\,3627, in particular throughout the bar and most of the spiral arms.  The \co{} detections are particularly strong at four peaks; in the central starburst region, at both ends of the bar, and a peak to the south between one of the bar ends and the south-western arm.  Atomic gas on the other hand is mostly absent in the bar except at the north-western end, but traces most of the spiral arms.  As a result the map shows dense/warm gas is the dominant gas phase at the four peak locations (in the case of the central starburst region overwhelmingly so), but not throughout the rest of the bar and spiral arms.  The map strongly resembles the \corat{} map for this galaxy, indicating that cooler and/or lower density molecular gas is the dominant phase outside the four peaks.

\begin{figure} 
  \plotone{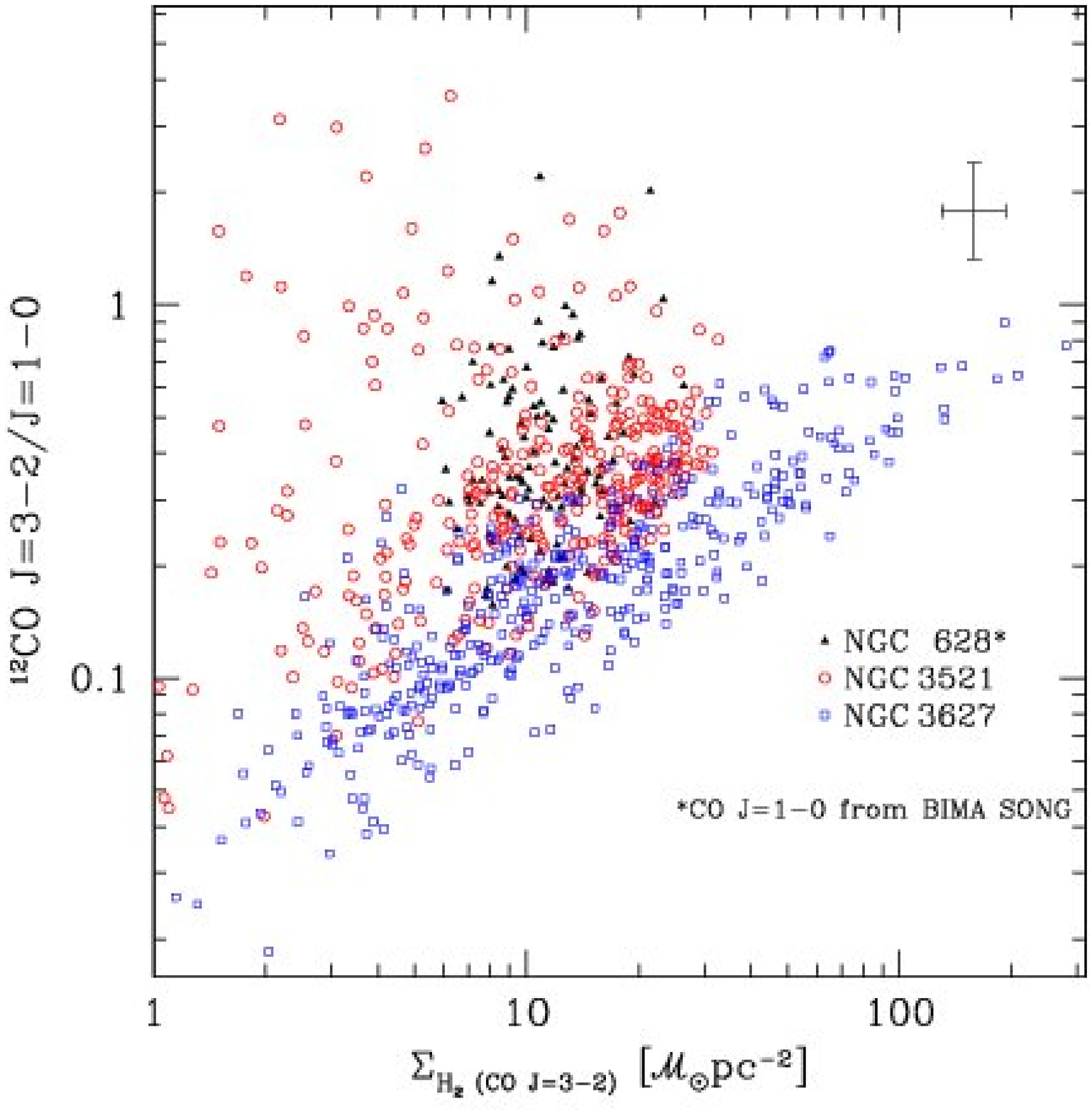}
\caption{Plot of the pixel-to-pixel variation in the \corat{} ratio versus the dense H$_2$ (CO\,$J$=3-2) gas surface density for all three galaxies.  \corat{} is calculated by dividing our integrated \co{} intensity map (main beam temperature scale) by the corresponding integrated \cooz{} intensity (from \citet{kun07}, except for NGC\,628 (*) where the BIMA SONG \cooz{} data were used instead).  \molsd{} is calculated from our NGLS \co{} observations using the assumption for dense gas (\corat{} = 0.6, see text), and has been corrected for helium fraction and inclination \citep[using the inclinations from][]{wal08,deb08}.  NGC\,628 is marked with solid black triangles, NGC\,3521 with open red circles, and NGC\,3627 with open blue squares (these symbols are used for all subsequent plots).  We have only included pixels where the signal-to-noise for the NGLS \co{} data is greater than 3.  The error bars indicate a typical calibration uncertainty.
\label{fig:corats}}
\end{figure}

Fig.~\ref{fig:corats} shows the pixel-to-pixel variation in the \corat{} ratio versus H$_2$ gas surface density (from \co{}, inclination and helium corrected) for all three galaxies.  Throughout the vast majority of all three galaxies the ratio is less than 0.5.  There is a discernible trend for NGC\,3627, with higher surface density regions having higher ratios.  This trend is less obvious for NGC\,3521, and not clear at all from the limited NGC\,628 data.  In general, while regions with low H$_2$ (CO\,$J$=3-2) gas surface density can have a wide range in \corat{}, regions of high H$_2$ (CO\,$J$=3-2) gas surface density only show high ratios.  This fits with our assumption that \co{} tends to trace warmer, denser regions than \cooz{}.
Given that starburst regions in galaxies are often observed to have higher \corat{} ratios \citep{mei01}, it is somewhat surprising that the starburst galaxy NGC\,3627 generally has the lowest ratio of these three galaxies at a given molecular gas surface density.

\begin{figure} 
  \plotone{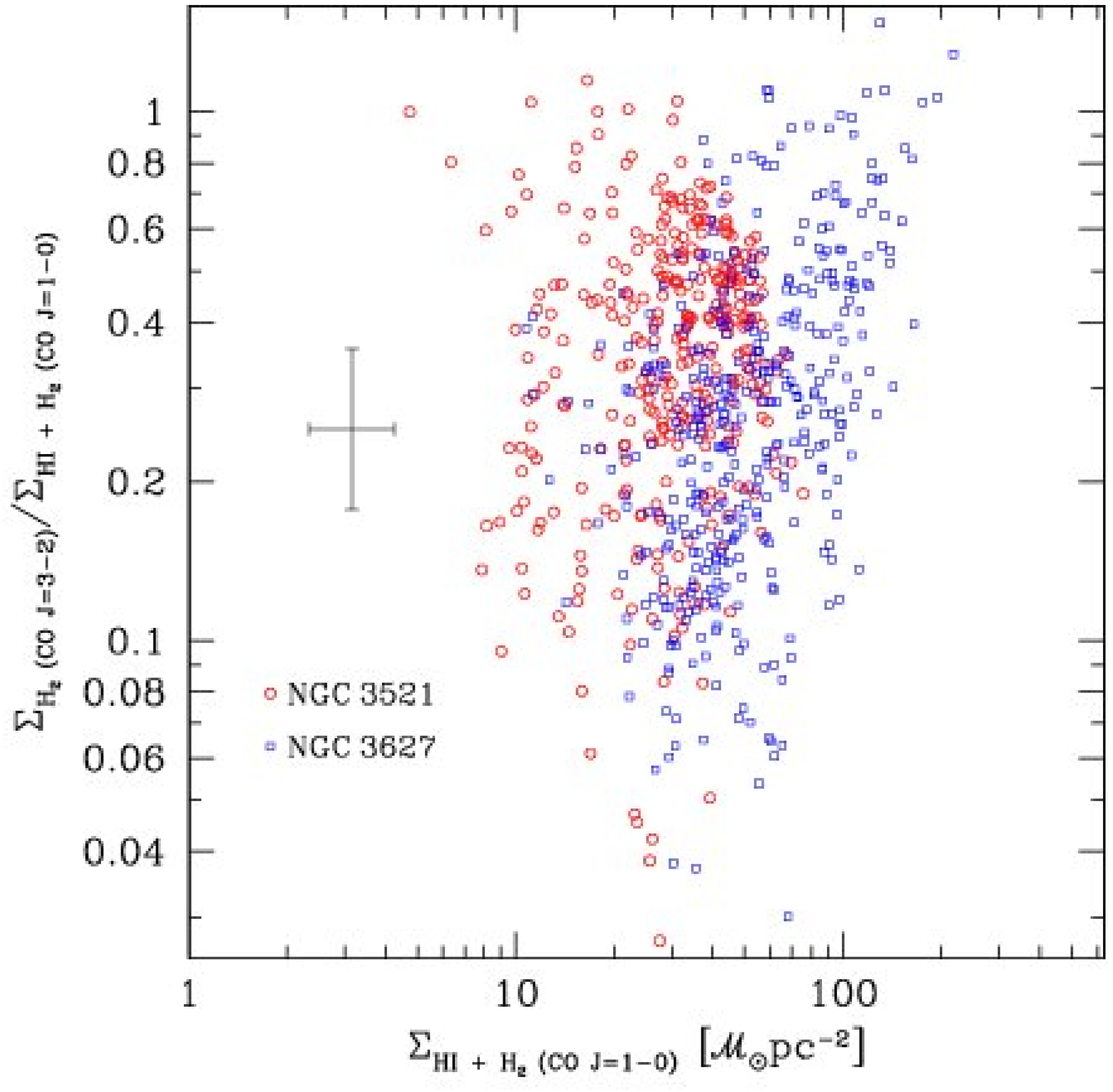}
\caption{Plot of the pixel-to-pixel variation in the H$_2$ (CO\,$J$=3-2) gas surface density fraction (\molsd{} divided by \himolsd{}) verses total gas surface density (\hi{} mass $+$ H$_2$ [CO\,$J$=1-0] mass) for the two galaxies where we have \citet{kun07} \cooz{} maps available.  Symbols are as in Fig.~\ref{fig:corats}.  All gas surface densities were corrected for helium content, and we applied an inclination correction where appropriate \citep[using the inclinations from][]{wal08,deb08}.  We have only included pixels where the signal-to-noise for the NGLS \co{} data is greater than 3.  The error bars indicate a typical calibration uncertainty.
\label{fig:h2frac}}
\end{figure}

Fig.~\ref{fig:h2frac} shows the pixel-to-pixel variation in the H$_2$ (CO\,$J$=3-2) gas surface density fraction, \moltotsd{}, versus the total gas surface density \himolsd{} (inclination and helium corrected) for the two galaxies where we have \citet{kun07} \cooz{} maps available.  Only pixels where the signal-to-noise for the NGLS \co{} data is greater than 3 were included.  The absence of points below $\sim$10\Msun{}pc$^{-2}$ is most likely a result of the detection limits of the three surveys, the signal-to-noise cutoff that we subsequently applied, and being biased towards pixels where \co{} was detected in the NGLS.  With the large scatter in the points it is difficult to discern any trend between the global gas surface density and the fraction of dense/warm molecular gas, and in general the plot suggests there is no relation between gas surface density and gas composition.  This may be consistent with results such as \citet{ler08} with regards to the influence general ISM gas densities have on denser structures more closely associated with star formation, provided we can confirm it with measurements from other galaxies at a later stage.

\section{Gas Dynamics}
\label{sec:resdyne}

We compared the NGLS \co{} velocity fields and velocity dispersion maps for our three galaxies to the THINGS \hi{} velocity fields and velocity dispersion maps \citep{wal08}.  The THINGS \hi{} data cubes were first convolved with a Gaussian beam to the resolution of the JCMT \co{} maps (14\farcs5), and then collapsed into moment maps, as described in \S~\ref{sec:data}.  To compare velocity fields we simply subtracted the NGLS \co{} fields from the THINGS \hi{} fields.  To compare velocity dispersions we divided the NGLS \co{} velocity dispersion maps by the \hi{} velocity dispersion maps to get the ratio of the velocity dispersions.  The resulting velocity field difference and velocity dispersion ratio maps are shown in the lower two panels of Figs.~\ref{fig:n0628rat}, \ref{fig:n3521rat}, and \ref{fig:n3627rat}.  It should be noted that our linear resolution elements for the three galaxies are of the order of several hundred parsecs, relatively large compared to the internal structures of the galaxies, and therefore the gas traced by \hi{} and \co{} at the same location in the maps may be physically some distance apart.

Over the regions in NGC\,628 where we detect \co{}, the velocity field differences are generally small (absolute differences almost entirely $<6$\kms{}) in comparison to the other two galaxies, no doubt in part due to the low inclination of this galaxy (and hence small component of the rotation along the line of sight).  There does appear to be a slight gradient across the field, with the northern (receding) end of the galaxy having \hi{} velocities marginally greater than \co{} velocities, and the southern (approaching) end showing slightly lower \hi{} velocities than \co{} velocities.  The extent to which this gradient is real is not clear, and would require a more extensive tracer of the molecular gas to expand the area where we can compare the velocities over the very extended \hi{} disk.  The \co{} velocity dispersion is significantly below that of the atomic gas throughout the entire region where we detect \co{}, with few regions above 50\% of the \hi{} velocity dispersion (the median ratio of the dispersions is 34\%).  In general we expect the molecular gas to have lower velocity dispersions because it is confined to a colder, thinner disk.

There are large regions of disagreement between the atomic and molecular velocity fields for the highly inclined galaxy NGC\,3521, with peak differences of up to $\pm 40$\kms{} in the regions to the north and south of the galaxy center.  Unlike the gradient seen for NGC\,628, the differences appear to be more clump-like, and are in the opposite sense to that galaxy (higher \co{} velocities on the receding side, and vice versa).  Examination of a position-velocity slice along the major axis through the THINGS \hi{} cube for NGC\,3521 show significant quantities of low velocity atomic gas.  This gas following a non-circular rotation pattern, which is not detected in the \co{} data cube, appears to account for the large differences in the \hi{} and \co{} velocity fields.  \citet[][and references therein]{deb08} discuss removal of such low velocity gas from the moment maps for THINGS, so that velocity fields best represent the rotational pattern of the galaxy.  Presumably, if we compared an \hi{} velocity field that better reflected the circular rotation with our \co{} velocity field, there would be fewer large differences.  Fig.~\ref{fig:n3521rat}d shows the \co{} to \hi{} velocity dispersion ratios for NGC\,3521.  The \co{} velocity dispersion is again much lower than for the atomic gas, of similar order to NGC\,628.  The median ratio of the velocity dispersions is 30\%, though again these velocity dispersions ratios are likely affected by the non-circular \hi{} gas and may be more similar if only \hi{} in circular rotation were used.

Fig.~\ref{fig:n3627rat}c shows the velocity field differences for NGC\,3627.  The molecular and atomic velocity fields for this galaxy are well matched, with the exception of the region around the nucleus and the southern end of the bar where the \co{} velocities are over 30\kms{} higher in some parts.  This is the receding side of the galaxy, though it is notable that this region has relatively little \hi{} emission.  The bar would also strongly affect the dynamics in this region.  Fig.~\ref{fig:n3627rat}d shows the \co{} to \hi{} velocity dispersion ratios for NGC\,3627.  The \co{} velocity dispersion is closer to the \hi{} velocity dispersions for this galaxy compared to the others, though most of the galaxy still has \co{} velocity dispersions $<75$\% of the \hi{} values.  The median ratio of the velocity dispersions is 42\%.  Most notable is the nucleus where the \co{} velocity dispersion is in fact greater than \hi{} values.  As noted in \S~\ref{sec:resratio} the molecular gas dominates the central starburst region, making up almost all of the gas (95\% to 99\%) at the nucleus, so it is likely that the atomic velocity dispersion here is based on low signal-to-noise data.

\section{Star Formation}
\label{sec:ressf}

\subsection{Star Formation Rate and Gas Depletion Time}
\label{sec:ressfmaps}

To make a comparison between the present star formation rate (SFR) and our measurements of dense/warm molecular gas derived from NGLS \co{} observations we have produced maps of the rate of star formation from the available ancillary data.  We followed the same procedures as \pI{} to produce SFR maps, and maps of the gas depletion time for H$_2$ (CO\,$J$=3-2).  Star formation rates were derived from the available SINGS H$\alpha$ and 24~$\mu$m maps using the procedure derived by 
\citet[][eqn. 7]{cal07}, 
\begin{equation}  
\begin{split}
{\rm SFR}({\cal M}_{\sun}\,{\rm yr}^{-1}) = &\ 5.3\times10^{-42} [L({\rm H}\alpha)_{\rm obs} + \\ & (0.031\pm0.006)L(24\mu {\rm m})],
\label{eqn:calsfr}
\end{split}
\end{equation}
where the H$\alpha$ and 24~$\mu$m are both in erg s$^{-1}$, and $L(24\mu {\rm m})$ is expressed as $\nu L(\nu)$.  This calculation assumes an initial mass function (IMF) consisting of two power laws, unlike the similar derivation in \citet{ken98a} that uses a Salpeter IMF.  
By using both H$\alpha$ and 24~$\mu$m data we account for both unobscured and obscured star formation within the galaxies.

\begin{figure} 
  \includegraphics[width=85mm]{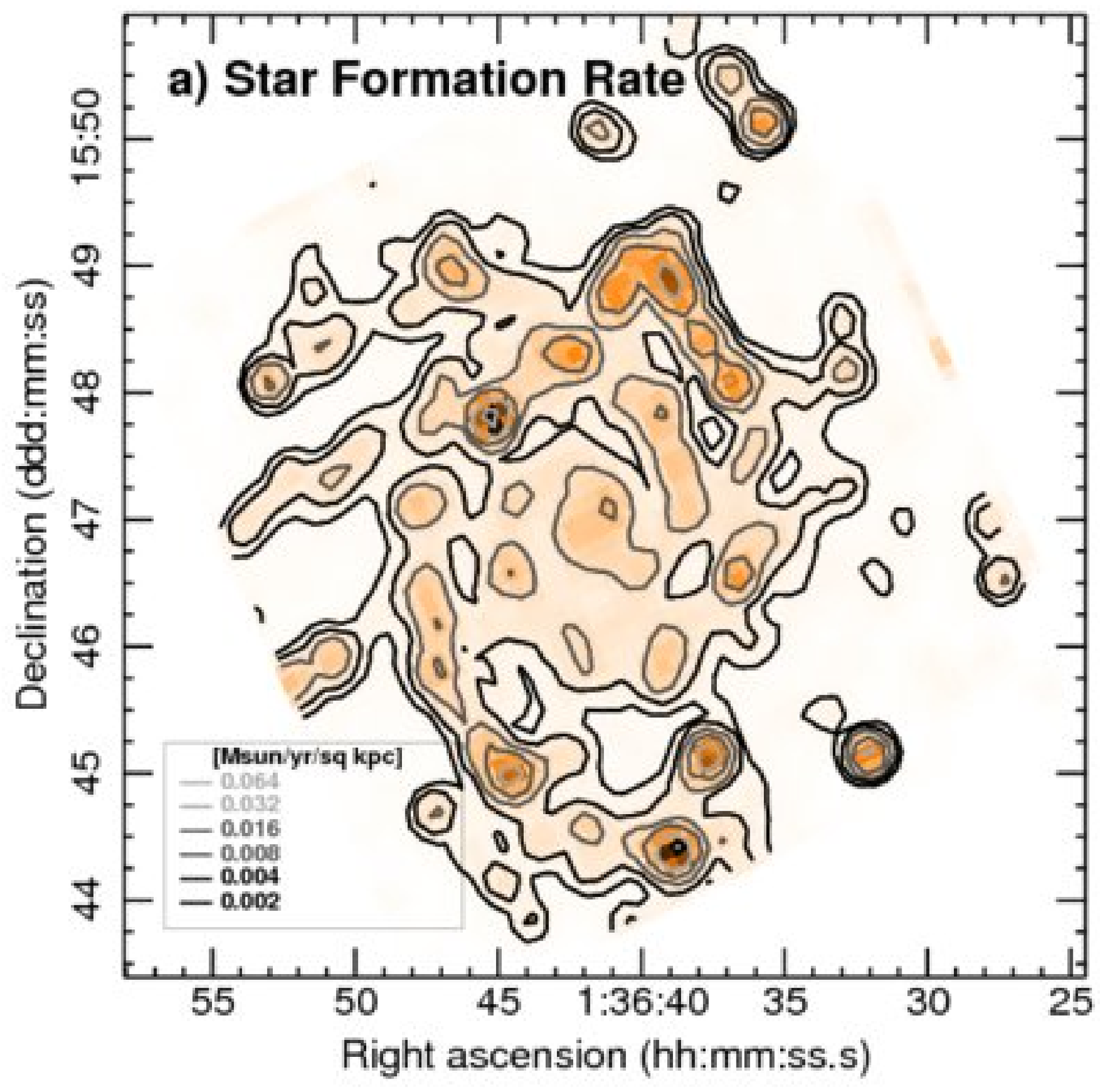}
  \includegraphics[width=85mm]{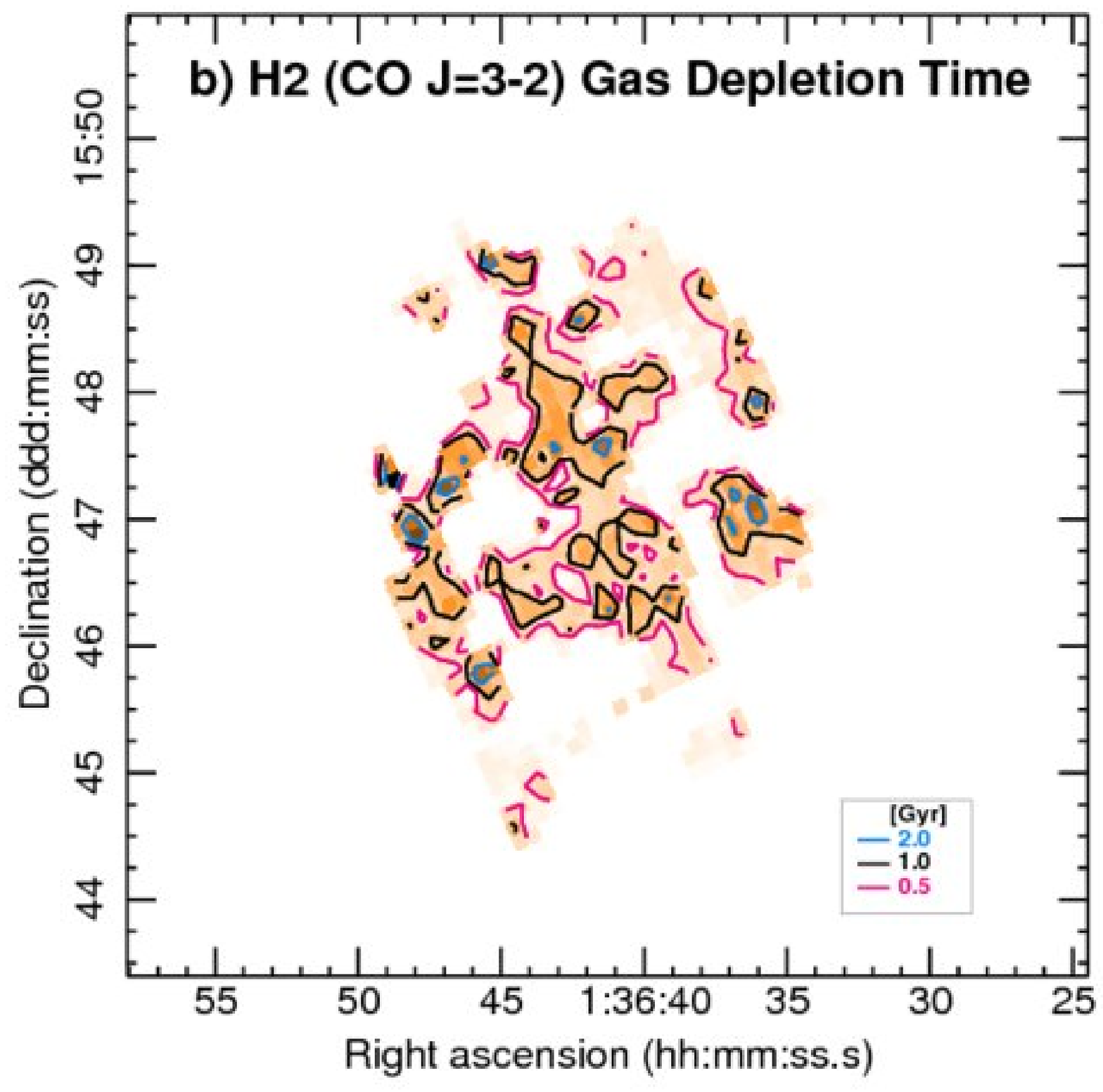}
\caption{a) Star formation rate map for NGC\,628, derived from the SINGS H$\alpha$ and 24~$\mu$m images \citep{ken03}, convolved to the JCMT beam at 850~$\mu$m.  Contour levels are 0.002, 0.004, 0.008, 0.016, 0.032, and 0.064~${\cal M}_{\sun}\,{\rm yr}^{-1}\,{\rm kpc}^{-2}$.  b) H$_2$ (CO\,$J$=3-2) gas depletion time (gas mass divided by SFR) for NGC\,628 in Gyr.  The magenta, black, and cyan contours mark where the gas depletion time is 0.5, 1.0, and 2.0~Gyr, respectively.
\label{fig:ngc0628sf}}
\end{figure}

The gas depletion time, which is also the inverse of the star formation efficiency (SFE), is simply the mass of the gas in question divided by star formation rate.  This assumes that the gas involved will eventually be used up in the process of forming stars, and that no other process is replenishing this gas.  While these may hold somewhat for the total gas population over the lifetime of the galaxy (if there is no external influence), they are unlikely to be the case for the dense/warm molecular gas we trace with \co{}.  If we are only considering how quickly the current supply of gas is being consumed for star formation, and not the longer term implications of when that gas might run out, this should not be a concern.  Using our calculation for the dense/warm molecular gas mass from the NGLS \co{} maps (H$_2$ [CO\,$J$=3-2]), we have produce maps of the depletion time of the gas presumably closest to the star formation.

The maps of SFR and H$_2$ (CO\,$J$=3-2) gas depletion time for NGC\,628 are shown in Fig.~\ref{fig:ngc0628sf}.  Star formation in this galaxy appears to be mostly concentrated around the spiral arm structure.  In the center of the galaxy the SFR is generally lower compared to the sites of strong star formation in the surrounding arms.  Accordingly this region has a relatively long gas depletion time ($>$1~Gyr) in comparison to much of the rest of the galaxy, though some sites along the arms have similar depletion times.

\begin{figure} 
  \includegraphics[width=85mm]{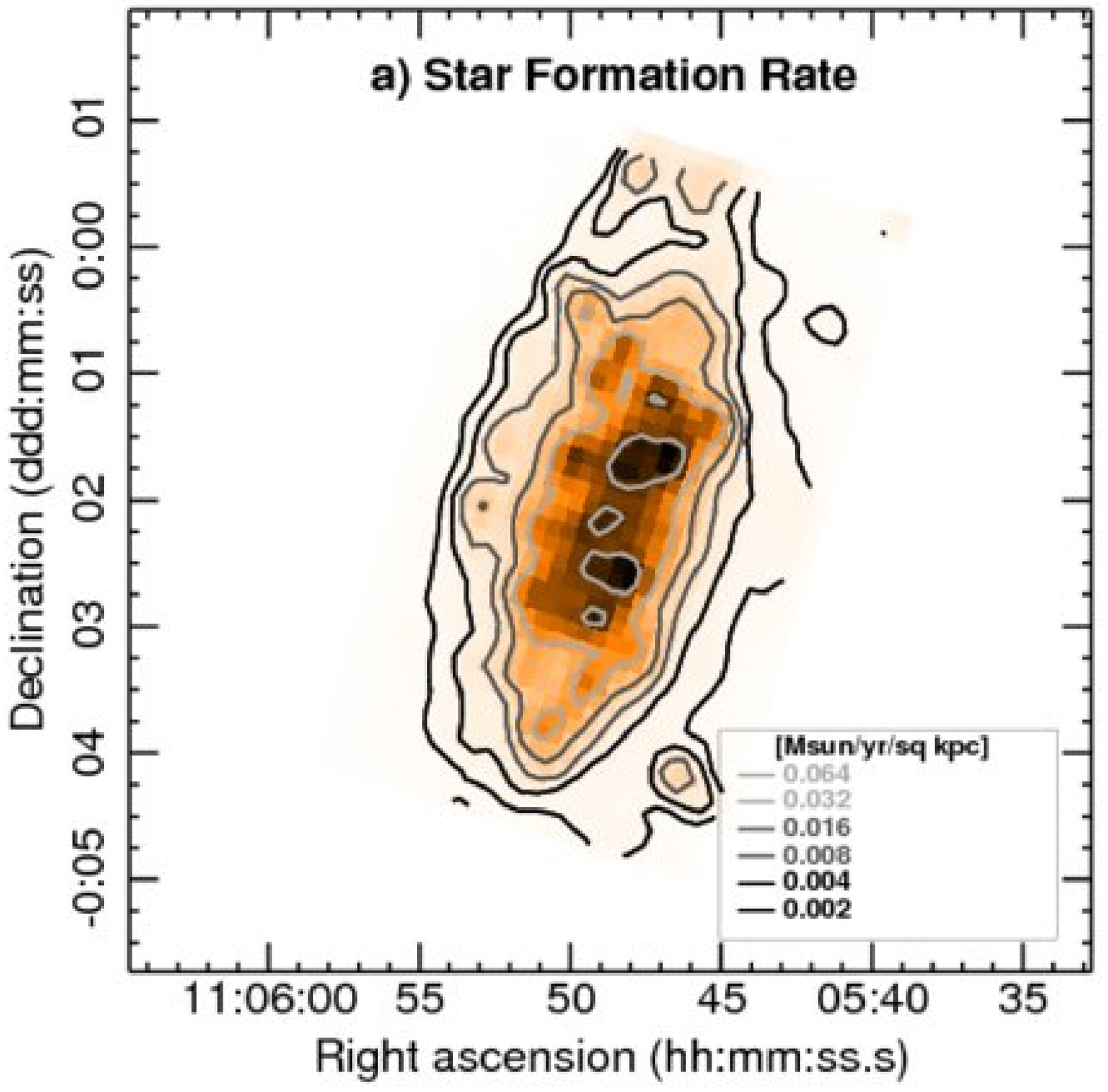}
  \includegraphics[width=85mm]{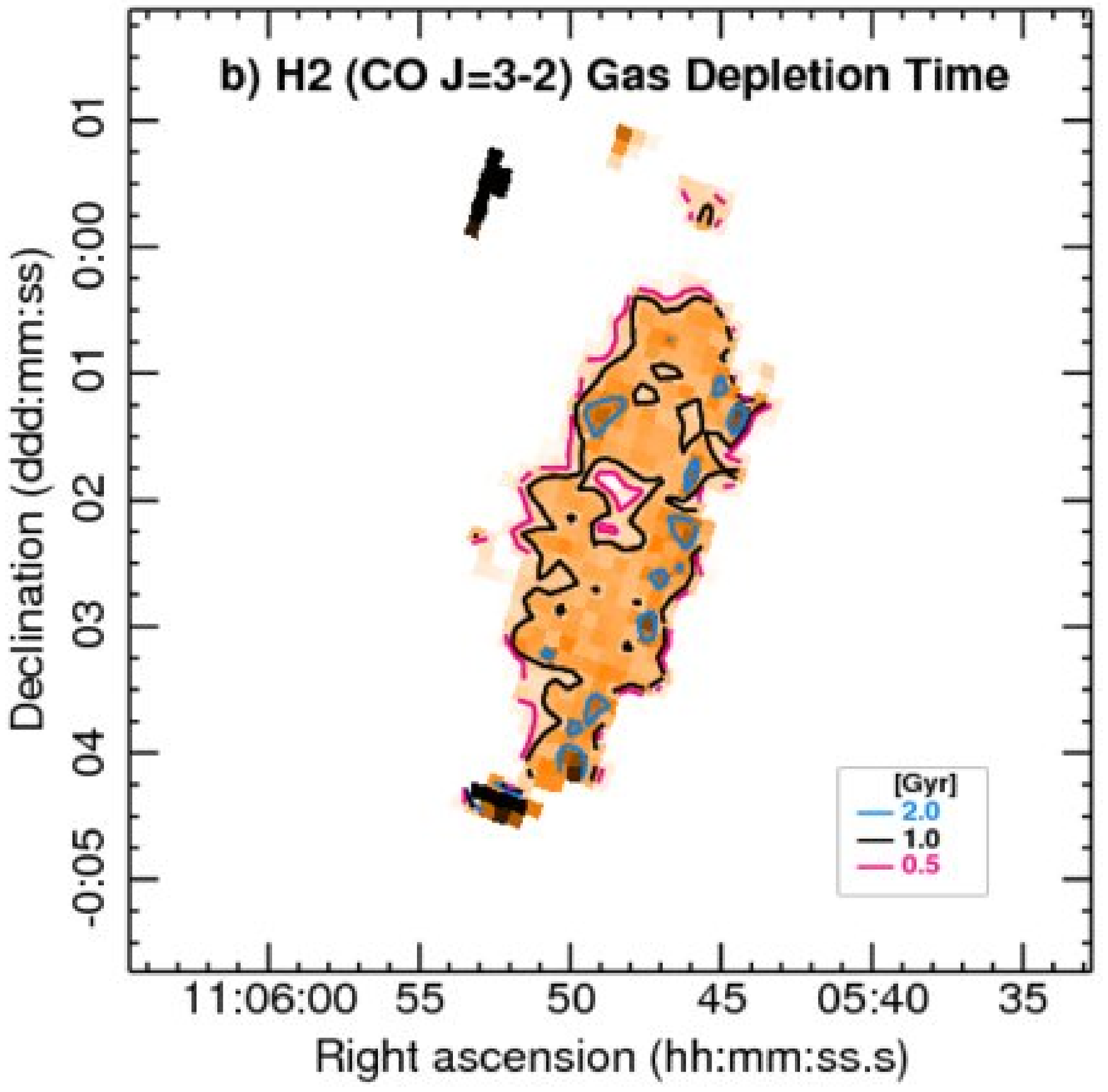}
\caption{a) Star formation rate map for NGC\,3521, derived from the SINGS H$\alpha$ and 24~$\mu$m images \citep{ken03}, convolved to the JCMT beam at 850~$\mu$m.  Contour levels are 0.002, 0.004, 0.008, 0.016, 0.032, and 0.064~${\cal M}_{\sun}\,{\rm yr}^{-1}\,{\rm kpc}^{-2}$.  b) H$_2$ (CO\,$J$=3-2) gas depletion time (gas mass divided by SFR) for NGC\,3521 in Gyr.  The magenta, black, and cyan contours mark where the gas depletion time is 0.5, 1.0, and 2.0~Gyr, respectively.
\label{fig:ngc3521sf}}
\end{figure}

Fig.~\ref{fig:ngc3521sf} shows the SFR and H$_2$ (CO\,$J$=3-2) gas depletion time maps for NGC\,3521.  Star formation appears to be strongly concentrated in the center of this galaxy, though there are likely strong projection effects in this near edge-on galaxy.  The strongest SFR peaks are in the inner $\sim$1\arcmin{} or so ($\sim$3~kpc), the same region in the inner parts of the \co{} intensity map that shows a slight depression.  This distribution is reflected in the gas depletion time map, where the depletion time in the center is less than 0.5 Gyr in some places, whereas some surrounding regions have longer times of 1-1.5 Gyr.  Otherwise the depletion times for dense H$_2$ gas in NGC\,3521 are relatively uniform at $\sim$1.0$\pm$0.5~Gyr.

\begin{figure} 
  \includegraphics[width=85mm]{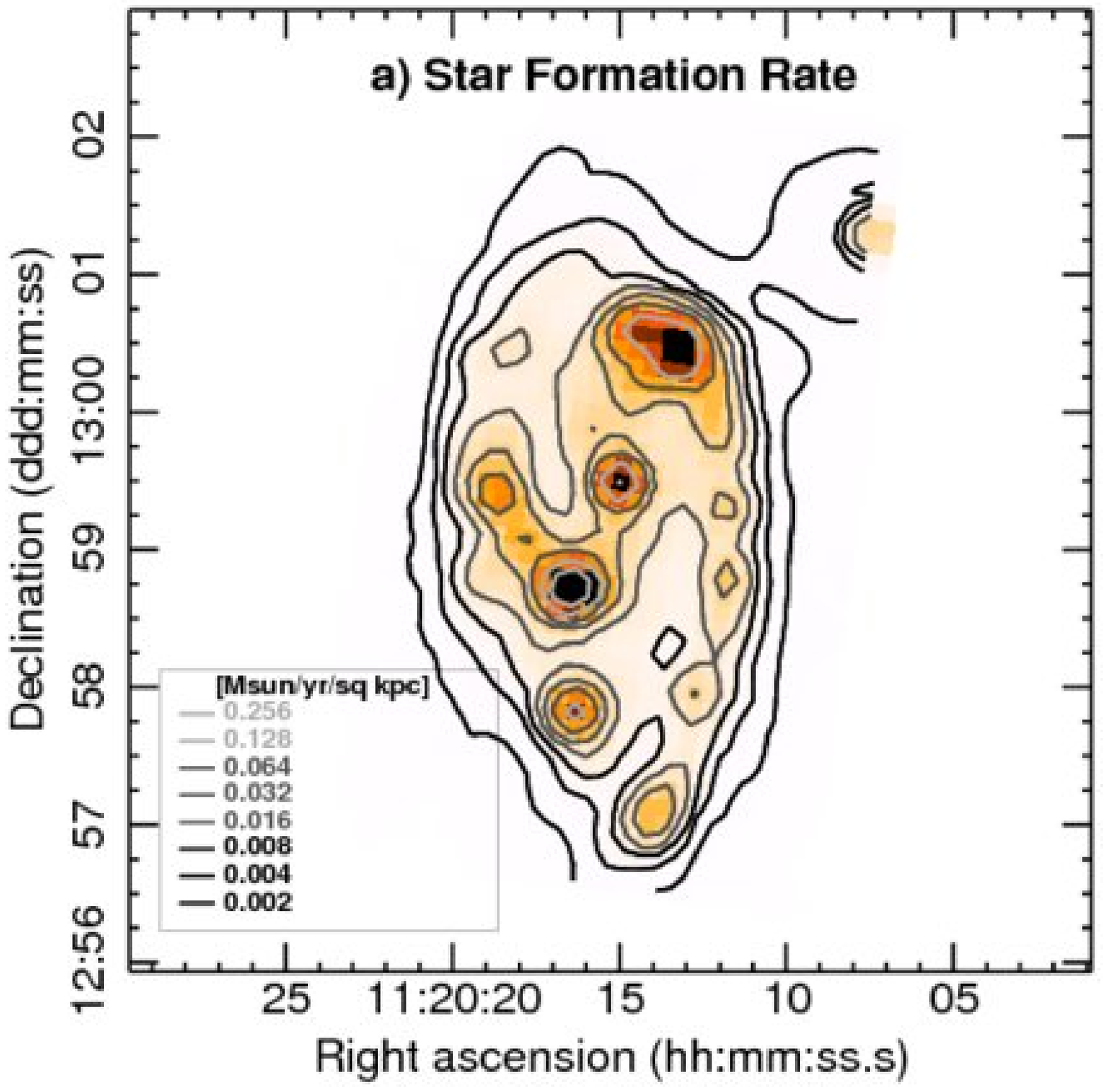}
  \includegraphics[width=85mm]{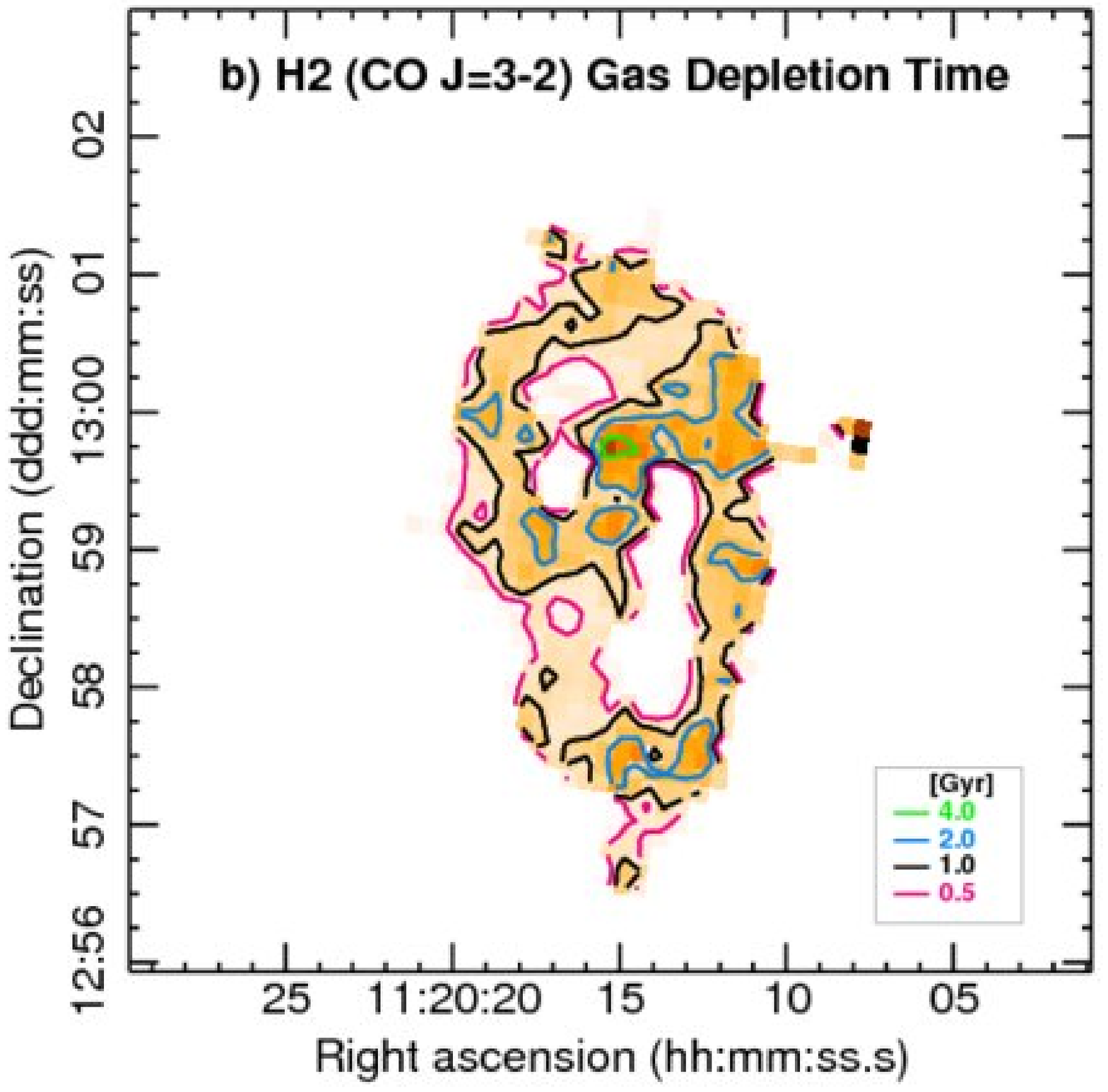}
\caption{a) Star formation rate map for NGC\,3627, derived from the SINGS H$\alpha$ and 24~$\mu$m images \citep{ken03}, convolved to the JCMT beam at 850~$\mu$m.  Contour levels are 0.002, 0.004, 0.008, 0.016, 0.032, 0.064, 0.128, and 0.256~${\cal M}_{\sun}\,{\rm yr}^{-1}\,{\rm kpc}^{-2}$.  b) H$_2$ (CO\,$J$=3-2) gas depletion time (gas mass divided by SFR) for NGC\,3627 in Gyr.  The magenta, black, cyan, and green contours mark where the gas depletion time is 0.5, 1.0, 2.0, and 4.0~Gyr, respectively.
\label{fig:ngc3627sf}}
\end{figure}

Fig.~\ref{fig:ngc3627sf} shows the SFR and H$_2$ (CO\,$J$=3-2) gas depletion time maps for NGC\,3627.  There is intense star formation activity at both ends of the bar in this galaxy, as well as in the center of the bar.  Other sites of strong star formation are scattered along the spiral arms.  What appears to be a star formation peak showing up on the north-western corner of the field, outside the region where we see \co{} and other gas tracers, is not related to NGC\,3627 but is an artifact from an over-exposed foreground star on the H$\alpha$ image.  The gas depletion time map shows that throughout the spiral arms the depletion time is similar to the other two galaxies ($<$2~Gyr).  However, in the middle of the bar around the central starburst region, despite the strong star formation here, the depletion time rapidly lengthens due to the high surface density of H$_2$ (CO\,$J$=3-2) here, and is over 4~Gyrs in the center pixel.  

With the exception of regions like the center of NGC\,3627, the typical H$_2$ (CO\,$J$=3-2) gas depletion times throughout all three galaxies is of the order of 1~Gyr.  Median gas depletion times are 1.1~Gyr for NGC\,628, 1.3~Gyr for NGC\,3521, and 1.2~Gyr for NGC\,3627 (mean values are 1.2~Gyr, 1.8~Gyr, and 1.4~Gyr, respectively).  These are similar to the mean global gas depletion times found for three of the Virgo spiral galaxies studied in \pI{} (1.1 to 1.7~Gyr).  As we discussed in that paper, these times are comparable to those found for the spiral galaxies studied in \citet{ler08} and \citet{big08}, but much shorter than a Hubble time, though as already mentioned they should not be interpreted as the time that the fuel for star formation will be exhausted.  The relative consistency of the global gas depletion times for dense/warm molecular gas between spiral galaxies with very different structures, gas composition, and environment is likely more an indication of the transitional nature of such gas in spirals rather than the properties of the galaxy itself.

\subsection{Star Formation Efficiency Relationships}
\label{sec:ressfsfe}

\begin{figure*} 
  \includegraphics[width=85mm]{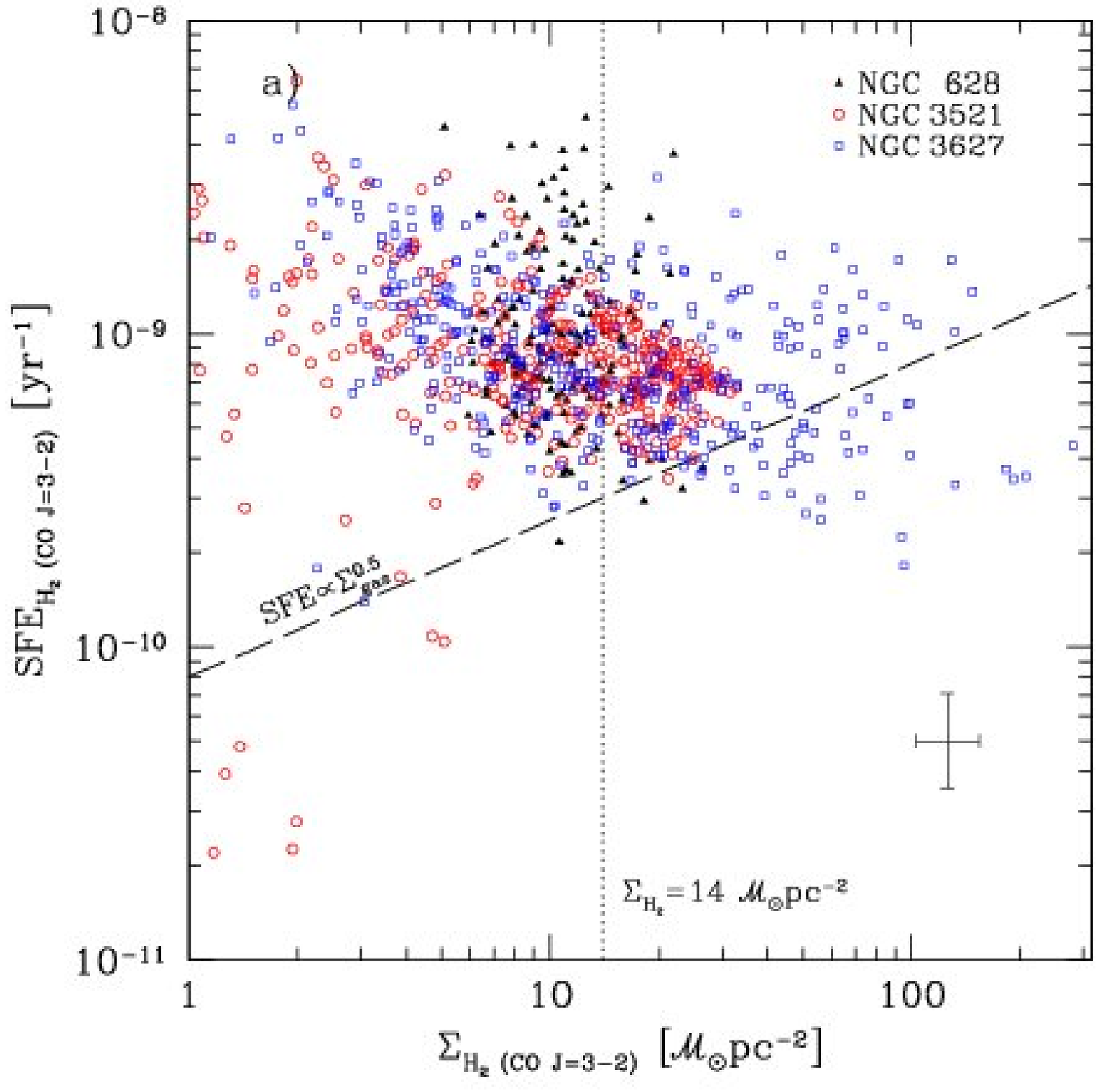}
  \includegraphics[width=85mm]{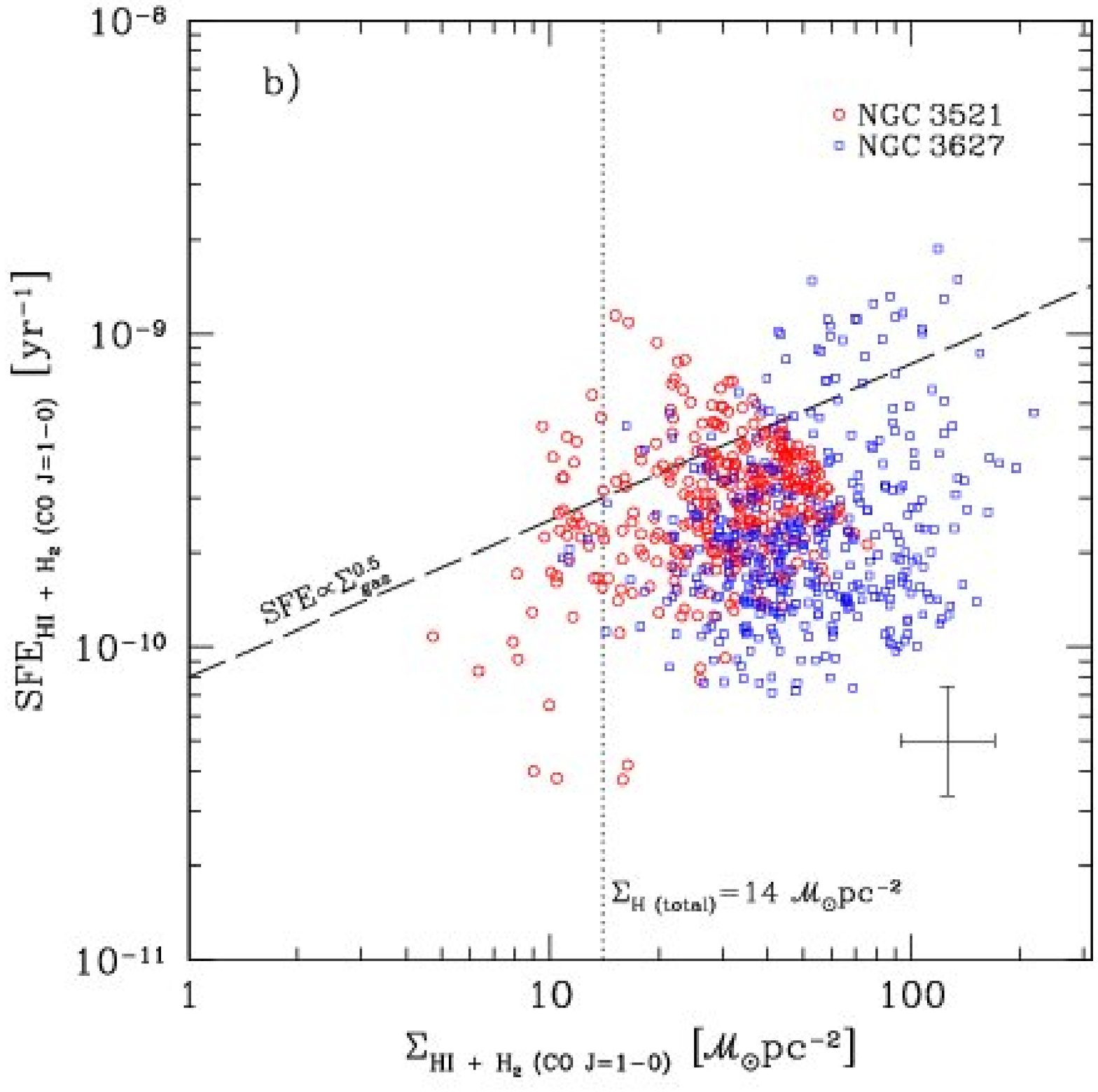}
\caption{Pixel-to-pixel plots of star formation efficiency (inverse of the gas depletion time) versus gas surface density (symbols as in Fig.~\ref{fig:corats}).  a) The H$_2$ (CO\,$J$=3-2) gas star formation efficiency (star formation rate per unit of H$_2$ [CO\,$J$=3-2] gas) versus H$_2$ (CO\,$J$=3-2) gas surface density for all three galaxies.  b) The total gas (\hi{} $+$ H$_2$ [CO\,$J$=1-0]) star formation efficiency versus total gas surface density ($\Sigma_{{\rm HI + H_2 (CO\,J=1-0)}}$) for the two galaxies where we have \citet{kun07} \cooz{} maps available.  In both panels, the lines correspond to the same markings in \citet{ler08} Fig.~5.  The vertical dotted line marks $\Sigma_{gas} = 14$\Msun{}pc$^{-2}$, the \hi{}-to-H$_2$ transition surface density in spirals \citep[$\Sigma_{gas} = 14 \pm 6$\Msun{}pc$^{-2}$][]{ler08}.  The dashed line shows the relationship SFE $\propto \Sigma_{gas}^{0.5}$, an approximation to the Schmidt-Kennicutt star formation law.  All gas surface densities were corrected for helium content, and we applied an inclination correction where appropriate \citep[using the inclinations from][]{wal08,deb08}.  For both plots we have only included pixels where the signal-to-noise for the NGLS \co{} data is greater than 3.  The error bars indicate a typical calibration uncertainty.
\label{fig:sfeden}}
\end{figure*}

Using the \hi{} data from THINGS and other ancillary data from related surveys (SINGS, $^{12}$CO\,$J$=2-1 from HERACLES, and GALEX UV observations), \citet{ler08} have performed an extensive analysis of the star formation efficiency within a sample of nearby galaxies.  This included examinations of various star formation laws, such as the well know Schmidt-Kennicutt law \citep{ken98a,ken98b}.  In Fig.~\ref{fig:sfeden} we have shown two plots based on the SFE versus $\Sigma_{gas}$ plots shown in \citeauthor{ler08} Fig.~5.  We have reproduced these plots for our galaxies on a pixel-to-pixel basis \citep[azimuthally smoothed data are used in ][]{ler08}.  
For reference we have included the same lines shown in \citeauthor{ler08} in both panels, marking the \hi{}-to-H$_2$ transition surface density in spirals (vertical dotted line at $\Sigma_{gas} = 14$\Msun{}pc$^{-2}$, only relevant where \hi{} is included), and the relation SFE $\propto \Sigma_{gas}^{0.5}$ (an approximation to the Schmidt-Kennicutt star formation law).

The left panel (Fig.~\ref{fig:sfeden}a) shows the H$_2$ (CO\,$J$=3-2) gas star formation efficiency (star formation rate per unit of dense H$_2$ gas) versus the H$_2$ (CO\,$J$=3-2) gas surface density for all three galaxies.  The H$_2$ (CO\,$J$=3-2) gas surface density used in both values was estimated using the assumption for dense/warm molecular gas mentioned earlier (\corat{} = 0.6), and hence we are only looking at the efficiency of star formation in the molecular gas regions most closely associated with star formation.  The right panel (Fig.~\ref{fig:sfeden}b) on the other hand show the behavior of the more global gas population.  It shows the total gas (\hi{} $+$ H$_2$ [CO\,$J$=1-0]) star formation efficiency versus total gas surface density ($\Sigma_{{\rm HI + H_2 (CO\,J=1-0)}}$).   Hence much of the gas traced in this plot is much less directly linked to star formation than in the first panel (and does not use our NGLS \co{} data), though the plot is more directly comparable to the results of \citet{ler08}.

Both plots have large scatter and it is difficult to discern clear trends in either.  The left panel shows that the star formation efficiency of H$_2$ (CO\,$J$=3-2) gas is relatively high (as indicated in the previous section by the short gas depletion times).  The efficiency is relatively constant with gas density, and if anything becomes less efficient when the H$_2$ (CO\,$J$=3-2) gas surface density is higher, and it does not follow the line of the Schmidt-Kennicutt law.  However, given the order of magnitude peak-to-peak scatter of the distribution it is difficult to conclude whether the efficiency is constant or declining with gas density.  In the right panel we see the star formation efficiency is about an order of magnitude lower than the first plot, as might be expected when the total gas content is taken into account.  The absence of points at lower densities is likely due to the bias towards regions where \co{} was detected.  There is no clear trend in this plot due to the large scatter and narrow range of gas surface density.  Overall there is little evidence that either plot is following a Schmidt-Kennicutt star formation law.

\begin{figure} 
  \plotone{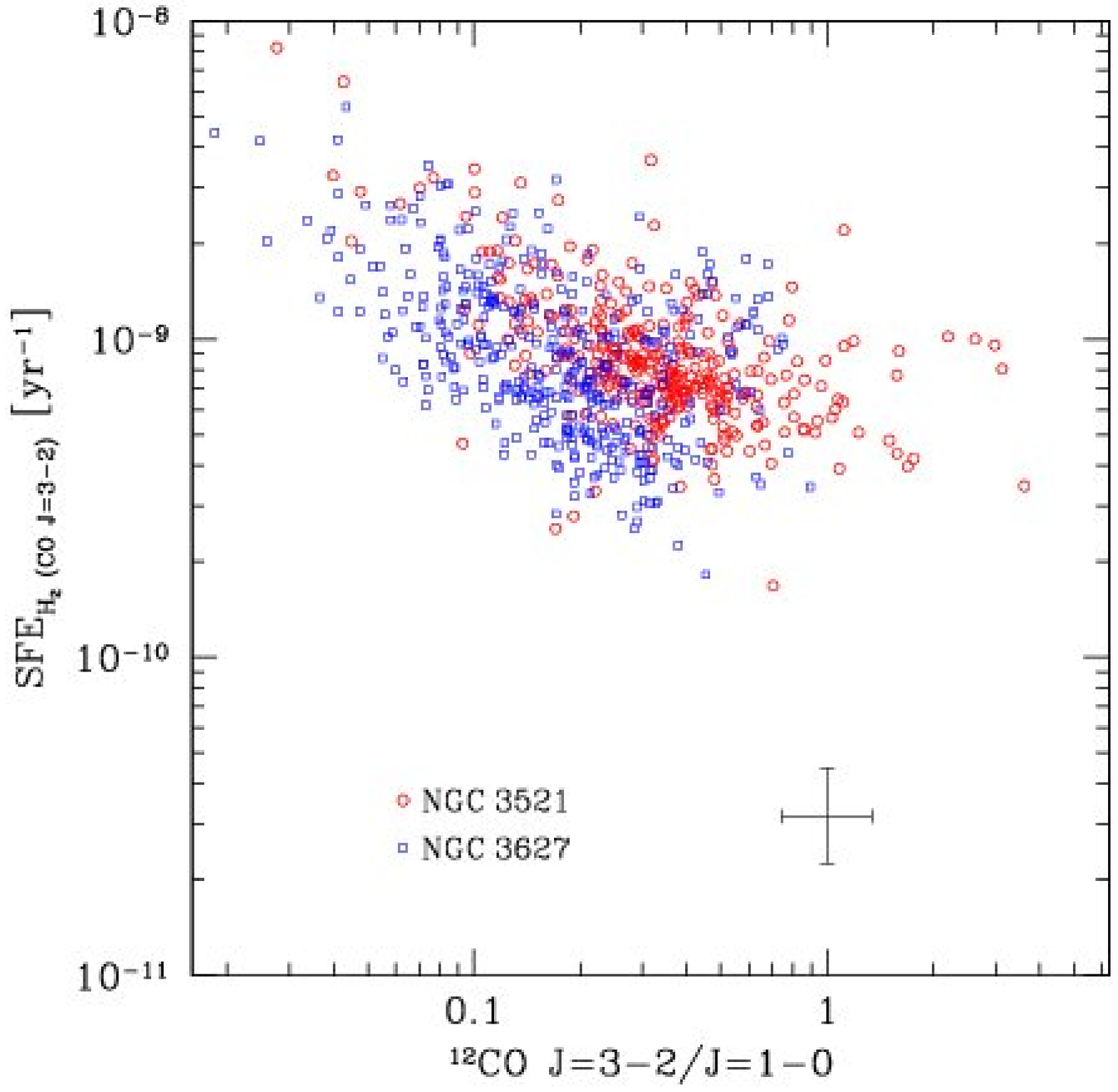}
\caption{Plot of the H$_2$ (CO\,$J$=3-2) gas star formation efficiency (star formation rate per unit of H$_2$ [CO\,$J$=3-2] gas) versus \corat{} for the two galaxies where we have \citet{kun07} \cooz{} maps available (symbols as in Fig.~\ref{fig:corats}).  \corat{} is calculated by dividing our integrated \co{} intensity map (main beam temperature scale) by the corresponding integrated \cooz{} intensity \citep[from][]{kun07}.  The H$_2$ (CO\,$J$=3-2) gas surface density for the star formation efficiency calculation was estimated using the assumption for dense gas (\corat{} = 0.6, see text).  All gas surface densities were corrected for helium content, and we applied an inclination correction where appropriate \citep[using the inclinations from][]{wal08,deb08}.  We have only included pixels where the signal-to-noise for the NGLS \co{} data is greater than 3.  The error bars indicate a typical calibration uncertainty.
\label{fig:sfecorat}}
\end{figure}

Fig.~\ref{fig:sfecorat} shows a plot of SFE$_{\rm H_2 (CO\,J=3-2)}$ versus the \corat{} ratio for the two galaxies where we have \citet{kun07} \cooz{} maps available.  The scatter is again large but maybe a bit tighter than the previous plots, and for the most part the SFE appears to be flat or slightly declining with increasing \corat{}.  
\citet{mur07} also looked at the relationship between SFE$_{\rm H_2}$ and \corat{} for M\,83 (NGC\,5236), and found a clear trend for higher SFE at higher \corat{} ratios.  Their azimuthally smoothed data only cover a small range in \corat{} compared to our pixel-to-pixel measurements, and their SFE$_{\rm H_2}$ range is close to our scatter in that value, but our results do not seem to support this trend.  The most likely explanation from our plots is that the star formation efficiency in theses two galaxies is independent of the \corat{} ratio.

\begin{figure} 
  \plotone{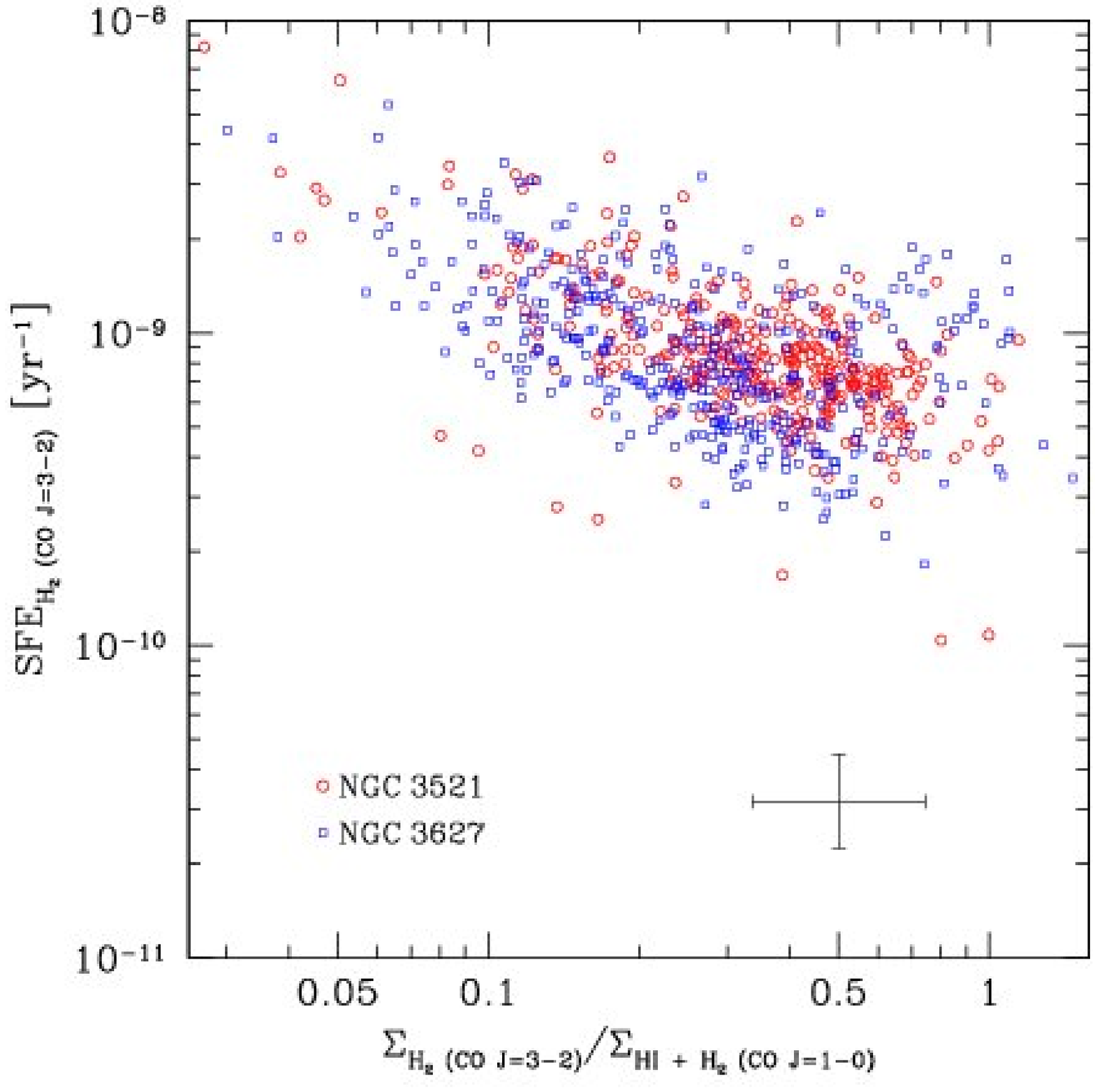}
\caption{Plot of the H$_2$ (CO\,$J$=3-2) gas star formation efficiency (star formation rate per unit of H$_2$ [CO\,$J$=3-2] gas) versus the H$_2$ (CO\,$J$=3-2) gas surface density fraction (\molsd{} divided by \himolsd{}) for the two galaxies where we have \citet{kun07} \cooz{} maps available (symbols as in Fig.~\ref{fig:corats}).  The H$_2$ (CO\,$J$=3-2) gas surface density for the star formation efficiency calculation was estimated using the assumption for dense gas (\corat{} = 0.6).  All gas surface densities were corrected for helium content, and we applied an inclination correction where appropriate \citep[using the inclinations from][]{wal08,deb08}.  We have only included pixels where the signal-to-noise for the NGLS \co{} data is greater than 3.  The error bars indicate a typical calibration uncertainty.
\label{fig:sfeh2frac}}
\end{figure}

Fig.~\ref{fig:sfeh2frac} shows a plot of SFE$_{\rm H_2 (CO\,J=3-2)}$ versus the H$_2$ (CO\,$J$=1-0) gas mass surface density fraction, \moltotsd{}, for the two galaxies where we have \citet{kun07} \cooz{} maps available.  As with the previous plots there is a large scatter in the points close to a magnitude in star formation efficiency.  But as with the previous plot, and the first panel of fig.~\ref{fig:sfeden}, as the surface density of the dense/warm molecular gas increases the star formation efficiency remains constant or becomes less efficient.  This could indicate that in regions of dense and/or warm molecular gas the efficiency is either independent of the gas density, or become slightly less efficient as the density increases.  Investigations with larger samples with extensive multiwavelength data are needed to confirm if either is the case.

\section{Conclusions}
\label{sec:conclusions}

We have used the JCMT to map the \co{} emission from three non-cluster spiral galaxies, completed as part of the ongoing Nearby Galaxies Legacy Survey.  These galaxies are all included in the SINGS survey \citep{ken03}, and  as such have excellent published data sets covering their dust and ISM properties.  Combined with the new NGLS \co{} data, these data set also allow us to probe a range of galaxy properties including differences in the distribution of \co{}, \cooz{}, and \hi{} emission, galaxy dynamics, and relationships between star formation rates and warm/dense molecular gas.

These galaxies all have moderate to strong \co{} detections over large areas of the fields observed by the survey, showing resolved structure and dynamics in their warm/dense molecular gas disks.  We find some large differences between the \co{} and \hi{} velocity fields of some galaxies, in particular NGC\,3521.  These differences appear to be due to the presence of large quantities of low velocity \hi{} gas in the THINGS data cubes.

Global gas depletion times for dense/warm molecular gas in these galaxies, as well as those examined in \pI{} and other surveys, are relatively consistent with each other (in the range 1 to 2~Gyr).  This is despite very different structures, gas compositions, and environments.  This is likely more an indication of the transitional nature of such gas in spirals rather than the properties of the galaxy itself.

Similar to the results from the \citet{ler08}, we do not see any correlation of the star formation efficiency with the gas surface density, with at best a slight decline in efficiency in high gas density regions traced by \co{}.  Additionally, we find that the star formation efficiency of the dense/warm molecular gas traced by \co{} is flat or slightly declining as the \corat{} ratio increases, in contrast to the correlation found in the study by \citet{mur07} into the starburst galaxy M83.  A similar weak trend is also seen when examining star formation efficiency versus the H$_2$ (CO\,$J$=1-0) gas mass surface density fraction.  All three of these could indicate that in regions of dense and/or warm molecular gas the efficiency is either independent of the gas density, or become slightly less efficient as the density increases.

\acknowledgments

We thank the anonymous referee for helping to clarify several issues in this article.  We would like to thank Fabian Walter for making the THINGS \hi{} data cube for NGC\,3521 available to our survey after we initially experienced trouble obtaining it from the web site.  The James Clerk Maxwell Telescope is operated by The Joint Astronomy Centre on behalf of the Science and Technology Facilities Council of the United Kingdom, the Netherlands Organisation for Scientific Research, and the National Research Council of Canada. The research of J.I., K.S., and C.D.W. is supported by grants from NSERC (Canada).  Travel funding for observatory trips by B.E.W. and T.W. was supplied by the National Research Council (Canada).  This research has made use of the NASA/IPAC Extragalactic Database (NED) which is operated by the Jet Propulsion Laboratory, California Institute of Technology, under contract with the National Aeronautics and Space Administration.  
We acknowledge the usage of the HyperLeda database (http://leda.univ-lyon1.fr).  
This work made use of THINGS, `The \hi{} Nearby Galaxy Survey' \citep{wal08}.

\begin{figure} 
  \plotone{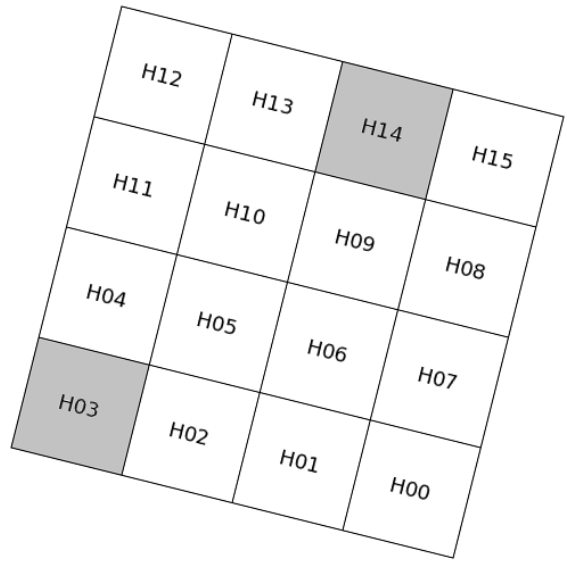}
\caption{The basic layout of the sixteen receptors in the HARP-B array.  Each receptor sits in the center of a square.  The layout is shown tilted by 14\fdg04 clockwise, one of the angles relative to the scan axis used for scanning.  Depending on the part of the sky in which observations are taken, it may also be orientated 14\fdg04 the other way, and the use of a K mirror in the telescope introduces further orientation flips of 90\degr{}.  The two grey receptors (H03 and H14) were missing or unusable during all observation.
\label{fig:harp}}
\end{figure}

\appendix

\section{NGLS Raster Map Observations and their Reduction Strategy}
\label{sec:dred}

This section describes the observing and data reduction methods used for the NGLS HARP-B \co{} raster observations of NGC\,628, NGC\,3521, and NGC\,3627.  This description is also relevant for the reduction of other raster map data taken for the NGLS, such as the Virgo spiral galaxies in \pI{} (NGC\,4254, NGC\,4321, NGC\,4569, and NGC\,4579).  It can also be generalized for jiggle map data observations (which often require less or no flagging).

After a series of Science Verification runs between April and November 2007, the NGLS started taking science data using the 16 element HARP-B array with the backend spectrometer ACSIS \citep{smi03}.  Since that time there has been approximately one dedicated JLS run per month that the three surveys share.  Observations for the three galaxies we are examining here, NGC\,628, NGC\,3521, and NGC\,3627, were taken over multiple JLS runs between November 2007 and March 2008.

The James Clerk Maxwell Telescope (JCMT) is a 15~meter single dish submillimeter telescope on Mauna Kea, Hawai'i.  It is currently the largest single dish submillimeter telescope in the world.  HARP-B is a 16-element sparsely filled array of heterodyne receptors operating in the submillimeter window centered at 850~$\mu$m (350~GHz).  The receptors are arranged in a 4$\times$4 pattern and separated by 30\arcsec{}, approximately two beam-widths ($\sim$14\farcs5 at this wavelength).  See Fig~\ref{fig:harp} for the basic layout of the array (with two receptors that were missing during observing marked).  To produce a fully sampled map we need to set up a Minimum Schedulable Block (MSB) that either scans or jiggles the array around the field.  For all three galaxies in this paper, which have fields larger than the array footprint of 2\arcmin{}, we used a raster scanning technique for mapping.

The default HARP-B raster scanning mode is to rotate the array by $\pm$14\fdg04 to the scan axis and to scan along the long axis of the field, which produces scan rows at 7\farcs2761 intervals.  On the outer edges of the field coverage is sparse where half the array receptors start already within the target field, so we added an additional 60\arcsec{} to the start and end along the direction of the scan to account for this.  There are significant variations in signal-to-noise between receptors in the array, which can produce striping in raster maps where different rows have different sensitivity.  To reduce this effect and produce maps with more even sensitivity coverage we used a `basket weave' technique, scanning along both the major and minor axes of the field.  
Additionally, we found in the science verification phase of the survey that missing and bad receptors in the array caused large variations in our completeness across the target field, and even holes in our coverage where missing/flagged scans crossed in the basket weave.  To counter this problem, we used a half-step scan method, only moving half an array step (58\farcs2) between scan rows/columns.  In this way, each position was covered by two receptors in both scan directions.  However, using the half-step method does require that an additional 60\arcsec{} be added at either end {\em perpendicular} to the scan direction so that all points within the target field have the same integration time.  With the previous addition of 60\arcsec{} along the scan axis, we are adding 60\arcsec{} around all four sides of the field for both scan directions.

Each galaxy's MSB was set up with a sample time per position of 2.5~seconds, so with the half-step basket weave pattern each pointing will have 10~seconds integration time per MSB (if none of the receptors that contribute to that pointing are flagged or missing).  We repeated these MSBs up to seven times until we reached the survey's target RMS (19~mK after binning to 20\kms{} resolution).  The backend spectrometer ACSIS was set up with a bandwidth of $\sim$1~GHz and a resolution of 488.28 kHz (0.423\kms{} at the frequency of the observations), centered at the systemic velocity for the galaxy being observed (where the \co{} line rest frequency is 345.79~GHz).  Appropriate pointing and calibration observations were taken during observing nights.  Individual setup and observing details for the three galaxies in this paper are in summarized Table~\ref{tab:obs}.

Data reduction and most of the analysis were carried out with recent JAC maintained versions of the Starlink software collection (initially the Humu version, and later Lehuakona).  Most of the Starlink tasks we used in the reduction were in the KAPPA (most reduction and analysis tasks) and SMURF (cube making) packages, with some tasks in CUPID (clump finding) and CONVERT (file exporting) also used.  The visualization programs GAIA and SPLAT were used to inspect the raw and processed data.  Additionally, analysis tasks in MIRIAD were used after the creation of cubes and moment maps.  The general reduction procedure was to inspect the raw data files, flag poor data where necessary, combine the raw data into a cube, subtract a baseline from the cube, and collapse it into moment maps.  Details of these steps follow.

Throughout the period that the survey observations were taken, the state of the 16 receptors in the array varied.  Two receptors (H03 and H14) were unavailable for the duration of the project and for the most part were automatically flagged at the telescope, while several other receptors had various, mostly intermittent problems that led to bad baselines, spikes, and false lines.  We used the task chpix (KAPPA) to flag such receptors where required.  Additionally, almost all scans contained a narrow interference spike that was also flagged.  This spike occurred at the same frequency on all receptors (with varying strength) throughout a scan, and would remain at that frequency over the course of a night or even over several runs.  Some scans taken for NGC\,3521 were excluded when pointing observations taken straight after showed significant pointing errors ($>$5\arcsec{}), while all data for NGC\,628 taken on 11 January 2008 were excluded due to poor baselines and high noise on virtually all receptors.  Other observations that did not meet the survey quality assurance criteria (such as high $T_{sys}$ or many receptors with bad baselines) were included, but often with heavy flagging of poor receptors.

After flagging, the raw data were combined into cubes using the task makecube (SMURF).  We produced cubes with 7\farcs2761 pixels, and trimmed the velocity axis to the central $\sim$800\kms{} of the bandpass.  We used a ${\rm sinc}(\pi x){\rm sinc}(k\pi x)$ (`SincSinc') kernel in makecube for the pixel weighting function.  The resulting cube was then trimmed to remove the leading and trailing scan ends outside the target field where our scan coverage is incomplete.  After removing a simple first order polynomial baseline (fit to the line-free ends of the bandpass) and binning to $\sim$20\kms{}, this cube was examined for further bad baselines and to check if we had reached our target depth for the survey.  If necessary the process of flagging and cube making was repeated until we were satisfied that the poor baselines had been removed.

To produce a more accurate fit to the baseline level, we used a method in which we masked out emission in the cubes, and then fit third order polynomial baselines to the unmasked data.  To make the baseline mask, we first smoothed the cube with a [3 pixel, 3 pixel, 25 channel] box filter, and then used the task mfittrend (KAPPA) to make a mask of all of the features in the smoothed cube.  In automatic mode, the `method' parameter in mfittrend allows the choice of several masking methods that have varying success at picking up broad and narrow line features.  As the galaxies we are looking at contain a mixture of narrow and broad \co{} lines, it was important to produce a mask that contains all emission features.

Individual methods can miss some emission, such as the default `Single' method that picks up narrow lines well, but can often miss broad features.  To produce a better mask, we combined the results of two of these methods, `Single' and `Global' (which does a better job on broad lines, but picks up more noise especially at the line-free ends of the bandpass).  We altered this combined mask to leave the line-free regions at the ends of the bandpass unmasked (determined by considering the published width of the \hi{} line and visual examination of the cube).  Additionally, for NGC\,3521 and NGC\,3627, which have broad lines in their centers, we included a block mask over a small central region for the full width of the line so that only the line-free ends would be used for the baseline fit.  The block mask for NGC\,3521 was split into overlapping northern and southern components that cover different velocity ranges.  The final combined masks were applied to the original cube, and a third-order polynomial was fit to the masked cube to produce the baseline fit.  These baselines were subtracted from the original cube to make the final baseline subtracted cube.

Moment maps of the galaxies were made from the baseline subtracted cubes using the Starlink implementation of the clumpfind algorithm \citep{wil94}, the Cupid task findclump, to identify emission above 3$\sigma$.  Sensitivity varied across the field, particularly in the regions outside the target field (see above), so before running it through the moment map procedure we first scaled the cubes by the noise level in order to better identify real clumps.  To do this, we produced a map of the standard deviation in the line-free channels of the cube (using the baseline mask we previously made), and divided the cube by this map.  This scaling was removed again after producing the maps so that the units were correct.

As with the baseline mask steps, we made a smoothed cube with a [3,3,25] box filter.  Using the baseline mask previously made for the baseline fit, we calculated the RMS in the line free regions of the cube.  We then used findclump (Cupid) on the smoothed cube to find all emission greater than 3$\sigma$, and create a mask to exclude the regions without emission.  The clump mask was then applied to the original baseline subtracted cube, and the result was collapsed into two dimensional moment maps using collapse (KAPPA).  We produced maps of the integrated intensity (moment 0), the intensity weighted velocity co-ordinate (moment 1, the velocity field), and the intensity weighted dispersion (moment 2, the velocity dispersion).  We converted the \co{} data from corrected antenna temperature ($\rm T_A^*$) to the main beam temperature scale by dividing by $\eta_{MB}=0.6$.  This correction is different from the value of 0.67 used in \pI{}, and all \co{} measurements from that paper used here are corrected for this difference.  The resulting moment maps are discussed in \S~\ref{sec:resmaps}.


\begin{thebibliography}{}
\bibitem[Bigiel et al.(2008)]{big08} Bigiel, F., Leroy, A. K., Walter, F., Brinks, E., de Blok, W. J. G., Madore, B., \& Thornley, M. D., 2008, \aj, 136, 2846
\bibitem[Bosma(1981)]{bos81} Bosma, A.\ 1981, \aj, 86, 1825
\bibitem[Calzetti et al.(2007)]{cal07} Calzetti, D. et al. 2007, \apj, 666, 870
\bibitem[Chemin et al.(2003)]{che03} Chemin, L., Cayatte, V., Balkowski, C., Marcelin, M., Amram, P., van Driel, W., Flores, H., 2003, \aap, 405, 89
\bibitem[Combes \& Becquaert(1997)]{com97} Combes, F., \& Becquaert, J.-F.\ 1997, \aap, 326, 554
\bibitem[de Blok et al.(2008)]{deb08} de Blok, W. J. G., Walter, F., Brinks, E., Trachternach, C., Oh, S-H., \& Kennicutt, R. C., Jr., 2008, \aj, 136, 2648
\bibitem[de Vaucouleurs et al.(1991)]{dev91} de Vaucouleurs, G., de Vaucouleurs, A., Corwin, H. G., Buta, R. J., Paturel, G., \& Fouqu\'{e}, P., 1991, Third Reference Catalogue of Bright Galaxies, Springer, New York
\bibitem[Freedman et al.(2001)]{fre01} Freedman, W.~L., et al.\ 2001, \apj, 553, 47
\bibitem[Greve et al.(2005)]{gre05} Greve, T. R., et al. 2005, \mnras, 359, 1165
\bibitem[Helfer et al.(2003)]{hel03} Helfer, T. T., Thornley, M. D., Regan, M. W., Wong, T., Sheth, K., Vogel, S. N., Blitz, L., Bock, D. C.-J., 2003, \apjs, 145, 259
\bibitem[Iono et al.(2009)]{ion09} Iono, D., et al.\ 2009, \apj, 695, 1537
\bibitem[Israel(2009a)]{isr09a} Israel, F.~P.\ 2009, \aap, 493, 525
\bibitem[Israel(2009b)]{isr09b} Israel, F.~P.\ 2009, \aap, 506, 689
\bibitem[Kamphuis \& Briggs(1992)]{kam92} Kamphuis, J., \& Briggs, F., 1992, \aap, 253, 335
\bibitem[Karachentsev et al.(2004)]{kar04} Karachentsev, I. D., Karachentseva, V. E.,
    Huchtmeier, W. K., \& Makarov, D. I., 2004, \aj, 127, 2031
\bibitem[Kennicutt et al.(1998a)]{ken98a} Kennicutt, R. C., Jr., et al., 1998a, \araa, 36, 189
\bibitem[Kennicutt et al.(1998b)]{ken98b} Kennicutt, R. C., Jr., et al., 1998b, \apj, 498, 541
\bibitem[Kennicutt et al.(2003)]{ken03} Kennicutt, R. C., Jr., et al., 2003, \pasp, 115, 928
\bibitem[Kennicutt et al.(2008)]{ken08} Kennicutt, R. C., Jr., Lee, J. C., Funes, S. J., J. G., Sakai, S., \& Akiyama, S., 2008, \apjs, 178, 247
\bibitem[Knapen et al.(1993)]{kna93} Knapen, J.~H., Cepa, J., Beckman, J.~E., Soledad del Rio, M., \& Pedlar, A.\ 1993, \apj, 416, 563
\bibitem[Kuno et al.(2007)]{kun07} Kuno, N., et al. 2007, \pasj, 59, 117
\bibitem[Leroy et al.(2008)]{ler08} Leroy, A. K., Walter, F., Brinks, E., Bigiel, F., de Blok, W. J. G., Madore, B., \& Thornley, M. D., 2008, \aj, 136, 2782
\bibitem[Meier et al.(2001)]{mei01} Meier, D.~S., Turner, J.~L., Crosthwaite, L.~P., \& Beck, S.~C.\ 2001, \aj, 121, 740
\bibitem[Muraoka et al.(2007)]{mur07} Muraoka, K. et al. 2007, \pasj, 59, 43
\bibitem[Oka et al.(2007)]{oka07} Oka, T., Nagai, M., Kamegai, K., Tanaka, K., \& Kuboi, N.\ 2007, \pasj, 59, 15 
\bibitem[Regan et al.(2002)]{reg02} Regan, M. W., Sheth, K., Teuben, P. J., Vogel, S. N., 2002, \apj, 574, 126
\bibitem[Reuter et al.(1996)]{reu96} Reuter, H.-P., Sievers, A. W., Pohl, M., Lesch, H., \& Wielebinski, R., 1996, \aap, 306, 721
\bibitem[Schaye \& Dalla Vecchia(2008)]{sch08} Schaye, J., \& Dalla Vecchia, C., 2008, \mnras, 383, 1210
\bibitem[SINGS team(2007)]{sin07} SINGS Team, 2007, Spitzer Infrared Nearby Galaxies
Survey Fifth Data Delivery User's Guide, SSC, Pasadena
\bibitem[Smith et al.(2003)]{smi03} Smith, H., et al., 2003, Proceedings of the SPIE, Vol. 4855, Millimeter and Submillimeter Detectors for Astronomy, ed. Phillips, T. G., \& Zmuidzinas, J., 338
\bibitem[Solomon et al.(1987)]{sol87} Solomon, P.~M., Rivolo, A.~R., Barrett, J., \& Yahil, A.\ 1987, \apj, 319, 730
\bibitem[Strong et al.(1988)]{str88} Strong, A. W., et al. 1988, A\&A, 207, 1 
\bibitem[Tosaki et al.(2007)]{tos07} Tosaki, T., Miura, R., Sawada, T., Kuno, N., Nakanishi, K., Kohno, K., Okumura, S.~K., \& Kawabe, R.\ 2007, \apjl, 664, L27
\bibitem[Walter et al.(2008)]{wal08} Walter, F., Brinks, E., de Blok, W. J. G., Bigiel, F., Kennicutt, R. C., Jr., Thornley, M. D., \& Leroy, A. K., 2008, \aj, 136, 2563
\bibitem[Williams et al.(1994)]{wil94} Williams et al., 1994, \apj 428, 693
\bibitem[Wilson et al.(1999)]{wil99} Wilson, C. D., Howe, J. E., \& Balogh, M. L. 1999, \apj, 517, 174
\bibitem[Wilson et al.(2009)]{wil09} Wilson, C.~D., et al.\ 2009, \apj, 693, 1736
\bibitem[Young et al.(1995)]{you95} Young, J.~S., et al.\ 1995, \apjs, 98, 219 
\bibitem[Zhang et al.(1993)Zhang, Wright, \& Alexander]{zha93} Zhang, X., Wright, M., \& Alexander, P., 1993, \apj, 418, 100


\end{thebibliography}
\end{document}